\documentclass[trackchanges, twocolumn]{aastex7}
\usepackage{amsmath}
\usepackage{makecell}
\usepackage{upgreek}

\begin{document}
\title{Shadow-Induced Warps in Protoplanetary disks
}

\author[orcid=0000-0002-8537-9114,gname=Shangjia, sname='Zhang']{Shangjia Zhang}
\altaffiliation{NASA Hubble Fellowship Program (NHFP) Sagan Fellow}
\affiliation{Department of Astronomy, Columbia University, 538 W. 120th Street, Pupin Hall, New York, NY, 10027, USA}

\email[show]{sz3342@columbia.edu}
\correspondingauthor{Shangjia Zhang}

\author[orcid=0000-0003-3616-6822]{Zhaohuan Zhu}
\affiliation{Department of Physics and Astronomy, University of Nevada, Las Vegas, 4505 S. Maryland Pkwy, Las Vegas, NV, 89154, USA}
\affiliation{Nevada Center for Astrophysics, University of Nevada, Las Vegas, Las Vegas, NV 89154, USA}
\email{zhaohuan.zhu@unlv.edu}

\author[0000-0002-9285-2184]{Callum W. Fairbairn}
\affiliation{Institute for Advanced Study,
1 Einstein Drive,
Princeton NJ 08540, USA}
\email{cfairbairn@ias.edu}

\begin{abstract}

Shadows are commonly observed in protoplanetary disks in near-infrared and (sub)millimeter images, often cast by misaligned inner disks or other obscuring material. While recent studies show that shadows can alter disk dynamics, only the case symmetric across the midplane (e.g., from a polar-aligned inner disk) has been studied. Here we study shadows cast by an inner disk with a $30^\circ$ mutual inclination using 3D radiation-hydrodynamical simulations. Given the same shadow shape and amplitude, the $30^\circ$ inclined shadow leads to a much stronger accretion compared with the polar case, reaching $\alpha \sim$ 1, because the disk is squeezed twice in one azimuth, leading to shocks and strong radial flows near the midplane. The outer disk develops a warp: the inner disk region tilts toward {alignment with} the shadow, while the outer, exponentially tapered disk tilts and twists in a different direction, inclined $\sim$ 32$^\circ$ relative to the inner region. Locally isothermal simulations with a prescribed temperature structure reproduce the effect, confirming that it is thermally driven. Fourier-Hermite analysis shows that it is the m=1, n=1 temperature perturbation that drives the warp by launching bending waves, with the tilting response of the disk approximately proportional to the modal amplitude. This mode always exists unless the shadow is coplanar or polar. Given a fixed temperature contrast, the m=1,n=1 mode peaks at $\sim$15$^\circ$ mutual inclination, but still contributes substantially across 3$^\circ$ to 30$^\circ$. Shadows cause disk warps--{they are} not only a consequence of them. We discuss testable predictions for current and future ALMA and NIR observations.

\end{abstract}



\section{Introduction}
\label{sec:intro}
Protoplanetary disks consist of gas and dust orbiting pre-main-sequence stars. For most regions of a disk over its lifetime, the dominant heating source is irradiation from the central star(s). This produces a well-established thermal structure: a superheated surface layer overlying a cooler midplane \citep{calvet91, chiang97, dalessio98}. Unless the inner disk is depleted of dust, as in transition disks, stellar photons directly heat only the surface layers and cannot penetrate to the midplane. Instead, the midplane remains in shadow and is warmed indirectly by radiation reprocessed and re-emitted from the surface. This structure is vividly illustrated in edge-on systems: when the Hubble Space Telescope first imaged and confirmed the existence of such disks, dark midplane lanes in silhouettes flanked by bright reflection nebulae were already evident \citep{McCaughrean96, Burrows96}.



Strikingly, recent high-resolution imaging with the James Webb Space Telescope has revealed that many edge-on disks display lateral asymmetries between their upper and lower reflection nebulae (e.g., 15 out of 20 cases in the sample of \citealt{villenave24}), suggesting asymmetric stellar irradiation across the shadowed midplane. Such asymmetry is perhaps unsurprising in light of the extensive observations of more face-on disks, where non-coplanar shadows are frequently seen \citep[e.g.,][]{benisty22}. Narrow shadows—two lanes separated by 180$^\circ$—have been identified in systems such as HD 142527 \citep{avenhaus17, hunziker21}, HD 100453 \citep{benisty17}, RX J1604.3–2130 A \citep{pinilla15}, DoAr 44 \citep{avenhaus18}, SU Aur \citep{ginski21}, GG Tau A \citep{keppler20}, CV Cha, SY Cha \citep{ginski24}, HD 135344B \citep{stolker16}, and CQ Tau \citep{uyama20, safonov22}. Wide shadows—broad obscurations covering roughly half the disk—have been observed in V1247 Ori \citep{ohta16, kraus17}, ZZ Tau IRS \citep{hashimoto24}, TW Hya \citep{debes23}, HD 139614 \citep{muro-arena20}, HD 169142 \citep{bertrang18}, HD 143006 \citep{benisty18}, PDS 66 \citep{wolff16}, HD 163296 \citep{rich19}, MWC 758 \citep{grady13, benisty15, ren18}, and V1098 Sco \citep{williams25}.

Illumination asymmetry can arise either from asymmetric attenuation—such as a misaligned inner disk \citep{marino15}—or from intrinsically asymmetric central emission, such as stellar cool spots and accretion hotspots \citep{wood98}. In the past decade, the misalignment scenario has received greater attention. In relatively face-on disks, small misalignments produce broad shadows, whereas larger misalignments give rise to two narrow shadow lanes \citep{marino15, facchini18}. In edge-on disks, the degree of misalignment instead governs the lateral brightness asymmetry between the two reflection nebulae {\citep{juhasz17, kimmig25}}.

A natural question arising from these observations is how asymmetric irradiation influences disk evolution—a topic that has received increasing attention in the context of misaligned disks. Polar shadows have been shown to launch spiral arms, drive accretion, transport angular momentum, carve concentric gaps and rings, trigger vortices \citep[][]{montesinos16, montesinos18, cuello19, su24, zhang24b, zhu25, ziampras25}, and even excite disk eccentricity \citep{qian24}.

These phenomena already constitute a zoo of dynamical consequences awaiting tests from rich scattered-light and ALMA observations \citep[e.g.,][]{andrews20, bae22}—especially from the ever more sensitive kinematic datasets such as exoALMA \citep{teague25}. Yet, as we will show in this paper, shadows can drive even more intriguing dynamical effects that have so far gone unnoticed because all existing 2D and 3D models assume vertical symmetry with respect to the disk midplane {(this effectively means either polar or coplanar configurations between the inner disk casting the shadow, and the outer disk)}. For a generic inclined shadow—neither coplanar nor polar—the thermal response must instead be evaluated over the full solid angle.

We investigate this problem using 3D radiation–hydrodynamical simulations with shadows cast by a $30^\circ$ inclined inner disk. This configuration naturally generates an antisymmetric temperature perturbation across the midplane, which in turn drives a large-scale warp. To our knowledge, this represents the first discovery of a thermally driven warp in radiation–hydrodynamical simulations. Section \ref{sec:methods} describes our methods, Section \ref{sec:results} presents the main results, and Section \ref{sec:discussion} discusses the conditions for this mechanism to occur (Section \ref{sec:condition}), existing puzzles (Sections \ref{sec:why1} and \ref{sec:why2}), and its observational implications (Sections \ref{sec:observation}). We conclude in Section \ref{sec:conclusion}.







\section{Methods}
\label{sec:methods}

{We focus on the outer disk regions where shadows are cast. Our main radiation-hydrodynamical simulations (Sections \ref{sec:ray_tracing} and \ref{sec:opacity}) model a transition disk with a cavity of $\sim$160 au and an exponential outer cutoff, resulting in a density peak at 160 au (Section \ref{sec:transition_disk}). The simulation domain extends from 21.6 au to 1260 au, providing ample space to capture both the inner cavity and the outer cutoff regions. We also perform pure hydrodynamical simulations with prescribed temperature structures to elucidate the physical mechanisms driving disk warping in the more complex radiation-hydrodynamical runs (Section \ref{sec:prescribed_temperature}). These include simulations with the same grid setup as the radiation-hydrodynamical model, as well as full-disk runs adopting power-law radial density profiles without an inner cavity or exponential cutoff (Section \ref{sec:full_disk}).}
These models are summarized in Table~\ref{tab:models}.

All simulations are carried out with the Athena++ code \citep{stone20}.  For the radiation--hydrodynamical simulations, we use the implicit radiation module \citep{jiang14, jiang21}, which solves the time-dependent and angle-dependent radiative transfer equations with implicit methods. This approach allows us to accurately capture radiation transport across both optically thin and optically thick regimes, including effects such as shadowing and beam crossing.  

It should be noted that throughout the paper we employ three different coordinate systems. Our simulations are performed in spherical polar coordinates $(r,\theta,\phi)$. The disks are set up and in part analyzed (e.g., measuring accretion rate) in cylindrical coordinates $(R,\phi,z)$. Other quantities, such as angular momentum vectors, are calculated in Cartesian coordinates $(x,y,z)$.

Since the disk develops a warp with varying tilt and twist across radii, we analyze each spherical radius in a locally rotated frame in which the angular momentum vector defines the new ${z}'$ axis. Quantities in this transformed frame (e.g., velocity components) are denoted with a prime.

\begin{deluxetable*}{llllll}
\tabletypesize{\scriptsize}
\tablecaption{Summary of simulations.\label{tab:models}}
\tablehead{
  \colhead{Run ID} &
  \colhead{Disk Type} &
  \colhead{Tilt of Shadow} &
  \colhead{Amplitude} &
  \colhead{Boundary Condition} &
  \colhead{Figures}
}
\startdata
\cutinhead{Radiation-hydro}
\texttt{R30} & Transition disk & $30^\circ$ & --- & \makecell[l]{Modified outflow} & \ref{fig:warp_surface_evolution},\ref{fig:warp_properties_evolution},\ref{fig:accretion},\ref{fig:phi_theta_view},\ref{fig:cart_slices} \\
\texttt{R90} & Transition disk & $90^\circ$ & --- & \makecell[l]{Modified outflow} & \ref{fig:accretion},\ref{fig:phi_theta_view} \\
\cutinhead{Pure hydro (prescribed $T$)}
\texttt{H30} & Transition disk & $30^\circ$ & 60\%  & \makecell[l]{Modified outflow} & \ref{fig:hydro_warp_evolution}, \ref{fig:compare_accretion_rad_hydro}\\
\texttt{HF30Mo} & Full disk & 30$^\circ$ & 20\%  & \makecell[l]{Modified outflow} & \ref{fig:warp_evolution_30deg_different_bc} \\
\texttt{HF30R} & Full disk & 30$^\circ$ & 20\% & Reflecting & \ref{fig:warp_prop_evolution_hydro_smaller}, \ref{fig:residual_warp_surface_hydro}, \ref{fig:tilt_period} \\
\texttt{HF7R} & Full disk & 7.5$^\circ$ & 20\%  & Reflecting & \ref{fig:warp_prop_evolution_hydro_smaller}, \ref{fig:similarity_7p5_m1n1} \ref{fig:residual_warp_surface_hydro}, \ref{fig:tilt_period} \\
\texttt{HFm1n1R0p054} & Full disk & $m=1,n=1$ & 5.4\%  & Reflecting & \ref{fig:warp_prop_evolution_hydro_smaller}, \ref{fig:similarity_7p5_m1n1}, \ref{fig:tilt_period} \\
\texttt{HFm1n1R0p01} & Full disk & $m=1,n=1$ &  1\%  & Reflecting & \ref{fig:inner_outer_twist}, \ref{fig:tilt_period}\\
\texttt{HFTm1n1R0p01} & Full disk + outer cutoff & $m=1,n=1$ &  1\%  & Reflecting & \ref{fig:inner_outer_twist}, \ref{fig:tilt_period} \\
\enddata
\end{deluxetable*}

\subsection{Shadowing in Ray-Tracing}
\label{sec:ray_tracing}

Our radiation-hydrodynamical simulations are designed to model transition disks, since these are the types of disks in which shadows are most frequently observed \citep{benisty22}. A typical transition disk consists of an often unresolved inner disk and a resolved outer disk \citep{vandermarel23}. Our setup dynamically tracks the outer disk, while the effects of the inner disk are simply prescribed by an attenuation model, which blocks the stellar irradiation. We modify the ray-tracing of the stellar irradiation from the previous model of \citet{zhang24b} for polar shadows by considering the attenuation of a realistic inner disk between 5-10 au. {At this size, the inner disk remains unresolved in most ALMA observations, or at best is marginally resolved within one to two beam widths \citep{francis20}. At sub-au scales, the inner disk structure can be reconstructed from NIR interferometry \citep{bohn22, setterholm25, codron25}, yet the detailed structure is unclear. In our simulation, we adopt a specific size and density profile for the inner disk, but this choice primarily serves to capture the generic behavior of the optical depth structure caused by the inner disk, i.e., highest at the disk midplane and becoming optically thin several scale heights above it.} In addition, we also generalize the shadow by the inner polar disk \citep{zhang24b} to a more generic shadow by any inclined inner disk {and focus on a 30$^\circ$ misalignment.}

The shadow is included as part of the attenuated stellar irradiation, which is given by
\begin{align}
\mathbf{F}_*(r, \theta, \phi) = &\left(\frac{R_*}{r}\right)^2 \sigma_b T_*^4 \nonumber \\ 
&\times \exp\big(-\tau_*(r,\theta,\phi, t)\big) \nonumber \\
&\times \mathcal{R}(\theta,\phi, t)\hat{\mathbf{r}},
\label{eq:irradiation}
\end{align}
where the first line gives the unattenuated stellar flux, with $r$ the distance to the star, and $T_*$ and $R_*$ the stellar surface temperature and radius, respectively, for which we adopt solar values. $\sigma_b$ denotes the Stefan–Boltzmann constant.  
The second line accounts for attenuation by the outer disk itself, where $\tau_*$ is the optical depth at a given cell, calculated by ray-tracing the line between the star and that position {at each hydrodynamical timestep} \citep{zhang24}.  
The third line represents additional attenuation due to the shadow cast by the inner disk {that is gradually introduced}.



The ramp-up function $\mathcal{R}(\theta,\phi, t)$ helps to connect the shadow-free initial condition to the full attenuation when the shadow reaches its full amplitude $\exp(-\tau_{*,\mathrm{bc}}(\theta,\phi))$ by
\begin{align}
    \mathcal{R}(\theta,\phi, t) = 1 - [1 - \exp(-\tau_{*,\mathrm{bc}}(\theta,\phi))]\,S(t),
\end{align}
where $\tau_{*,\mathrm{bc}}(\theta,\phi)$ is the optical depth contributed by the inner disk, and
\begin{align}
    S(t) = \sin^2\!\Bigg[\frac{\pi}{2}\frac{\min\{\max[0,\,t-t_\mathrm{relax}],\,t_\mathrm{grow}\}}{t_\mathrm{grow}}\Bigg],
    \label{eq:smooth_transition}
\end{align}
is a taper function to gradually introduce the shadow from $\tau_{\mathrm{*,bc}}=0$ (or $\mathcal{R} = 1$, no attenuation from the inner disk) to its full amplitude, {since the shadow does not appear instantaneously in observations. Numerically, this avoids stiff changes at a single timestep.} The parameter $t_\mathrm{relax}$ controls when to turn on the shadow and $t_\mathrm{grow}$ controls how quickly the shadow grows to its maximum amplitude. $t_\mathrm{relax}$ is set to 18 P$_0$ and $t_\mathrm{grow}$ to 10 P$_0$, where P$_0$ ($\approx$ 253 yr) is the orbital period at reference radius R$_0$ (= 40 au). Note that in the rest of the paper, the time origin is reset so that $t=0$ when the shadows initially start to grow, after the initial disk hydrostatic relaxation $t_\mathrm{relax}$.

When the shadow is fully turned on, the inner disk's optical depth,
\begin{align}
    \tau_{*,\mathrm{bc}}(\theta, \phi) = &\tau_{*,\mathrm{bc}}^{\max} \exp\!\left[-\frac{\Big(\theta_s (\theta, \phi) - \tfrac{\pi}{2}\Big)^2}{\sigma^2}\right],
    \label{eq:amplitude}
\end{align}
which represents the vertical dependence of the optical depth that can be fitted with a Gaussian-like functional form. Here, $\tau_{\mathrm{*,bc}}^{\max}$ is the radial optical depth integrated between 5 and 10 au at the inner disk midplane. The best-fit parameters for the optical depth integrated in the radial direction {along the $\theta_s$ direction (Equation \ref{eq:new_polar_angle}) at its full amplitude, $\tau_{*,\mathrm{bc}}(\theta_s)$}, are $\tau_{\mathrm{*,bc}}^{\max} = 6000$ and $\sigma = 0.09$. The optical depth reaches unity when $|\theta_s - \pi/2| = 0.27$, i.e., three times $\sigma$. To calculate the inner disk attenuation, we assume the inner disk follows a power-law surface density profile,
\begin{equation}
    \Sigma_g = \Sigma_{g,0} \left(\frac{R}{R_0}\right)^{-1},
    \label{eq:full_disk_density}
\end{equation}
with reference radius $R_0 = 40$ au and normalization $\Sigma_{g,0} = 3$ g cm$^{-2}$. The inner disk is confined between 5 and 10 au. The temperature scales as $T \propto R^{-0.5}$, and the disk aspect ratio is $h/r = 0.1$ at 40 au, corresponding to $h/r \sim 0.07$ at 10 au, where $h$ is the gas scale height. We assume that the small dust has the same spatial distribution as the gas. The adopted dust-to-gas ratio and opacity are described in Section \ref{sec:opacity}.

The angle $\theta_s$ is defined as the polar angle $\theta$ of the shadow with respect to the inner disk's angular momentum direction, obtained through an Euler rotation. Specifically, we assume that the inner disk’s $z$-axis is tilted by an angle $i$ toward the $y$-axis while keeping the $x$-axis fixed, yielding
\begin{equation}
    \theta_s = \arccos \!\left( \sin i \, \sin \theta \, \sin \phi + \cos i \, \cos \theta \right).
    \label{eq:new_polar_angle}
\end{equation}
For a shadow cast by a coplanar inner disk, such as in a standard two-temperature disk \citep{chiang97}, $i=0^\circ$, $\theta_s = \theta$, while for a shadow cast by a polar inner disk \citep{zhang24b}, $i=90^\circ$, $\theta_s = \arccos(\sin\theta\sin\phi)$. In the current work, we adopt a generic case with $i=30^\circ$ (\texttt{R30}). We also run a $i=90^\circ$ case (\texttt{R90}) for reference. 

\subsection{Radiation Transport}
\label{sec:opacity}
Our radiation transport approach closely follows the methodology described in \citet{zhang24}. Below we will briefly recap the setup, but refer the reader to this previous work for more details.
For radiation-hydrodynamical simulations, we use the DSHARP composition \citep{birnstiel18} and a power law MRN dust size distribution \citep[$n(a) \propto a^{-3.5}$,][]{mathis77} to calculate opacity. We assumed that only small grains determine the temperature distribution due to their high opacity at the peak of the stellar spectrum and the fact that mm-sized particles are settled at the thin midplane in most of the ALMA observations. {Therefore, we consider grains with sizes between 0.1 and 1~$\mu$m, which account for $f_\mathrm{s} = 0.02184$ of the total dust mass, assuming a dust size distribution with a minimum grain size of $a_\mathrm{min} = 0.1~\mu$m and a maximum grain size of $a_\mathrm{max} = 1$~mm.} The mass ratio between all the dust and gas is assumed to be 0.01. {Consequently, small grains account for 2.184$\times$10$^{-4}$ of the gas mass}. The opacity used for ray-tracing at the stellar effective temperature is $\kappa_* = 3995~\mathrm{cm}^2~\mathrm{g}^{-1}$. In contrast, the opacities relevant for disk thermal emission are temperature-dependent Planck- and Rosseland-mean opacities and typically on the order of $\sim 10~\mathrm{cm}^2~\mathrm{g}^{-1}$ for characteristic disk temperatures (see Figure 1 of \citet{zhang24}). All these values are normalized to the total dust mass.

While stellar irradiation is prescribed as in Equation \ref{eq:irradiation} through radial ray-tracing, the radiation transport of disk thermal emission is handled using the discrete ordinate method, in which rays are discretized into angular bins. This effectively introduces two frequency groups in the simulation. One for stellar irradiation (UV/optical) and one for disk thermal emission (IR/mm). We adopt the discretization scheme optimized for curvilinear coordinates (\texttt{angle\_flag} = 1) and set \texttt{nzeta} = 2 and \texttt{npsi} = 2. Here, \texttt{nzeta} samples angles from $0$ to $\pi/2$ in the $\upzeta$ (polar) direction, and \texttt{npsi} samples angles from $0$ to $\pi$ in the $\uppsi$ (azimuthal) direction. The angles $\upzeta$ and $\uppsi$ correspond to the polar and longitudinal directions with respect to the local coordinate system, meaning that they vary spatially across the grid \citep{jiang21}.

\subsection{Temperature Setup for Pure Hydrodynamical Simulations}
\label{sec:prescribed_temperature}
We also carry out hydrodynamical simulations with fixed temperature structures that mimic the inner disk's shadow. {We adopt an adiabatic equation of state with orbital cooling following the prescription of \citet{zhang24}. The gas internal energy relaxes toward the prescribed temperature on a timescale of $\beta_c \Omega^{-1}$. With $\beta_c = 10^{-6}$, the disk cools essentially instantaneously, making it effectively locally isothermal.} 

Motivated by the disk’s thermal response to inner disk attenuation, we adopt a temperature shape function similar to the role of the third line of the Equation~\ref{eq:irradiation}. On a vertically isothermal, radially power-law temperature background, we prescribe a temperature modification $f(\theta_s, t)$,
\begin{align}
    T(r, \theta, \phi, t) = &\, T_{\mathrm{bg}}(R) \nonumber\\
         & \times f(\theta_s, t).
         \label{eq:prescribedT}
\end{align}
The background profile is
\begin{equation}
    T_{\mathrm{bg}}(R) = T_0 \left(\frac{R}{R_0}\right)^{-s},
    \label{eq:bkgT}
\end{equation}
where $s$ is the radial temperature slope. We adopt $s = 0.5$.
The temperature modification is
\begin{equation}
    f(\theta_s, t) = 1 + A(t)\, g(\theta_s),
    \label{eq:temperature_modification}
\end{equation}
where $A(t)$ is a time-dependent amplitude function,
\begin{equation}
    A(t) = A_0 \sin^2\!\Bigg[\frac{\pi}{2}\frac{\min\{\max[0,\,t-t_\mathrm{relax}],\,t_\mathrm{grow}\}}{t_\mathrm{grow}}\Bigg],
    \label{eq:amplitude2}
\end{equation}
similar to the form defined in Equation \ref{eq:smooth_transition}. The perturbation can be activated at time $t_{\mathrm{relax}}$ and grows to its full amplitude within $t_{\mathrm{grow}}$.

The temperature shape function $g(\theta_s)$ describes the transition from the cold midplane to the superheated surface:
\begin{equation}
\begin{aligned}
g(\theta_s) = 
\begin{cases}
-1, \\[-1ex]
\quad \text{if } |\theta_s - \frac{\pi}{2}| < \sigma_s, \\[1ex]
-\cos\left[ \pi \frac{|\theta_s - \frac{\pi}{2}| - \sigma_s}{\sigma_t} \right], \\[-1ex]
\quad \text{if } \sigma_s \leq |\theta_s - \frac{\pi}{2}| < \sigma_s + \sigma_t, \\[1ex]
1, \\[-1ex]
\quad \text{if } |\theta_s - \frac{\pi}{2}| \geq \sigma_s + \sigma_t
\end{cases}
\end{aligned}
\label{eq:angular_dependence}
\end{equation}
where $\sigma_s$ is the width of the constant-temperature shadow region, and $\sigma_t$ is the width of a smooth transition from the cold midplane to the hot surface. We adopt $\sigma_s$ = 0.05 and $\sigma_t$ = 0.35, allowing a smoother and more gradual transition than in the radiation-hydrodynamical simulations (where the temperature changes sharply at the $\tau=1$ surface located 0.27 radians away from the midplane). Beyond this angle, the hot surface also has a constant temperature. {The reason for prescribing a smooth temperature transition is to test whether the warp developed in the radiation-hydrodynamical simulations (Section \ref{sec:warp_accretion}) is induced by the sharp temperature gradient. By adopting a much smoother transition, we still find a similar warp evolution (Section \ref{sec:thermally_induced}), thereby ruling out this hypothesis.} The amplitude $A_0$ can take values between $-1$ and $1$. When $A_0 = 0$, the disk is vertically isothermal, i.e., it follows the background temperature T$_{bg}(R)$. For $A_0 > 0$, the midplane temperatures are reduced due to the inner disk attenuation. When the shadow is coplanar to the outer disk midplane, this temperature profile is similar to the one used in previous studies \citep[e.g.,][]{yun25a, yun25b}. When the shadow is polar, the temperature profile becomes the shadow shape prescribed in \citet{zhu25}. Although not explored in the current paper, negative amplitudes ($A_0 < 0$) are still meaningful in this context, as they would correspond to a colder surface and hotter midplane in an actively accreting disk. {One can also prescribe any other forms of temperature structure not limited to Equation \ref{eq:angular_dependence} to include both stellar irradiation heating and viscous heating.} We have four simulations under this setup, with $A_0 = 60\%$ and $20\%$ and inclinations of $i = 30^\circ$ and $7.5^\circ$ (\texttt{HF30}, \texttt{HF30Mo}, \texttt{HF30R}, and \texttt{HF7R}, as listed in Table \ref{tab:models}). {Since we run simulations with vertically isothermal setups during t$_{\mathrm{relax}}$, the snapshot taken  before introducing the shadow can be used as references for cases without shadows (i.e., $A_0 = 0\%$).}

{Once the shadow reaches its full amplitude, the temperature structure is fixed in time. This is a reasonable approximation when the outer disk is optically thin to stellar irradiation, as in our simulations, where self-shadowing is negligible. However, when the outer disk becomes optically thick to stellar irradiation, this prescription becomes less realistic, as the formation of gaps, rings, spirals, vortices, and warps can all modify the attenuation received by regions located behind these substructures. We briefly discuss this as a future direction using radiation-hydrodynamical simulations at the end of Section~\ref{sec:why2}.}

At a given radius, any perturbation on a 2D cylindrical surface ($\phi, z$) can be decomposed into Fourier–Hermite components. Because the disk responds differently to individual Fourier-Hermite modes (see Section \ref{sec:m1n1_mode}), we perform additional simulations using these {individual components ($g_{mn}$)} to identify which mode is responsible for driving the warp. For a perturbation expressed in Fourier–Hermite form, the shape function becomes
\begin{equation}
    g_{mn}(R,\phi,z) = \textrm{He}_n\!\left(\frac{z}{h(R)}\right)\, \sin(m\phi),
    \label{eq:mnshape}
\end{equation}
where $\textrm{He}_n$ is the $n$-th probabilist's Hermite polynomial. For an inclined shadow that is neither coplanar nor polar, the perturbation always contains the $m=1,n=1$ mode, which we will show is the key driver of the warp. For this mode,
\begin{equation}
    g_{11}(R,\phi,z) = \frac{z}{h(R)} \, \sin\phi,
    \label{eq:m1n1shape}
\end{equation}
since He$_1$($z$) is simply $z$.
In this case, at one scale height above the midplane and at $\phi=\pi/2$, the shape function $g_{11}$ reaches unity, or equivalently the temperature modification $f(\phi=\pi/2, z=h) = 1 + A_{11}(t)$ (we use $A_{mn}(t)$ to distinguish the modal amplitude from the full shadow's $A(t)$). The upper and lower disk surfaces experience opposite temperature perturbations ($f(\pi/2,-h) = 1 - A_{11}(t)$), and any two azimuthal points separated by 180$^\circ$ also have opposite temperature perturbations (e.g., $f(3\pi/2,h) = 1 - A_{11}(t)$). We have three simulations under this setup, with $A_{0,11}$ = 5.4\% and 1\% (\texttt{HFm1n1R0p054}, \texttt{HFm1n1R0p01}, \texttt{HFTm1n1R0p01} as listed in Table \ref{tab:models}). The case of $A_{0,11}$ = 5.4\% is chosen since it corresponds to the amplitude of $m=1,n=1$ component in the \texttt{HF30Mo} and \texttt{HF30R} simulations.

\subsection{Transition Disk Setup}
\label{sec:transition_disk}
The outer disk setup follows \citet{zhang24b}. On top of a power-law surface density, the inner disk is carved with a central cavity, and the outer edge tapers off with an exponential cutoff. The cavity radius is set to 160 au. The outer disk is optically thin to stellar irradiation given the opacity and dust to gas ratio adpoted in Section \ref{sec:opacity}. Without the large cavity, the optical depth to stellar irradiation often exceeds unity. For example, a tilted inner disk between 5–10 au with the same $\Sigma_{g,0}$ has $\tau_{\mathrm{*,bc}} = 6000$ (Section \ref{sec:ray_tracing}), and therefore casts shadows onto the outer disk. The cavity also provides sufficient space to study the dynamics within the cavity, ensuring that accretion is less affected by the inner boundary condition at 21.6 au. The gas surface density profile is given by:

\begin{align}
    \Sigma_\mathrm{g} =& \Sigma_\mathrm{g,0} \mathrm{(R/R_0)^{-1}} \nonumber\\ \times & \Big[\frac{1}{2}
    \mathrm{tanh}\Big(\frac{\mathrm{R-160\ au}}{\mathrm{20\ au}}\Big) + \frac{1}{2}\Big] \nonumber\\ \times & \ \mathrm{exp(-R/100\ au)},
	\label{eq:transition_disk_density}
\end{align}
where $\Sigma_\mathrm{g,0}$ is the gas surface density at a reference radius of $\mathrm{R_0}$ = 40 au. Following \citet{zhu12}, $\Sigma_\mathrm{g,0}$ is set to $3\ \mathrm{g\ cm^{-2}}$. If there were no cavity, the disk mass would be 0.01 M$_\odot$. {This disk mass is on the lower end of the gas masses estimated from the dust continuum \citep{curone25}, molecular line emissions \citep{zhangk25, trapman25}, and disk kinematics \citep{lodato23, martire24,longarini25} in recent surveys of large disks. We intentionally adopt a lower mass to ensure that the outer disk remains optically thin to stellar irradiation, allowing us to use a prescribed temperature structure to interpret the radiation-hydrodynamical simulations, as discussed in Section~\ref{sec:prescribed_temperature}. We expect more complex and interesting dynamics to occur in higher-mass disks, as further discussed at the end of Section~\ref{sec:why2}.}

The outer disk's rotational axis (vertical direction) is initialized to align with the z-axis. We assume a vertically isothermal and a power-law radial temperature structure, from which the vertical density and velocity structures can be calculated accordingly. The disk is initially in vertical hydrostatic equilibrium, but these initial conditions will transition to a new equilibrium state according to the stellar irradiation once the simulation starts. This applies to both the radiation–hydrodynamical simulations (\texttt{R30}, \texttt{R90}) and the pure hydrodynamical simulation (\texttt{H30}). The difference is that $t_{\mathrm{relax}}$ plays a more important role in the radiation–hydrodynamical runs for the disk to reach an equilibrium state, whereas in the pure hydrodynamical case the system remains close to its initial condition within $t_{\mathrm{relax}}$. A more detailed setup of the initial conditions can be found in \citet{zhang24}.

Our 3D simulations use a resolution of $192 \times 128 \times 320$ cells in $(r,\theta,\phi)$ for both pure hydrodynamical and radiation-hydrodynamical simulations. The radial grid is logarithmically spaced from $0.54$ to $31.5\ R_0$ ($R_0=40$ au), corresponding to $21.6$–$1260$ au from the inner to outer boundaries. Compared to the model in \citet{zhang24b}, the outer boundary is extended by a factor of two. The polar direction spans $0.21$–$2.93$ radians (i.e., $\sim 80^\circ$ above and below the midplane), while the azimuthal direction covers the full $0$–$2\pi$ range.

For the hydrodynamic boundaries, we apply modified outflow conditions at the inner, outer, upper, and lower boundaries. Outflowing quantities are copied to ghost cells as in standard outflow boundaries, while inflow is suppressed by setting the normal velocity in the ghost cells to zero. Radiation is allowed to freely stream out of the domain; if radiation enters the domain, it is assumed to have the background temperature of 10 K, typical of molecular clouds. Periodic boundary conditions are adopted in the azimuthal ($\phi$) direction.

In addition to the two radiation hydrodynamical runs with $i=30^\circ$ and $i=90^\circ$ (\texttt{R30}, \texttt{R90}), we also perform a pure hydrodynamical simulation of a transition disk with identical initial conditions (\texttt{H30}). In this case, we use the prescribed temperature structure described in Equation \ref{eq:prescribedT}.

\subsection{Full Disk Setup}
\label{sec:full_disk}
To understand why thermal perturbations warp the disk, we adopt a simplified density profile, described by a power-law (Equation \ref{eq:full_disk_density}).
Since the hydrodynamical simulations can be scaled arbitrarily by setting $\Sigma_{g,0}$ and $R_0$, The only non-scalable parameter is the disk aspect ratio, $h/r$. We set $h/r = 0.1$ at $R=R_0$, consistent with the radiation--hydrodynamical runs.  

To demonstrate the effect of an exponential cutoff on warp evolution, we also perform simulations with  
\begin{equation}
    \Sigma_g = \Sigma_{g,0} \left(\frac{R}{R_0}\right)^{-1}
    \exp\left(-\frac{R}{R_c}\right),
    \label{eq:full_disk_exp_density}
\end{equation}
i.e., the expression of Equation~\ref{eq:transition_disk_density} without an inner cavity. We choose $R_c = R_0$ for the full disk model (\texttt{HFTm1n1R0p01}).

For all pure hydrodynamical simulations with full disks (\texttt{HF30Mo}, \texttt{HF30R}, \texttt{HF7R}, \texttt{HFm1n1R0p054}, \texttt{HFm1n1R0p01}, \texttt{HFTm1n1R0p01}), the radial domain spans two decades, from $0.1$ to $10\,R_0$, while the vertical domain covers only $\pm0.35$ radians ($\pm20^\circ$) about the midplane to reduce computational cost. We use 460$\times$72$\times$620 cells in ($r,\theta,\phi$) which provides an effective resolution of 10 cells/$h$ in the radial and polar directions -- double that of the radiation-hydrodynamical runs.  

For most runs with full disks (\texttt{HF30R}, \texttt{HF7R}, \texttt{HFm1n1R0p054}, \texttt{HFm1n1R0p01}, \texttt{HFTm1n1R0p01}), we adopt reflecting boundary conditions to conserve mass and angular momentum. We also test modified outflow boundaries to examine their influence on the results (\texttt{HF30Mo}).

{For a tilt of $\sim$7$^\circ$, as in the \texttt{HFm1n1R0p054} model, the disk surface at approximately 2.5 scale heights reaches the vertical boundary, which still encloses roughly 99\% of the vertically integrated mass. Therefore, the majority of the disk remains within the computational domain, and the warp evolution should be qualitatively reliable. However, caution should be taken, as the choices of vertical and radial extent, boundary conditions, and density floor can affect angular momentum conservation and wave propagation, thereby quantitatively influencing the results.}

\section{Results}
\label{sec:results}
We begin by using our radiation–hydrodynamical simulations to illustrate the dynamical consequences of inclined shadows, namely the development of warps and enhanced accretion (Sections \ref{sec:warp_accretion}, \ref{sec:flow_structure}). We then demonstrate, with locally isothermal hydrodynamical simulations that prescribe the temperature structure, that these effects are driven by shadow-induced thermal perturbations (Section \ref{sec:thermally_induced}). Next, we identify the specific temperature mode $(m=1,n=1)$ responsible for exciting the warp (Section \ref{sec:m1n1_mode}), {and show that the exponential density taper leads to a relative twisting between the high-density outer disk and the low-density tapered region--well beyond the disk size typically defined by ALMA observations (e.g., the region enclosing 90\% of the total flux; Section \ref{sec:twist})}. Finally, we establish a scaling relation between the mode amplitude and the disk tilt (Section \ref{sec:scaling}), and show that in full-disk models the tilt undergoes periodic modulation (Section \ref{sec:period}).

\begin{figure*}
    \centering
  \includegraphics[width=1.0\linewidth]{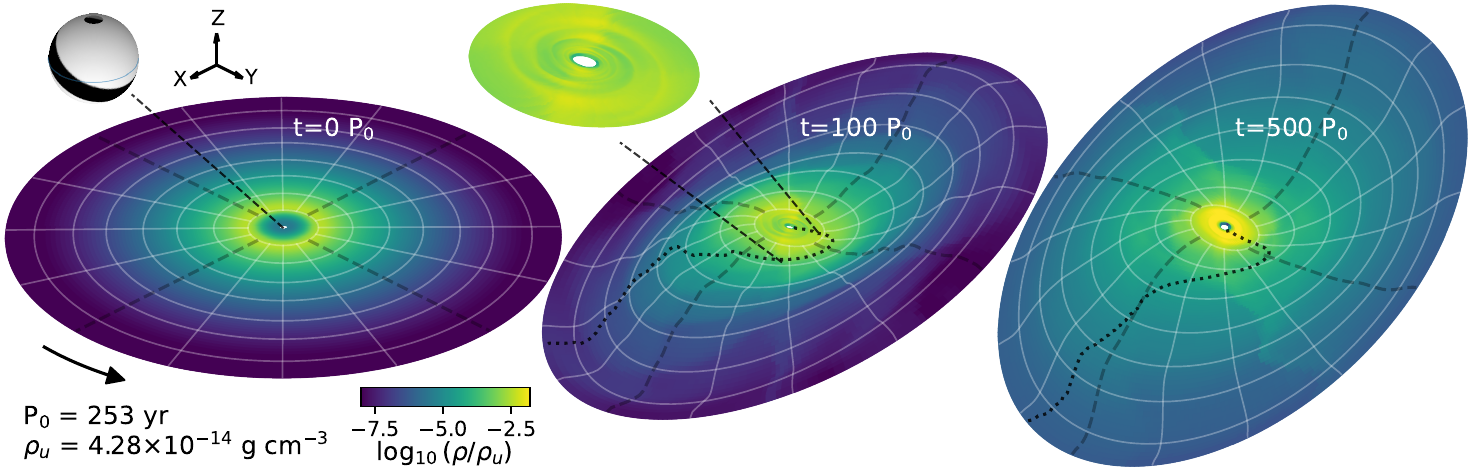}
    \caption{Time evolution of the 3D hydrodynamical model (\texttt{R30}), showing the density on the midplane surface at $t=0$, $100$, and $500\ P_0$ (left to right, $P_0 \approx$ 253 yr). The stellar irradiation field is prescribed as a spherical distribution with a shadow lane inclined by $30^\circ$ relative to the original disk midplane (sketch in the top left corner, Equation~\ref{eq:irradiation}). The disk develops a warp in response to the shadow and also launches two-armed spirals (visible in the zoomed-in panel). The inner and outer disks warp in different directions, traced by the dotted lines marking the twist angle $\gamma$ along different radii. Concentric grid lines are spaced every 160 au starting from $R=160$ au, and azimuthal divisions are spaced every $22.5^\circ$. The dashed gray lines are $\phi=0, \pi/2, \pi$, and $3\pi/2$ in the original coordinate. {Associated animation for this figure and a Blender rendering movie can be viewed online and downloaded (for a better resolution) at \url{https://doi.org/10.6084/m9.figshare.30535781.v2} and \url{https://doi.org/10.6084/m9.figshare.30531185.v1}.}}
    \label{fig:warp_surface_evolution}
\end{figure*}

\subsection{Warp and Strong Accretion}
\label{sec:warp_accretion}
\begin{figure}
    \centering
    \includegraphics[width=1.0\linewidth]{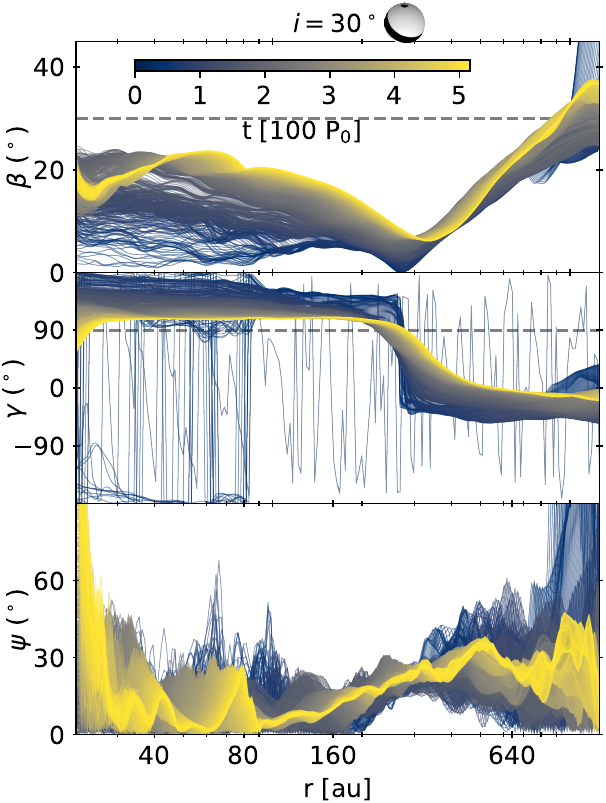}
    \caption{Time evolution of the 3D radiation–hydrodynamical simulation with $i=30^\circ$ (\texttt{R30}). Top: tilt. Middle: twist angles. Bottom: warp amplitude. Colors from blue to yellow denote increasing time. The first $100\ P_0$ ($P_0 \approx 253$ yr) are omitted to highlight the subsequent evolution. Otherwise, the twist plot would be dominated by oscillatory curves during the first $100\ P_0$. The horizontal dashed lines are the shadow's tilt (30$^\circ$) and twist angle (90$^\circ$).}
    \label{fig:warp_properties_evolution}
\end{figure}

{At t = 0 P$_0$ (after the relaxation phase t$_\mathrm{relax}$), the disk is still perfectly aligned with the $z$-axis.} Introducing a $30^\circ$ inclined shadow drives a global warp. When the shadow is initially imposed, the disk midplane is aligned with the $z$-axis, but it begins to tilt shortly thereafter. The inner disk tilts in the direction that aligns with the shadow, while the outer disk warps in a different direction. This evolution is illustrated in Figure~\ref{fig:warp_surface_evolution}, which shows the warped midplane and midplane density at $t=0$, $100$, and $500\ P_0$ ($P_0 \approx 253$ yr) after the shadow is applied. We define the instantaneous midplane as the plane perpendicular to the local angular momentum vector, measured as a function of radius (Figure~\ref{fig:warp_properties_evolution}). The shadow attenuation is shown in the left inset of Figure \ref{fig:warp_surface_evolution}, where the shadow lane is clearly visible. Dashed lines mark $\phi=0^\circ, 90^\circ, 180^\circ,$ and $270^\circ$ in the original coordinate system, while dotted lines trace the twist at each radius.

At the same time, the inner cavity becomes more filled in, indicating strong accretion. The middle inset of Figure \ref{fig:warp_surface_evolution}, which zooms into the inner 160 au, clearly shows the presence of two spiral arms which drives accretion, consistent with the findings of \citet{zhang24b}. The warp and accretion continues to evolve: by $t = 500\ P_0$ ($\approx 0.13$ Myr), the inner disk has accumulated more material, while the outer disk exhibits an even larger tilt.  

We quantify the warp using three standard 
parameters in warp disk studies in Figure \ref{fig:warp_properties_evolution}, all derived from the angular momentum 
vector. We first define the unit angular momentum vector as
\(\hat{\mathbf{l}}(R) = (l_x, l_y, l_z) = \mathbf{L}(R)/|\mathbf{L}(R)|\). 
From this, the disk orientation is characterized by the tilt ($\beta$), twist ($\gamma$) angles, and warp amplitude ($\psi$):
\begin{equation}
\begin{aligned}
    \beta(r) &= \arccos\!\left(l_z\right) 
        , \\
    \gamma(r) &= \arctan2\!\left(l_y,\, l_x\right) 
        , \\
    \psi(r) &= \left| \frac{d\hat{\mathbf{l}}}{d\ln r} \right| \\
            &=  r \Bigg[\sin^2\beta\bigg(\frac{d \gamma}{dr}\bigg)^2+\bigg(\frac{d \beta}{dr}\bigg)^2\Bigg]^{1/2}
        .
\end{aligned}
\label{eq:beta_gamma_psi_defination}
\end{equation}
The tilt $\beta$ is the inclination of the disk midplane from the $z$-axis and traces how strongly the disk departs from the initial midplane. 
It increases to about $20^\circ$ between 20–160 au and exceeds $30^\circ$ 
in the outer disk beyond 640 au. The dip in $\beta$ coincides with a 
transition in the twist angle ($\gamma$), which measures the azimuthal orientation of the tilt in the $x$–$y$ plane. 
Within 300 au, $\gamma$ remains near $90^\circ$, consistent with the prescribed shadow orientation of the tilt towards the $y$-axis (horizontal dashed line). Beyond 300 au, 
the twist rotates toward $\sim 0^\circ$, signaling a reorientation of the outer 
disk. The warp $\psi$ quantifies the strength of the radial bending. It can also be calculated using the tilt and twist angle (bottom line of the Equation \ref{eq:beta_gamma_psi_defination}), which means that warps can consist of changing tilts, changing twists or a combination of both. Since $\beta$ and $\gamma$ are measured in the original coordinate system, the mutual inclination between any two annuli must be calculated from their local angular momentum vectors, i.e., which depends on both $\beta(r)$ and $\gamma(r)$. For reference, the mutual inclination between the annuli at 80 and 640 au is $32^\circ$, whereas the difference in tilts $\delta\beta$ alone is just $\sim$ 1$^\circ$. Figure \ref{fig:warp_properties_evolution} shows that the warp amplitude fluctuates with
radius but is generally stronger in the outer regions, where the disk is 
more strongly tilted. 

{Interestingly, the warp that emerges has a net change of angular momentum compared with the initial condition, for which $L_x=L_y=0$. This is clearly the case since the dominant twist angles in Figure \ref{fig:warp_properties_evolution}, between the inner and outer regions, are approximately orthogonal. Therefore, the horizontal projection of angular momentum does not cancel out. Since no external torques acts on the system\footnote{{The radiation pressure term, which is included in our calculations, could act as a source of external torque in the momentum equation; however, its contribution is negligible.}}, one might expect the total angular momentum to be conserved.  However, our disk is not entirely isolated since it can interact with our numerical boundaries. This will be discussed further in Section \ref{sec:why1}.}

\begin{figure}
    \centering
    \includegraphics[width=1.0\linewidth]{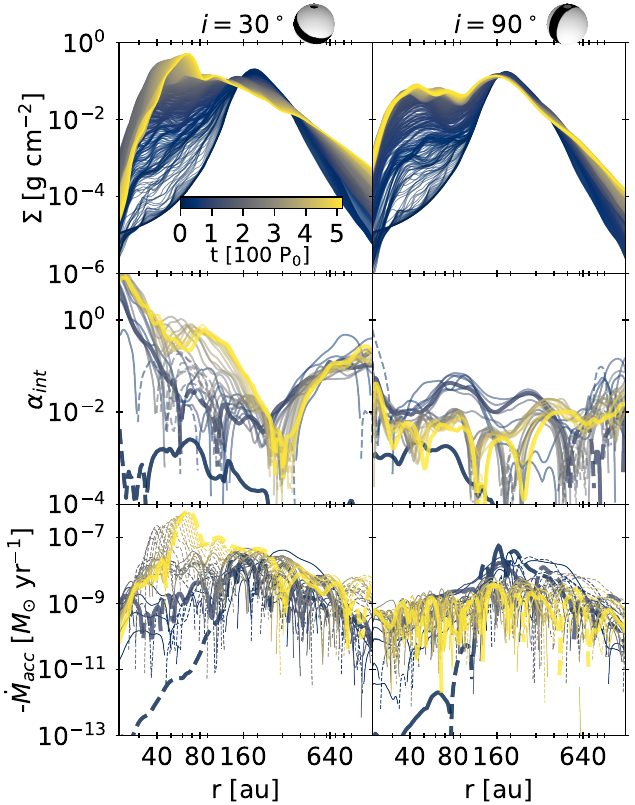}
    \caption{Time evolution of the 3D radiation–hydrodynamical simulations for $i=30^\circ$ (\texttt{R30}, left) and $i=90^\circ$ (\texttt{R90}, right). Top: surface density. Middle: vertically integrated Reynolds stress normalized by pressure. Bottom: mass accretion rate. Colors from blue to yellow denote increasing time, with $t=5$, $105$, and $505\ P_0$ highlighted by thicker lines ($P_0 \approx 253$ yr). Solid lines indicate positive values, while dashed lines indicate negative values.}
    \label{fig:accretion}
\end{figure}

We quantify the disk density evolution and accretion using Figure \ref{fig:accretion} and contrast the $30^\circ$ inclined shadow with the polar shadow studied in \citet{zhang24b}. Compared to the accretion driven by a polar shadow, the $30^\circ$ inclined shadow induces much stronger accretion, leading to an inner disk surface density that is an order of magnitude higher than in the polar shadow case (first row). This result is confirmed by the vertically averaged Reynolds stress normalized by the pressure, $\alpha_{\mathrm{int}}$ and the mass accretion rates shown in the second and third rows.
The $\alpha_{\mathrm{int}}$ is defined as
\begin{equation}
\alpha_{\mathrm{int}}=\frac{\int T_{R',\phi'}dz}{\int \langle P \rangle_{\phi',t} dz}\,,\label{eq:alphaint}
\end{equation}
where $T_\mathrm{R', \phi'} \equiv \langle \rho v_R' v_\phi' \rangle_{\phi',t} - \langle v_\phi' \rangle_{\phi',t} \langle \rho v_R' \rangle_{\phi',t}$, is the azimuthally  and time-averaged Reynold stress between $R'$ and $\phi'$ directions. Here, $\langle \rangle_{\phi',t}$ denotes averaging across the full 2$\pi$ in $\phi$ and time between t - 5 $P_0$ and t + 5 $P_0$. Note that the primes on these coordinates and velocities mean that these values are calculated in the rotated coordinate according to the local angular momentum vector. 

Thick lines mark values at $5$, $105$, and $505\ P_0$ ($P_0\approx 253$ yr), corresponding to the three snapshots in Figure 1. Solid curves indicate positive values, while dashed curves indicate negative values. In the inclined-shadow case, $\alpha_{\rm int}$ grows rapidly and can reach or even exceed unity inside $80$ au by $500\ P_0$ ($\approx 0.13$ Myr). In contrast, the polar-shadow case sustains only $\alpha \sim 10^{-2}$ throughout the evolution. Although lower, this value is still comparable to other major transport mechanisms such as the magnetorotational instability and gravitational instability.

We also integrated $\langle \rho v_R'
\rangle_{\phi',t}$ along the vertical direction to obtain azimuthally-averaged, time-averaged, and vertically integrated radial mass accretion rates ($\dot{M}{_{acc}}= 2\pi R\int \langle \rho v_R' \rangle_{\phi',t} dz'$) as functions of $R$. The mass accretion rate is likewise enhanced in the $30^\circ$ inclined–shadow case. At $R \sim 80$ au, $\dot{M}_{\rm acc}$ can reach $\sim 10^{-7} M_\odot {\rm yr}^{-1}$ by $500 P_0$, compared to only $\sim 10^{-9} M_\odot {\rm yr}^{-1}$ in the polar–shadow case. This trend is consistent with the measured $\alpha_{\rm int}$ values, since the accretion rate follows directly from the vertically integrated angular momentum equation,
\begin{align}
\dot{M}_{\rm acc}
&= -\frac{2\pi}{\partial R' v_{K}/\partial R'}
\frac{\partial}{\partial R'}\left( R'^2 \alpha_{\rm int}
\int \langle P \rangle_{\phi',t} dz' \right),
\label{eq:mdot}
\end{align}
where we assume $\langle v_\phi' \rangle_{\phi',t}$ equals the midplane Keplerian velocity $v_K$.

{Given the strong accretion, the inner disk can become more massive and change its inclination within the timescale of the simulation, which in turn would alter the shape of the shadow. This effect is not modeled here, as the inner disk is fixed and the outer disk is optically thin to stellar irradiation. The possibility of such feedback is discussed at the end of Section~\ref{sec:why2}.}

\subsection{Flow Structures}
\label{sec:flow_structure}
\begin{figure*}
    \centering
    \includegraphics[width=1.0\linewidth]{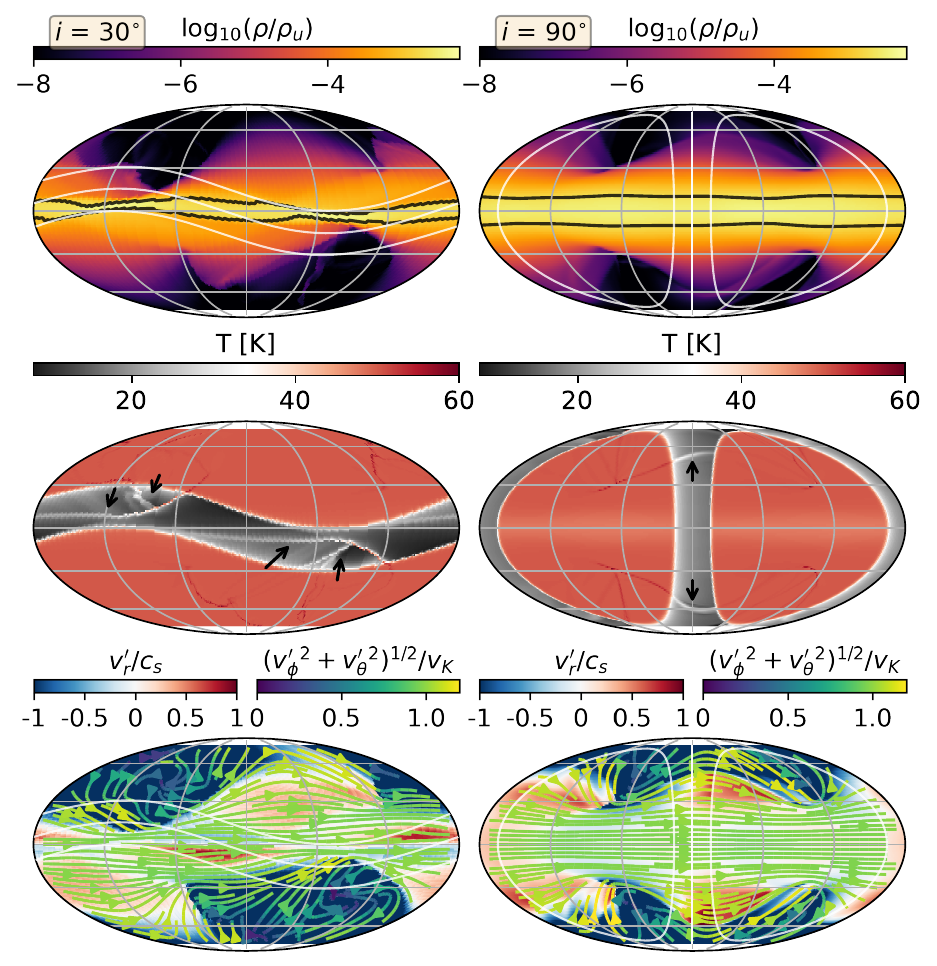}
    \caption{Slices of the radiation–hydrodynamical simulations at $r=4\ r_0$ (160 au) for $i=30^\circ$ tilted shadow (\texttt{R30}, left) and $i=90^\circ$ tilted shadow (\texttt{R90}, right) at $t=500\ P_0$ ($P_0 \approx 253$ yr). Top: density. Middle: temperature. Bottom: tangential velocity (streamlines) overlaid on the radial velocity background. Longitude lines are spaced every $60^\circ$, centered at $\phi'=\pi$; latitude lines are spaced every $30^\circ$. Although both shadows have the same intrinsic width, the polar shadow appears narrower due to projection effects. The center and edges (i.e., $\tau=1$ surfaces, at $\pm$ 0.27 of the midplane) of the shadow lanes in density and velocity plots are marked by white curves. In the density panels, the high density regions are defined as the regions within the black curves defined as the $\exp(-1/2)$ of the maximum density along the $\theta'$ direction. In the temperature panels, shock heated regions are marked by black arrows.}
    \label{fig:phi_theta_view}
\end{figure*}

To better illustrate the differences in flow structure between the $30^\circ$ inclined shadow (\texttt{R30}) and the polar case (\texttt{R90}), we plot various quantities on the $\phi'$–$\theta'$ (or $\phi$–$\theta$) plane at $r = 160$ au in Figure~\ref{fig:phi_theta_view}. The $30^\circ$ inclined–shadow case is shown in a rotated coordinate system aligned with the disk’s new angular momentum vector (local $z'$); without this rotation, the shadow belt would appear more tilted. In this rotated frame, rigid tilt is effectively removed. To illustrate the correspondence between the shadow and the density and velocity fields, we mark the center of the shadow lane as well as the $\tau_* = 1$ surfaces, given by $(\theta_s - \pi/2) = \pm 0.27$, as white curves on the density and velocity plots.

The vertical density structure of \texttt{R30} exhibits azimuthal variations, with rarefied expansions and compressed nodes, twice per orbit. This behavior is also active in other studies of nonlinearly warped disks, which identify strong breathing motions \citep{fairbairn21} and nozzle shocks \citep{kaaz23,kaaz25}. Even in the rotated frame perpendicular to the new angular momentum vector ($z'$), the densest regions are not perfectly aligned with the midplane and exhibit $m=2$ azimuthal modulations. In contrast, the polar–shadow case (\texttt{R90}) remains fully symmetric across the midplane and shows only small-scale $m=2$ modulations in the density due to the azimuthal pressure response to the two shadow lanes, as studied in \citet{zhang24b}. Overall, the midplane density variations in the polar case are much smaller than in the inclined–shadow case.
To quantify the scale height variations in the density fields of \texttt{R30} and \texttt{R90}, we plot the surface defined by $\exp(-1/2)$ of the maximum density along the $\theta'$ direction at each azimuth (black curves). In a vertically isothermal disk, this surface corresponds to one gas scale height, since $\rho = \rho_{\mathrm{mid}}\exp(-z^2/2h^2)$. Although in the \texttt{R30} case the vertical density distribution becomes non-Gaussian due to temperature perturbations, this surface still provides a simple way to trace high-density regions. The midplane of \texttt{R30} is much thinner than that of \texttt{R90}, with two nodes touching the shadow boundaries (the $\tau=1$ surfaces). We measure the narrowest nodes to be at $(z'|_{\exp(-1/2)\rho_{\max}}-z'|_{\rho_{\max}})/r = 0.03$ and the widest parts at $(z'|_{\exp(-1/2)\rho_{\max}}-z'|_{\rho_{\max}})/r = 0.12$, giving a factor of 3.6 contrast between these extremes. By comparison, \texttt{R90} shows only a small $m=2$ variation, with the narrowest part at $z'|_{\exp(-1/2)\rho_{\max}}/r = 0.16$ and the widest at $z'|_{\exp(-1/2)\rho_{\max}}/r = 0.18$, a contrast of just 1.125. In Appendix \ref{sec:scale_height_2}, we present an alternative definition of the gas scale height following \citet{fung19}. While this method yields different absolute heights, the level of azimuthal variation remains similar to that shown here. 

The vertical asymmetry arises from the reduced temperature within the shadowed region, which itself is asymmetric about the midplane. Within this cooled region, we find higher-temperature filaments (marked by black arrows) coincident with sharp density gradients, consistent with strong shock heating. These shocked regions have $T\approx$ 35 K compared with $T\approx$ 15 K inside the shadow at $r= 160$ au. Such shocks may be observable with ALMA through tracers like SO, as suggested in shadowed disks such as CQ Tau and MWC 758 \citep{zagaria25}.

In contrast, the polar shadow remains fully symmetric about the $z$-axis, with its two shadow lanes separated by $180^\circ$. However, the azimuthal structure does not exhibit perfect $m=2$ symmetry. The side entering the shadow (left of the central line of longitude, or the rightmost edge in the plot) is slightly hotter than the side leaving the shadow (right of the central line of longitude, or the leftmost edge in the plot) due to finite radiative cooling times \citep{cassasus18}. A similar effect occurs in the $30^\circ$ inclined shadow, where the side entering the shadow is consistently hotter in each hemisphere. The midplane of the polar case shows a temperature depression, indicating it is not fully optically thin and that the outer disk is subject to its self-shadow.

At $t = 500\ P_0$ ($\approx 0.13$ Myr), the self-shadowed midplane in the $30^\circ$ inclined case appears less distinct than in the polar case. This occurs because the high-density region shown in the first panel of Figure \ref{fig:phi_theta_view} is already hidden inside the $\tau_* = 1$ surface produced by the inner disk’s shadow through our ray-tracing (Equation \ref{eq:irradiation}). It remains unclear whether the shadow edges provide a sufficiently strong pressure gradient to confine the high-density region within the shadowed zone. However, if the accretion streams concentrate dust on a plane directly exposed to stellar irradiation, they will self-shadow the outer disk and substantially alter the shadow shape. Only full radiation–hydrodynamical simulations, such as those presented here, can capture this self-shadowing feedback loop.

The third row of Figure \ref{fig:phi_theta_view} shows streamlines on the $r=160$ au sphere overlaid on the radial velocity across this shell at t = 500 $P_0$ ($\approx 0.13$ Myr). In the $30^\circ$ inclined shadow model (\texttt{R30}), shocks occur where high-speed flows in the rarefied atmosphere collide with the dense midplane ({e.g., on the left of the upper surface}). These flows are deflected and converge with the midplane material, which primarily advects azimuthally at near-Keplerian speed. Even within the midplane itself, there are two regions (middle left and far right) of strong radial inflow. By contrast, in the polar shadow model (\texttt{R90}), the strongest radial flows are confined to the low-density surface layers, while the midplane maintains a more ordered Keplerian motion. This structural difference likely explains why the $30^\circ$ inclined model exhibits a much higher accretion rate: the midplane participates directly in radial transport, whereas in the polar case, the midplane remains dynamically stable up to $\pm15^\circ$ (2-3 scale heights) from the midplane. The much stronger accretion rate in the $30^\circ$ inclined model (\texttt{R30}) compared to \texttt{R90} arises from a similar mechanism identified by \citet{kaaz23, kaaz25}, where dissipation in nozzle shocks drives anomalously large effective $\alpha$. Essentially, orbital energy is impulsively dissipated during the compressive motions twice per orbit. In the $30^\circ$ inclined shadow model (\texttt{R30}), a close inspection of the radial flow shows that, within each hemisphere, the radial motions near the midplane and at the surface occur once over the full azimuth, with the upper and lower hemispheres having a $180^\circ$ phase difference. A similar modulation is present in the tangential velocity, where upward and downward motions alternate once per azimuth. This $m=1$ kinematic perturbation is a clear signature of sloshing motions, which can be interpreted as eccentric streamlines in anti-phase above and below the midplane \citep{fairbairn25}. By contrast, in the \texttt{R90} model the perturbations are symmetric across the midplane, both corresponding to $m=2$ modes.

\begin{figure*}
    \centering
    \includegraphics[width=1.0\linewidth]{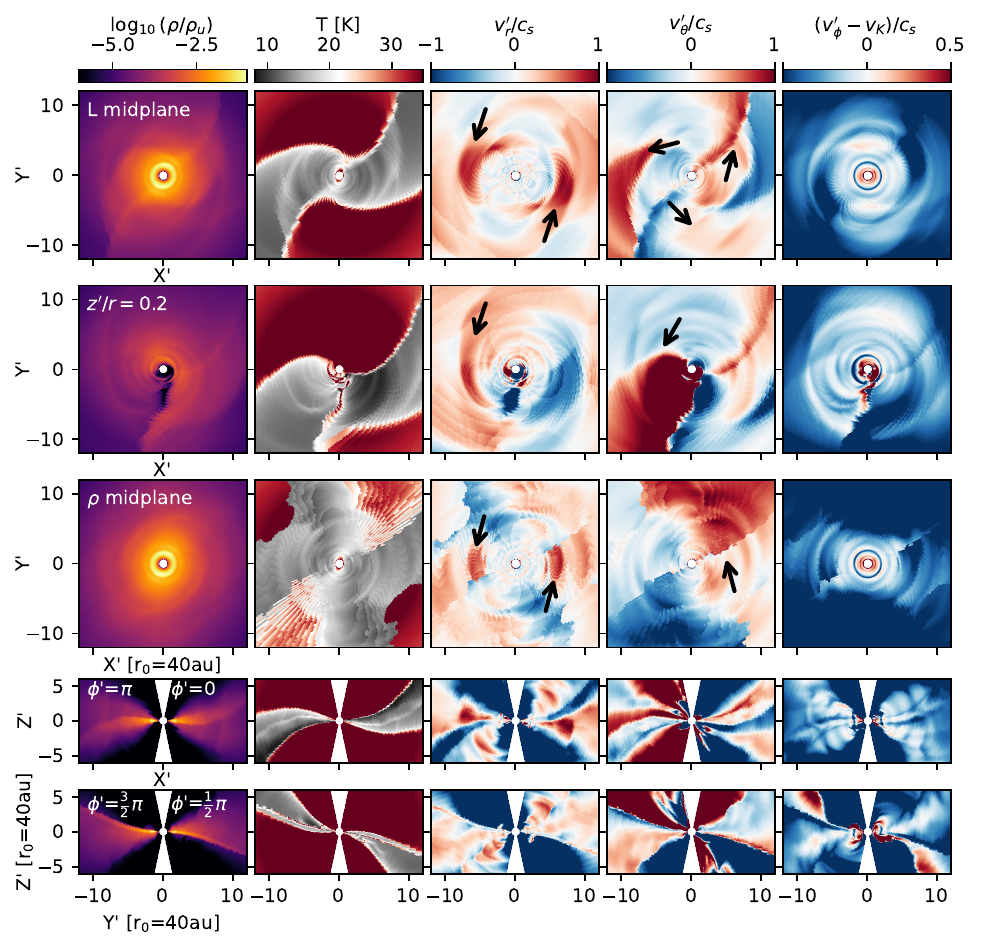}
    \caption{Slices of the 3D radiation–hydrodynamical simulation with $i=30^\circ$ (\texttt{R30}) at $t=500\ P_0$ ($P_0 \approx 253$ yr, so $t \approx 0.13$ Myr), shown in the transformed coordinate system $(r, \theta', \phi')$, aligned with the local angular momentum vector at each radius (Figure~\ref{fig:warp_properties_evolution}). Left to right: density, temperature, radial velocity, meridional velocity, and the deviation of the azimuthal velocity from Keplerian. All velocity components are projected into the transformed frame. Top to bottom: values at the new midplane ($\theta' = \pi/2$), at $z'/r = 0.2$ above the midplane, at the surface of maximum density, and vertical slices in the $y'-z'$ and $x'-z'$ planes. The disk rotates counter-clockwise. The arrows qualitatively indicate the azimuthal locations of velocity maxima, indicating the dominant modes that are present. Some of the grid-like patterns arise from the nearest-neighbor interpolation used in the coordinate transformation. To facilitate comparison with observations, line-of-sight velocities are presented in Figures \ref{fig:los_zr0} and \ref{fig:los_zr0p2}.}
    \label{fig:cart_slices}
\end{figure*}

To illustrate the flow in a more observationally relevant frame, we show slices in the rotated Cartesian system at the midplane, at the surface ($z'/r=0.2$), and in the meridional planes ($x'-z'$ and $y'-z'$) at $t=500\ P_0$ ($\approx 0.13$ Myr). Because the disk is warped, these transformed surfaces correspond to warped layers in the original coordinates. The plotted velocity fields thus represent the residual motions after rotating away the tilt associated with the net angular momentum of each radial annulus.

The first row of Figure \ref{fig:cart_slices} shows midplane quantities: density, temperature, and the perturbed velocity components $v_r'$, $v_\theta'$, and $v_\phi' - v_K$, normalized by the local sound speed. The density field displays both concentric rings and gaps as well as prominent $m=2$ spirals. An additional anti-diagonal overdensity may trace a shocked region. The temperature structure highlights the two shadow lanes separated by 180$^\circ$, consistent with previous studies \citep{su24, zhang24b, zhu25, ziampras25}, and explains why $m=2$ spirals are still excited even in the inclined shadow case. In the $v_r'$ panel, the $m=2$ spirals are evident (black arrows), while the $v_\theta'$ component shows a distinct $m=3$ spiral pattern (black arrows). In Section \ref{sec:m1n1_mode}, we demonstrate that this pattern corresponds to an $m=3, n=1$ mode identified through Fourier–Hermite decomposition. The azimuthal velocity also exhibits corresponding features, though the background reference level is more difficult to anchor, since the entire disk rotates with a sub-Keplerian profile due to radially decreasing pressure, with super-Keplerian regions only present at the central ring.

The second row shows the disk surface at 0.2 radians (two scale heights) above the midplane. The density distribution reveals several spiral features, concentrated primarily on one side of the disk. The temperature map shows that the top half of the disk is illuminated while the bottom half is shadowed. In the $v_r'$ panel, the flow is dominated by an $m=1$ spiral (black arrow). The $v_\theta'$ panel likewise shows $m=1$ motion in the lower-left quadrant (black arrow), with a weaker $m=2$ component appearing on the opposite side. If the disk were simply tilted but following Keplerian rotation in the new orbital plane, both $v_r'$ and $v_\theta'$ would vanish. The presence of $m=1$ patterns therefore indicates sloshing motions associated with the bending wave excited in the warped disk \citep{fairbairn21}. The $m=2$ pattern could be indirectly driven by the nonlinearly excited compressive vertical motions (breathing motions) which are $m=2$ and therefore lead to an enhanced dissipation and radial accretion twice per orbit \citep{fairbairn21}. Another interpretation is that these $m=2$ patterns are directly forced by the action of the $m=2, n=0$ Hermite component, which is strong at this shadow tilt near the midplane, as we will introduce in Figure \ref{fig:shadow_decomposition_heatmap}. Further study is needed to determine whether the two-armed spirals launched in the midplane by the $m=2$ temperature perturbation can propagate to the surface with a comparable amplitude. In other words, it remains unclear how much of the $m=2$ motion at the $z'/r = 0.2$ surface originates from the local breathing motion due to the warp, and how much is driven by $m=2$ perturbations propagating upward from the midplane.

{In observations, the line-of-sight velocity is a direct observable, while the three orthogonal velocity components must be inferred under assumptions about the disk’s emission geometry and symmetry. Appendix \ref{sec:los_r30} describes our calculation of the line-of-sight velocity. When the annuli ($\hat{z}'$) are inclined by less than 50$^\circ$ to the line of sight, the modal perturbations in $v_\theta'$ seen in Figure \ref{fig:cart_slices} are well preserved in the line-of-sight velocities (Figures \ref{fig:los_zr0} and \ref{fig:los_zr0p2}).}

The third row adopts an alternative definition of the midplane, aligned with the highest-density plane (maximum $\rho$ along $\theta'$) rather than $\rho$ at $\theta' = \pi/2$. In this frame, the $m=2$ spirals become more apparent in the density field. Interestingly, nearly all of the highest-density regions lie within the shadow, as indicated by the temperature map. This is already evident in the first panel of Figure \ref{fig:phi_theta_view} at $r=160$ au, where high-density regions are hidden in the shadow, but here we show that this occurs at all radii. We speculate that this arises because additional density is required to balance the pressure in low-temperature regions and/or because strong pressure gradients at the shadow edges confine the high-density regions. The $v_r'$ panel shows an $m=2$ pattern (black arrows), while the $v_\theta'$ panel is instead dominated by an $m=1$ component (black arrow).

The last two rows of Figure \ref{fig:cart_slices} show vertical slices of the disk along two orthogonal planes. In both views, the vertical density maximum consistently lies on one side of the disk, reaffirming the non-coplanar midplane seen in Figure \ref{fig:phi_theta_view}. The shadowed regions appear warped due to the coordinate transformation from the warped disk; in the original frame, the shadows form straight lanes. {This can be seen clearly in Figure~\ref{fig:cart_unprimed} in Appendix~\ref{sec:cart_unprimed}, where we show vertical slices of the density and temperature in the original (unprimed) simulation coordinates along the $x$- and $y$-axes.} While all five fields—density, temperature, and the three velocity components—exhibit pronounced asymmetries across the vertical ($z$) direction, they all retain $180^\circ$ rotational symmetry about the origin.

\subsection{Thermally Induced Warp}
\label{sec:thermally_induced}

\begin{figure}
    \centering
\includegraphics[width=1.0\linewidth]{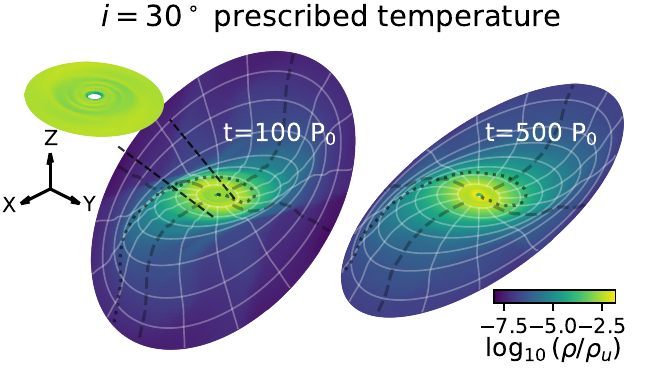}
\includegraphics[width=1.0\linewidth]{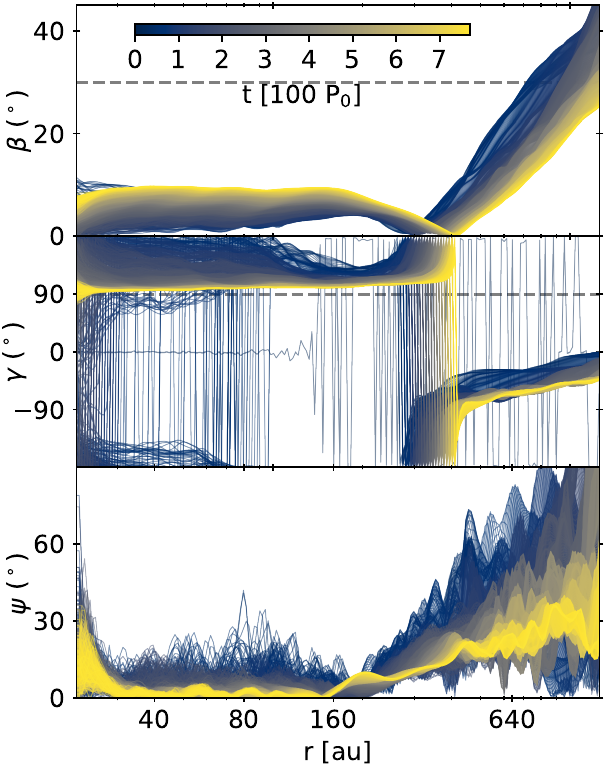}
    \caption{Warp evolution in a pure hydrodynamical simulation with a prescribed temperature distribution induced by a $30^\circ$ shadow (\texttt{H30} with modified outflow boundary condition). Top panel: surface evolution, analogous to Figure~\ref{fig:warp_surface_evolution}. Bottom three panels: evolution of tilt, twist, and warp, analogous to Figure~\ref{fig:warp_properties_evolution}.}
    \label{fig:hydro_warp_evolution}
\end{figure}

An immediate question is: what drives the warp? Motivated by the successful explanation of shadow-induced spirals and eccentricities through prescribed azimuthally varying temperatures \citep{su24, zhu25, qian24}, we hypothesize that the warp is likewise thermally driven. To test this, we perform pure hydrodynamical simulations with a prescribed temperature drop of 60\% ($A_0$ = 60\% in Equation \ref{eq:amplitude2}) in the shadowed region at all radii. The transition between the cold shadowed midplane and the superheated surface is smoothed ($\sigma_t$ = 0.35 in Equation \ref{eq:angular_dependence}) to minimize shocks and to isolate their role in warp and accretion evolution, while keeping all other parameters unchanged (\texttt{H30}) in Table \ref{tab:models}. 

Figure \ref{fig:hydro_warp_evolution} summarizes the evolution of this purely hydrodynamical model. The first row highlights a warp evolution closely resembling that seen in the radiation-hydrodynamical simulation in Figure~1. This model develops both a twist between inner and outer disks and a two-armed spiral pattern that drives accretion. The inner disk warps toward the shadowed direction, while the outer disk warps in the opposite direction.

The second to fourth rows quantify the warp properties, in direct comparison to Figure~2. The inner disk reaches an inclination of $\sim 10^\circ$, while the outer disk tilts by more than $30^\circ$. At $\sim 400$ au, the tilt reaches a minimum, coinciding with a twist transition from $90^\circ$ to $-45^\circ$. The warp amplitude is small in the inner disk but grows significantly in the outer disk. These trends mirror those of the radiation-hydrodynamical model in Figures \ref{fig:warp_surface_evolution} and \ref{fig:warp_properties_evolution}. A comparison of the disk density evolution and accretion is shown in Appendix Figure~\ref{fig:compare_accretion_rad_hydro}.

Some differences remain between the models. The tilt amplitude is smaller in the pure hydrodynamical simulation, and the transition between the inner and outer disks differs. In the radiation–hydrodynamical simulation, the twist transitions clockwise from the inner to the outer disk, whereas in the pure hydrodynamical simulation it transitions anticlockwise, as indicated by the dotted line on the disk surface (Figure \ref{fig:warp_surface_evolution} and Figure \ref{fig:hydro_warp_evolution} top panel) and the 1D evolution (Figure \ref{fig:warp_properties_evolution} and Figure \ref{fig:hydro_warp_evolution} third panel). This may result from the tilt reaching zero in the pure hydrodynamical case, where $\gamma$ becomes undefined. With a very small tilt only subtle changes in the warp can modify the twist directionality. Nevertheless, the overall agreement is notable given the differences between the two models. In particular, the prescribed temperature structure in the pure hydrodynamical model is fixed in space and time, with simplified orbital cooling ($\beta_c = 10^{-6}$), whereas the full radiation–hydrodynamical model has longer radiative cooling timescales (indicated by the azimuthally asymmetrical temperature ahead and behind the shadow in Figure \ref{fig:phi_theta_view}). The transition between shadowed and unshadowed regions is smoother in the pure hydrodynamical case and much sharper in the radiation–hydrodynamical case. Moreover, the self-shadowing effect is not captured in the pure hydrodynamical model. Despite these differences, the qualitative agreement across all three warp parameters strongly supports the conclusion that the inclined shadow and the resulting thermal structure drives the disk warp.

\subsection{The Mode that Induces Warp: m=1,n=1}
\label{sec:m1n1_mode}

\subsubsection{Decomposing Temperature Structure}

\begin{figure}
    \centering    \includegraphics[width=1.0\linewidth]{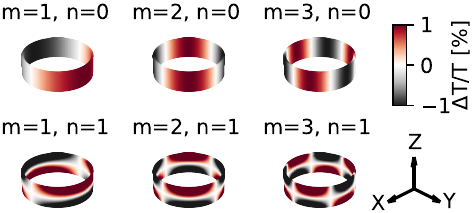}
    \caption{Examples of the decomposition of the temperature perturbation into azimuthal Fourier modes ($m$) and vertical Hermite modes ($n$) on a cylindrical surface in the $\phi$–$z$ plane.}
    \label{fig:decomposition_example}
\end{figure}

Knowing that the warp is thermally driven, we aim to identify the specific mode responsible for driving it. In studies of 3D warped disks and planet–disk interactions \citep{tanaka04, tanaka24, zhang06, fairbairn25}, disk quantities are often expanded as
\begin{equation}
\eta(R,\phi,z) = \sum_{m=0}^{\infty}\sum_{n=0}^{\infty}
\tilde{\eta}_{mn}(R)\ e^{i m \phi}\ \mathrm{He}_n\left(\frac{z}{h(R)}\right),
\label{eq:decomposition}
\end{equation}
where $\tilde{\eta}_{mn}(R)$ are the radial coefficients, $m$ is the azimuthal mode number, and $\mathrm{He}_n$ is the $n$-th order probabilist's Hermite polynomial. For reference, He$_0(z)$ = 1; He$_1(z)$ = $z$; and He$_2(z)$ = $z^2-1$. 

In warp disk theory, the warp is communicated by the propagation of the bending waves \citep{papaloizou95, lubow00}, which correspond to the $m=1, n=1$ mode. This is because slightly tilting a Keplerian annulus from the reference midplane induces variations once per orbit and density perturbations that are odd about the reference midplane. In our simulations, the presence of such a mode is also hinted at by the $m=1$ motions in each hemisphere and approximate $180^\circ$ phase shift between the two hemispheres seen in Figures \ref{fig:phi_theta_view}. Motivated by this, we decompose the shadow-induced temperature perturbations into a Fourier–Hermite series. Figure \ref{fig:decomposition_example} presents several of the lowest-order modes that are most relevant for interpreting the thermal induced features discussed in this paper. Two of these models have been studied recently, the $m=1,n=0$ mode that excites disk eccentricity \citep{qian24} and the $m=2, n=0$ mode that launches two spiral density waves \citep{zhu25}.

\begin{figure}
    \centering
\includegraphics[width=1.0\linewidth]{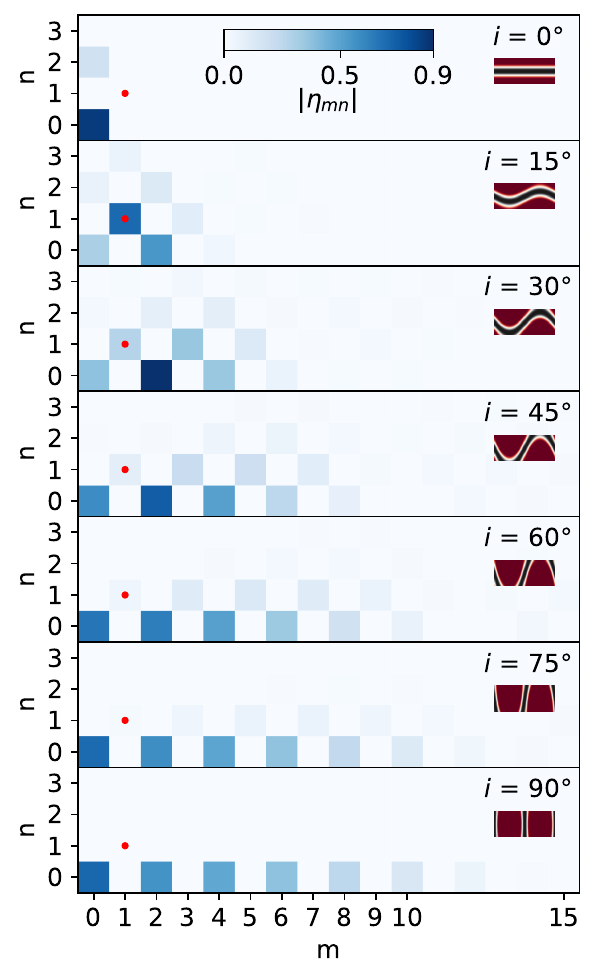}
    \caption{Contribution of Fourier-Hermite modes in a shadowed disk model transitioning from coplanar to polar configurations in 15$^\circ$ increments.
Modes are shown up to $n = 3$ and $m = 15$. The warp mode ($m = 1$, $n = 1$) is highlighted with red dots. Insets show the corresponding temperature distributions in the $\phi$–$z$ plane. $\eta_{mn} = A_{0,mn}/A_0$.}
\label{fig:shadow_decomposition_heatmap}
\end{figure}

In Figure \ref{fig:shadow_decomposition_heatmap}, we project the shadow temperature profile $g(\theta_s)$ introduced in Equation \ref{eq:angular_dependence} ($\sigma_s=0.05$, $\sigma_t=0.35$, and an amplitude $A_0$) onto Hermite–Fourier modes up to $n=3$ and $m=15$, for mutual inclinations between $0^\circ$ and $90^\circ$ in $15^\circ$ increments. In particular we expand the fractional temperature perturbation, after growing to its full amplitude, according to
\begin{align}
    \frac{T-T_{bg}(R)}{T_{bg}(R)} &= A_0 g(\theta_s) \nonumber \\&= \sum_{m=0}^{\infty}\sum_{n=0}^{\infty}
A_{0,mn}\ e^{i m \phi}\ \mathrm{He}_n\left(\frac{z}{h(R)}\right).
\label{eq:fraction_form}
\end{align}
We show the normalized modal strength against the shadow amplitude $\eta_{mn}$ = $A_{0,mn}/A_{0}$ in the figure, so that $\eta_{mn}$ is independent of the absolute amplitudes. For instance, if the full shadow's amplitude $A_0$ = 1\%, $\eta_{mn}=0.9$ indicates $A_{0,mn}$ = 0.9\% from that mode. Note that for $n=1$ mode, the temperature modification $f(\theta_s, t>t_{\rm grow})$ (Equation \ref{eq:temperature_modification}) reaches 1 + $A_{0,11}$ at one gas scale height ($h$) and $\phi=\pi/2$ according to Equation \ref{eq:decomposition}. At the midplane, the perturbation vanishes, while it becomes larger than 1 + $A_{0,11}$ beyond $h$. In contrast, for $n=0$ modes, the amplitude remains equal to $A_{0,m0}$ at all $z$. The inset in the upper right corner illustrates the shadow shape in $\phi-z$ plane, while the $m=1,n=1$ component is highlighted in red.

For $i=0^\circ$, the dominant nontrivial mode is $m=0,n=2$, corresponding to the symmetric two-layer vertical temperature structure that enhances vertical shear and modifies the vertical shear instability \citep{zhang24, yun25a, yun25b}. At $i=15^\circ$, the $m=1,n=1$ mode dominates, which we hypothesize excites a warp. At $i=30^\circ$, the $m=2,n=0$ mode becomes the strongest, driving two-armed spiral waves—consistent with the prominent $m=2$ spirals seen in our transition disk simulations. The second strongest component at $i=30^\circ$ is $m=3,n=1$ mode, which accounts for the $m=3$ spiral features visible in $v_\theta'$ in Figure \ref{fig:cart_slices}. The third strongest contribution is again the $m=1,n=1$, which is associated with the bending wave. At higher inclinations, the $m=2,n=0$ component remains the dominant non-axisymmetric mode, but the $m=1,n=1$ contribution never vanishes except in the exactly polar ($i=90^\circ$) case.

\begin{figure}
    \centering
    \includegraphics[width=1.0\linewidth]{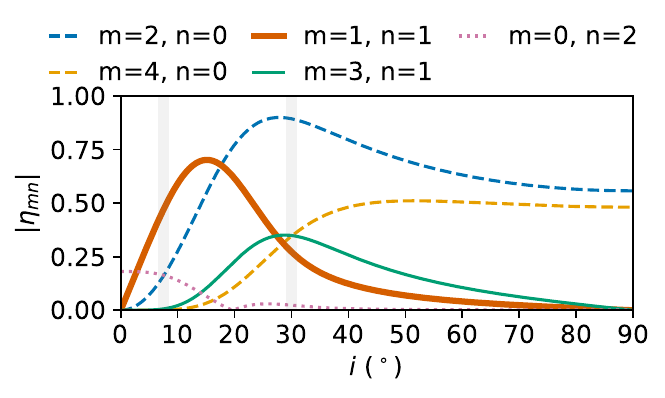}
    \caption{Mode contributions for selected $(m, n)$ modes shown in Figure \ref{fig:shadow_decomposition_heatmap}.
Odd-$m$ modes are plotted with solid lines, even-$m$ modes with dashed lines, and the ground mode with a dotted line. The warp mode ($m = 1$, $n = 1$) is highlighted in orange. Vertical lines mark inclinations of $i = 7.5^\circ$ and $30^\circ$, where simulations with prescribed temperature are performed. $\eta_{mn} = A_{0,mn}/A_0$.}
    \label{fig:select_mode_amplitude}
\end{figure}

Figure \ref{fig:select_mode_amplitude} quantifies these trends as a continuous function of mutual inclination. The $m=1,n=1$ mode (orange solid line) peaks near $i=15^\circ$, but still contributes more than 25\% of the total amplitude over the range $3^\circ \lesssim i \lesssim 30^\circ$. For reference, although the Solar System is often treated as nearly coplanar, the Sun’s equator is tilted by $\sim$ 7$^\circ$ relative to the Earth’s orbital plane. A commonly perceived small misalignment, 7$^\circ$, is already sufficient to generate significant $m=1,n=1$ components to warp the disk. The $m=2,n=0$ component (blue dashed line) rises steeply, reaches a maximum before $i=30^\circ$, and dominates (over 50\%) at larger inclinations. This explains why a 30$^\circ$ shadow can induce stronger accretion than a 90$^\circ$ shadow: the $m=2,n=0$ amplitude at 30$^\circ$ is about one-third larger than at 90$^\circ$. The $m=3,n=1$ mode (green line) peaks around $i=30^\circ$, consistent with the $m=3$ spiral features seen in $v_r'$ in Figure \ref{fig:cart_slices}. At higher inclinations, the $m=4,n=0$ mode emerges as the second-strongest contribution, though it may be observationally degenerate with $m=2,n=0$ since it also contributes to the $m=2$ spirals, while $m=4$ spirals are not clearly visible \citep[see Figure 8 in ][]{zhu25}. Finally, the $m=0,n=2$ mode dominates in the nearly coplanar regime ($i \lesssim 2^\circ$).

One can also consider the shadow projection in the transformed frame, which shifts into the primed warped coordinates at each instant of the evolving geometry. Within this perspective, the modal contributions can also change in time as the mutual inclination between the shadow and disk annuli evolve. 
For example, starting from an initial 30$^\circ$ mutual inclination between the shadow and the disk (\texttt{R30} and \texttt{H30}), the amplitude of the $m=1,n=1$ mode increases as the mutual inclination decreases, peaking around $i=15^\circ$ (Figure \ref{fig:select_mode_amplitude}), after which it declines with further reductions in mutual inclination. However, because no regions in the \texttt{R30} or \texttt{H30} models ever reach a mutual inclination smaller than 5$^\circ$ during the evolution (Figures \ref{fig:warp_properties_evolution} and \ref{fig:hydro_warp_evolution}), the $m=1,n=1$ mode amplitude remains higher throughout the simulation than it was at the initial condition.

\begin{figure*}
    \centering
    \includegraphics[width=1.0\linewidth]{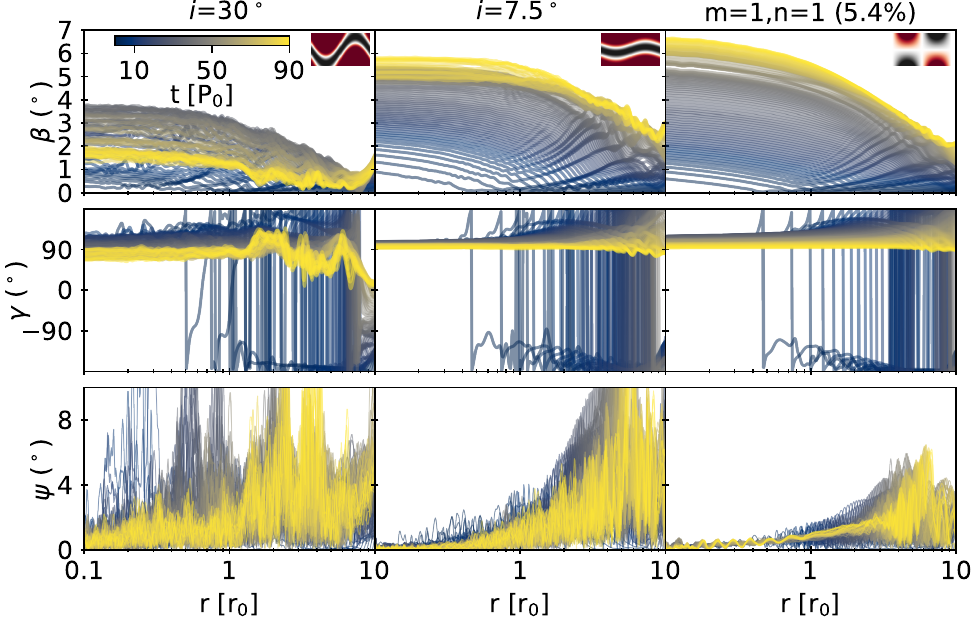}
    \caption{Time evolution of tilt (top), twist (middle), and warp (bottom) for three hydrodynamical models, similar to Figure \ref{fig:warp_properties_evolution}.
From left to right: 30$^\circ$ shadow (\texttt{HF30R}), 7.5$^\circ$ shadow (\texttt{HF7R}) with $A_{0}$ = 20\% amplitude, and $m=1, n=1$ shadow models with $A_{0,11}$ = 5.4\% amplitude (\texttt{HFm1n1R0p054}). The temperature perturbation forms are shown in the insets in the $\phi$–$z$ plane.}
    \label{fig:warp_prop_evolution_hydro_smaller}
\end{figure*}

\subsubsection{Isolating the Role of $m=1, n=1$ Mode in Warping}
We now test the role of the $m=1,n=1$ mode in warping the disk using dedicated hydrodynamical simulations. To maintain control and facilitate future comparison with linear theory, we employ higher resolution, power-law density profiles, reflecting boundary conditions, and a reduced shadow amplitude of $A_0$ = 20\% (\texttt{HF30R}). A comparison with the modified outflow boundary condition run (\texttt{HF30Mo}) is in Appendix \ref{sec:boundary_condition}. Aside from a 30$^\circ$ inclined shadow, we also perform a simulation with only the $m=1,n=1$ mode imposed with $A_{0,11}$ = 5.4\% temperature perturbation (\texttt{HFm1n1R0p054}), corresponding to its fractional contribution in the 30$^\circ$ inclined shadow with $A_{0}$ = 20\% total amplitude. Since the full 30$^\circ$ shadow also excites strong $m=2,n=0$, $m=1,n=3$, and $m=4,n=0$ modes (Figures \ref{fig:shadow_decomposition_heatmap} and \ref{fig:select_mode_amplitude}), we further include a 7.5$^\circ$ inclined shadow model in which the $m=1,n=1$ mode dominates with a weaker contribution of $m=2, n=0$ mode (\texttt{HF7R}). This setup allows us to disentangle the effects of the $m=1,n=1$ and $m=2, n=0$ components from other modes.

The evolution of the tilt, twist, and warp for these three models (\texttt{HF30R}, \texttt{HF7R}, \texttt{HFm1n1R0p054}) is shown in Figure \ref{fig:warp_prop_evolution_hydro_smaller}, with insets in the top-right corner illustrating their temperature perturbations in the $\phi$–$z$ plane. Immediately, we see that the pure $m=1,n=1$ mode indeed produces tilt and twist in the disk, consistent with our previous simulations, and that all three models exhibit qualitatively similar behavior. From left to right, we plot the $i=30^\circ$, $i=7.5^\circ$, and pure $m=1,n=1$ cases. The tilt profiles remain nearly constant within $r \lesssim r_0$ and decrease in the outer disk. In the $i=30^\circ$ model (\texttt{HF30R}), the maximum tilt reaches $\sim 4^\circ$ at 40 orbits before declining, whereas the $i=7.5^\circ$ and $m=1,n=1$ models (\texttt{HF7R}, \texttt{HFm1n1R0p054}) continue to grow, reaching $\sim 6$–$7^\circ$. This difference suggests that the evolution of the tilt in the $i=30^\circ$ case is more strongly affected by the forcing induced by other modes, while the $i=7.5^\circ$ case more closely follows the evolution of the pure $m=1,n=1$ mode. Note that the $i=7.5^\circ$ run (\texttt{HF7R}) is also affected by the presence of $m=2,n=0$ mode, since it is less tilted than \texttt{HFm1n1R0p054}, but it contains a stronger $m=1,n=1$ mode than $A_{0,11}$ = 5.4\% (vertical lines in Figure \ref{fig:select_mode_amplitude}). As we will explore in a subsequent paper, a formal linear analysis shows that the modal equations are coupled \citep[e.g.,][]{fairbairn25}, so that an $n=3$ forcing can still communicate with $n=1$ modes. The twist in all models propagate outward at a fraction of the sound speed, with the disks ultimately reaching $\gamma \sim 90^\circ$, in alignment with the shadow twist angle. The warp grows with radius in each case, although the $m=1,n=1$ model shows smaller fluctuations throughout the disk.

\begin{figure*}
    \centering
    \includegraphics[width=1.0\linewidth]{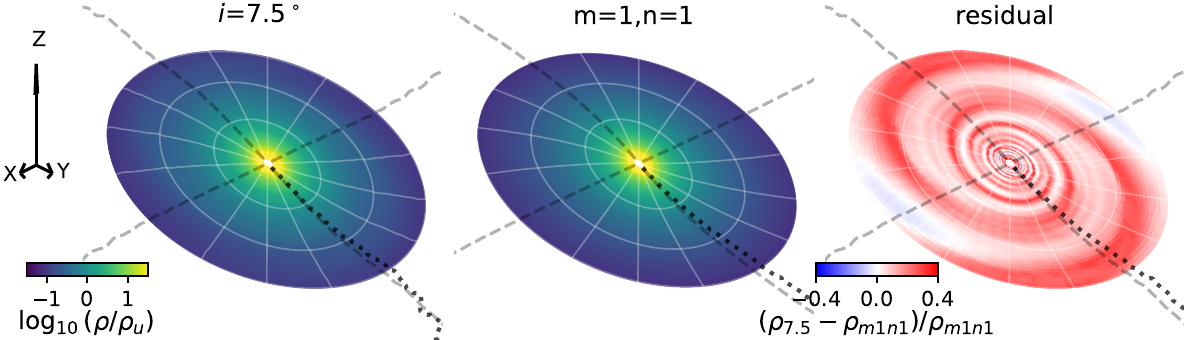}
    \caption{Left two panels: density on the warped surface for the $i=7.5^\circ$ (\texttt{HF7R}) and $m=1,n=1$ (\texttt{HFm1n1R0p054}) models at $t=90\ P_0$. This figure is analogous to Figure \ref{fig:warp_surface_evolution}, but with the $z$-axis stretched by a factor of five to emphasize the disk tilt. Right panel: fractional density residual between the two models, plotted on the warped surface of the $m=1,n=1$ model (\texttt{HFm1n1R0p054}). Only the inner 3 $r_0$ part of the disk is highlighted. Three concentric grid lines are at $r$ = 1, 2, and 3 $r_0$. Because the warped surfaces are nearly identical, plotting the residual on the $i=7.5^\circ$ surface (\texttt{HF7R}) yields an almost identical result.}
    \label{fig:similarity_7p5_m1n1}
\end{figure*}

We visualize the similarities and differences between the $i=7.5^\circ$ tilted shadow model \texttt{HF7R} and the $m=1,n=1$ mode \texttt{HFm1n1R0p054} in Figure \ref{fig:similarity_7p5_m1n1} by comparing their warped surfaces and density differences at t=90 $P_0$. The $z$-axis is stretched by a factor of five to emphasize the disk tilt. The two models show strong similarities: the tilt (seen as the elevated top-left and lowered bottom-right sides) and the twist (traced by the dotted lines) are nearly identical within $r < 3\ r_0$, confirming that the $m=1,n=1$ component drives the disk warp. The right panel shows the fractional density difference on the warped surface, which reveals prominent $m=2$ spiral residuals with amplitudes up to 40\%, arising from the $m=2,n=0$ contribution that remains in the $i=7.5^\circ$ model (\texttt{HF7R}).

In Appendix~\ref{sec:detailed_flow_pure_hydro} and Figure~\ref{fig:residual_warp_surface_hydro}, we present the residual flow structures of the \texttt{HF30R} and \texttt{HF7R} models after correcting for the warp, following the same procedure as in Figure~\ref{fig:cart_slices}. The features in both the density and velocity fields are well explained by the Fourier–Hermite analysis and have direct implications for observations.

\subsection{Inner-Outer Disk Twist by Density Cutoff}
\label{sec:twist}
\begin{figure}
    \centering
    \includegraphics[width=1.0\linewidth]{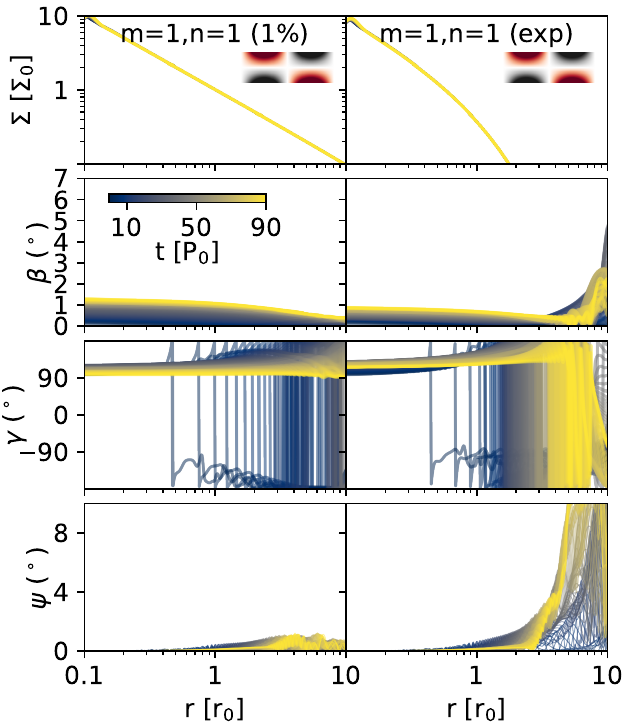}
    \caption{Similar to Figure \ref{fig:warp_properties_evolution}, time evolution of surface density (first column), tilt (second column), twist angle (third column), and warp amplitude (fourth column) for temperature perturbation in the form of $m=1, n=1$ with amplitude $A_{0,11}$ = 1\%, without and with exponentially cutoff in the outer disk (\texttt{HFm1n1R0p01} and \texttt{HFTm1n1R0p01}). The temperature perturbation forms are shown in the insets in the $\phi$–$z$ plane.}
    \label{fig:inner_outer_twist}
\end{figure}

We have so far identified that the $m=1,n=1$ temperature perturbation warps the disk, while other modes create substructures in the density and velocity fields. {One property that remains to be explained is the twist between the inner portion of the outer disk, which contains the majority of the mass, and the exponentially tapered region of very low density found in our transition disk models}. In contrast, this effect does not occur in the full disk models shown in Figure \ref{fig:warp_prop_evolution_hydro_smaller}. An obvious difference between these two models is the absence of an exponential cutoff in the full disk. To test this, we run two additional simulations with even smaller perturbations: $A_{0,11} = 1\%$ for the $m=1,n=1$ mode. One model is a full disk (\texttt{HFm1n1R0p01}) and the other includes an exponential cutoff (\texttt{HFTm1n1R0p01}). The results are shown in Figure \ref{fig:inner_outer_twist}. For the model without an exponential cutoff (\texttt{HFm1n1R0p01}), the $A_{0,11}$ = $1\%$ perturbation leads to about $1\%$ tilt in the inner disk, and the twist remains $\sim 90^\circ$ throughout the disk. In contrast, the model with the exponential cutoff (\texttt{HFTm1n1R0p01}) shows a slightly lower tilt in the inner disk but a much higher tilt in the outer disk. At the location of the dip in the tilt, the twist also deviates from $90^\circ$ and points toward -90$^\circ$ at the outer disk. The warp amplitude likewise grows to be much larger than in the no--cutoff case. {The greater warp amplitude in the tapered low density region could be an effect caused by bending waves: as the density becomes lower, the amplitude of the bending wave increases. The amplitude decreases again when the wave is reflected back to the denser region.} All of these properties echo what we find in the transition disk model (\texttt{R30} and \texttt{H30}). Thus, we conclude that the inner--outer disk twist originates from the exponential cutoff of the outer disk. By now, with our suite of simulations summarized in Table \ref{tab:models}, we have explained nearly all the key features seen in our fiducial radiation--hydrodynamical simulation (\texttt{R30}).

\subsection{Scaling Relation}
\label{sec:scaling}
As we have identified, it is the $m=1,n=1$ mode in the temperature perturbation that primarily leads to the warp. An immediate question is: what amplitude of tilt can such a perturbation cause? {With a few models (Table \ref{tab:models}) in hand, we can answer this tentatively yet semi-quantitatively.} In our simulations, an $A_{0,11}$ = 1\%, $m=1,n=1$ temperature perturbation (in the form of $f(\theta_s, t>t_{\rm grow}) = $ 1 + $A_{0,11}g_{11}(\theta_s)$, where $g_{11}(R,\phi,z) = (z/h) \, \sin\phi$,  Equation \ref{eq:fraction_form}) produces a tilt of $\sim 1^\circ$ (Figure \ref{fig:inner_outer_twist}), while an $A_{0,11}$ = $5.4\%$ perturbation leads to a tilt of about $6^\circ$ (Figure \ref{fig:warp_prop_evolution_hydro_smaller}). The fiducial radiation-hydrodynamical simulation has an $A_{0,11} \approx$ 14\% perturbation in the $m=1,n=1$ mode, resulting in a tilt of $\sim 20^\circ$ (Figure \ref{fig:warp_properties_evolution}). Thus, a rough scaling emerges: an $A_{0,11}$ = $1\%$ perturbation in the $m=1,n=1$ mode corresponds to a tilt of $\sim 1^\circ$. 

This relation, however, depends on the global disk setup. For example, the pure hydrodynamical transition disk model (\texttt{H30}) has the same shadow amplitude as the radiation-hydrodynamical runs (\texttt{R30}), but only reaches a tilt of $\sim 10^\circ$ (Figure \ref{fig:hydro_warp_evolution}). Likewise, the $i=30^\circ$ full-disk simulation (\texttt{HF30R}) shows a maximum tilt of $\sim 4^\circ$ (Figure \ref{fig:warp_prop_evolution_hydro_smaller}) instead of the $\sim 5.4^\circ$ predicted by the scaling. These deviations likely arise because in the 30$^\circ$ inclined cases, other modes—such as spiral density waves—dominate and interfere with the propagation of the bending wave. 

Nevertheless, all cases remain within a factor of two of this scaling relation, despite the presence or absence of an inner cavity, the exponential cutoff, different shadow shapes, and the excitation of additional modes. As a rule of thumb, we therefore conclude that an $A_{0,11} = 1\%$ $m=1,n=1$ temperature perturbation produces a disk tilt of $\sim 1^\circ$. In other words, the tilt scales approximately as $\beta \sim A_{0,11}$.

Except for the truncated outer disk, which twists in a different direction, the primary role of the shadow is to tilt the disk toward alignment with this preferred, attenuating plane. The tentative scaling relation therefore quantifies how much the shadow can reduce the mutual inclination between the outer disk and itself. The disk’s ability to align (or the amplitude of the $m=1,n=1$ mode) depends on the instantaneous mutual inclination, not just the initial condition. This scaling thus incorporates the time-varying inclination: as the disk tilts closer to the shadow, the $m=1,n=1$ forcing weakens, and the quoted scaling corresponds to the maximum tilt achieved, or equivalently, the minimum mutual inclination between the shadow and the outer disk. It remains unclear whether a strong temperature perturbation that is caused by a misaligned shadow could fully align the outer disk; in our simulations this does not occur, likely due to {conservation of angular momentum}.

\subsection{Periodicity in the Evolution of the Tilt}
\label{sec:period}

\begin{figure}
    \centering
    \includegraphics[width=1.0\linewidth]{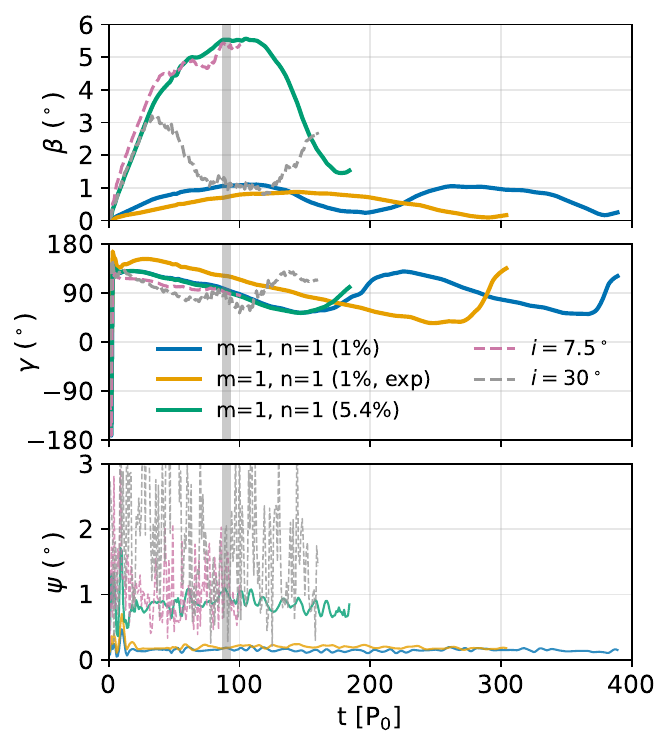}
    \caption{Long-term evolution of tilt (top), twist (middle), and warp (bottom) at $r = r_0$ across all full-disk models (blue: \texttt{HFm1n1R0p01}, orange: \texttt{HFTm1n1R0p01}, green: \texttt{HFm1n1R0p054}, dashed pink: \texttt{HF7R}, dashed gray: \texttt{HF30R}). The vertical line marks the time corresponding to the snapshots shown in Figures \ref{fig:similarity_7p5_m1n1} and \ref{fig:residual_warp_surface_hydro} and the end point of the evolution shown in Figures \ref{fig:warp_prop_evolution_hydro_smaller} and \ref{fig:inner_outer_twist}.}
    \label{fig:tilt_period}
\end{figure}

The decline of the tilt with time in Figure \ref{fig:warp_prop_evolution_hydro_smaller} for the $i=30^\circ$ case raises the question of whether the tilt eventually decays to zero and whether other simulations would also exhibit a decline at later times. To investigate this, we extend the full disk hydrodynamical simulations for longer times and plot the tilt, twist, and warp amplitude at $r=r_0$ in Figure \ref{fig:tilt_period}.  

All full disk models show clear oscillatory behavior: the tilt and twist vary periodically on timescales of 100--300 orbits. The blue solid line shows the $m=1,n=1$, $A_{0,11}=$ 1\% model (\texttt{HFm1n1R0p01}). The disk tilt peaks at $\sim 1^\circ$ near 110 orbits, decreases to nearly zero by 200 orbits, and then grows again. The twist decreases from $135^\circ$ to $\sim 45^\circ$ and recovers to $\sim 135^\circ$, completing a cycle. At maximum tilt, the twist aligns with the shadow direction at $\sim 90^\circ$. {The evolution timescale is the same across the whole disk.}

The $m=1,n=1$, $A_{0,11}$ = 5.4\% case (green solid line, \texttt{HFm1n1R0p054}) follows the same evolution, with tilt and warp amplitudes scaled up by a factor of 5.4. The twist evolution is nearly identical to the 1\% case (\texttt{HFm1n1R0p01}). When an exponential cutoff is applied (yellow solid line, \texttt{HFTm1n1R0p01}), the period lengthens to $\sim 300$ orbits while the tilt amplitude remains $\sim 1^\circ$, showing that disk truncation mainly affects the period.  

The $i=7.5^\circ$ case (\texttt{HF7R}, pink dashed line) closely tracks the $m=1,n=1$, 5.4\% case throughout the evolution. The $i=30^\circ$ case (\texttt{HF30R}, gray dashed line) initially follows the same trend but shows a more rapid tilt decay, reaching $\sim 1^\circ$ by 100 orbits before increasing again. This explains the lower tilt amplitude seen in Figure \ref{fig:warp_prop_evolution_hydro_smaller}. The period for this case is shorter, $\sim 100$ orbits, and the warp amplitude shows larger fluctuations, suggesting interference from other modes such as spiral density waves.  

While clear periodicity is established, its origin and controlling factors remain uncertain. Our results suggest that the disk’s outer boundary (e.g. whether including the exponential cutoff) influence the period. Furthermore, the radial temperature structure, which determines the bending wave speed, $c_s$/2 \citep{lubow00} may also set a characteristic timescale.
None of the transition disk models (\texttt{R30} and \texttt{H30}, Figures \ref{fig:warp_properties_evolution} and \ref{fig:hydro_warp_evolution}) show signs of declining tilt yet. Due to computational limitations, we cannot yet determine whether transition-disk models ultimately experience a decline in tilt. Their longer periods may be explained by the exponential cutoff, different temperature distributions, larger outer radii, or nonlinear effects. 
All these factors indicate that this periodicity is a property of the global wave propagation. {We consider it unlikely that this feature is a grid effect, as our resolution matches that of the high-resolution model (their \texttt{theta\_resolution}) in \citet{kimmig24}, where no significant oscillations are seen (see their Figure~8).} Future work is needed to clarify the drivers of this periodic evolution. 

\section{Discussion}
\label{sec:discussion}

\subsection{Conditions for the Warp}
\label{sec:condition}

Our simulations show that shadow-induced warping is robust as long as the $m=1, n=1$ temperature perturbation is present\footnote{Other $m=1$, $n=3, 5, 7, ...$ (odd modes) can also couple and contribute to $m=1,n=1$ mode \citep{fairbairn25}, though their amplitudes are much weaker than the $m=1,n=1$ mode (Figure \ref{fig:shadow_decomposition_heatmap}).}. Thus, the condition for warping reduces to the requirement that an $m=1, n=1$ temperature perturbation can be sustained. This requirement has been studied extensively and depends on the competition between disk advection and cooling \citep{cassasus18,zhu25}. The timescale of the disk advection is the shadow crossing time $t_{\rm cross}$, whereas the timescale of the the disk cooling is $t_{\rm cool}$. Once the disk receives asymmetric irradiation (e.g., from a misaligned shadow), it must thermally respond before azimuthal rotation averages out the asymmetry, thus  $t_{\rm cool} < t_{\rm cross}$. 

Focusing on the $m=1$ perturbation and a stationary shadow, the cooling criterion is
\begin{equation}
    t_{\rm cool} < t_{\rm cross} = \frac{2\pi}{\Omega},
\end{equation}
or equivalently,
\begin{equation}
    \beta_c \equiv t_{\rm cool}\Omega < 2\pi.
\end{equation}

This condition is generally satisfied in the outer regions of protoplanetary disks (beyond tens of au; \citealt{zhu15}). Importantly, it is also much less stringent than the cooling requirements for other hydrodynamical instabilities, such as the vertical shear instability, which requires $\beta_c \lesssim 0.1$ \citep{lin15}. By contrast, the inner disk is expected to have much longer cooling times when cooling proceeds primarily through dust thermal emission due to its higher optical depth. Future observations of key line coolants (e.g., [O \textsc{i}], \citealt{gorti11}) with the PRIMA observatory, combined with theoretical calculations of line-cooling efficiencies, will help constrain inner-disk cooling times and determine whether this warping mechanism can also operate at smaller radii.

In our simulations, the shadow is assumed to be stationary. Whether a rotating shadow can also warp the disk has not yet been explored; we will investigate this in a subsequent paper. Because bending waves propagate through the entire disk, how they globally manifest depends on boundary conditions (also see discussion in Section \ref{sec:why1}). Resonances within the disk can act as effective boundaries, reflecting waves while simultaneously exciting density waves. In our current models, the shadows are stationary, so no resonances occur. However, we suspect that the presence of resonances in a rotating shadow could alter, or even suppress, bending-wave propagation. {If the shadow rotates rapidly and the associated resonances lie within the disk, we expect spiral density waves to be launched at the Lindblad resonances \citep{montesinos18, su24, zhu25}. However, future simulations are needed to investigate the global morphology of the spirals produced by the $m=1, n=1$ and $m=3, n=1$ perturbations. Because these perturbations are antisymmetric across the $z$-direction, the resulting spiral patterns may differ from those of the previously studied $n=0$ modes.} If the shadow rotates slowly, or in a retrograde direction relative to the disk’s rotation, such resonances would not occur. Below, we summarize the conditions under which the stationary (non-rotating) shadow approximation is valid.

If the inner--outer disk misalignment arises from infall, the inner disk is not precessing, so the shadow is stationary and this condition is safely satisfied. If the inner disk is precessing due to an inclined binary or planet, the precession rates have been well studied in previous literature (e.g., see Equations 56 and 58 in \citealt{zhu25}). With reasonable assumptions on the disk profiles {(a surface density profile $\Sigma(R)\propto R^{-1}$ and the outer edge of the inner disk is half of the planet location, i.e., $R_{\mathrm{out}}/R_p=0.5$)}, the precession due to a misaligned planet is
\begin{equation}
    \frac{\omega_s}{\Omega_p} \sim 0.1 \frac{M_p}{M_*},
\end{equation}
where $\omega_s$ is the precession rate of the inner disk, $\Omega_p$ is the planet's orbital frequency, $M_p$ is the planet mass, and $M_*$ is the stellar mass. {$\omega_s$ is derived using the m=0 component of the planetary potential and has been previously derived in \citet{terquem98, bate02, lai14, zhu19}}. For a Jupiter-mass planet, $\omega_s / \Omega_p \sim 10^{-4}$, meaning that the shadow is essentially stationary relative to the outer disk. If the inner and outer disks are only mildly misaligned, the inner disk precesses in the opposite direction to the outer disk rotation, and no Lindblad resonances occur in the disk.

As for the binary scenario (i.e., the inner disk is circumbinary),
\begin{equation}
    \frac{\omega_s}{\Omega_{\mathrm{out}}} \sim 0.1 ,
\end{equation}
where $\Omega_{\mathrm{out}}$ is the orbital frequency at the inner disk's outer radius. {$\omega_s$ is derived in \citet{lubow18} where they apply the secular binary potential in the quadrupole approximation developed by \citet{farago10}}. Here the outer disk still has sufficient radial extent for the warp to develop before reaching the corotation radius, which may in fact lie beyond the truncated outer disk. In addition, the binary perturbation also tends to drive retrograde nodal precession, which would also suppress resonances in a prograde rotating outer disk.

Taken together, these scenarios suggest that the condition of a slowly moving inner-disk shadow is generally satisfied. Even if this condition is not met, the disk cannot remain in hydrostatic equilibrium with the shadow-induced asymmetric temperature structure and must respond dynamically. The exact signatures of such perturbations will be studied in a future paper. 
In summary, an inclined shadow can induce a temperature perturbation only if the cooling time is short enough for the gas to respond as it passes through the shadow. Thus, the combination of an inclined shadow and fast cooling is sufficient to generate temperature perturbations, which in turn inevitably drive dynamical perturbations.
Since in most plausible scenarios the inner disk precesses only slowly and the cooling is fast, the condition for warping in the outer disk should be met in the majority of cases. {Since the $m=1, n=1$ warping mode is significant for shadows inclined by $\lesssim$ 30$^\circ$, all the disks showing wide shadows can be subject to this warping mode, such as HD 143006 \citep{benisty18,codron25}, HD 139614 \citep{muro-arena20}, TW Hya \citep{debes23}, MWC 758 \citep{ren18}, V1098 Sco \citep{williams25}, and V1247 Ori \citep{kraus17}.}

\subsection{Why does an $m=1,\,n=1$ temperature perturbation warp the disk?}
\label{sec:why1}
Another puzzle that we have not fully explained is why the purely thermal perturbations in our simulations warp the disk in such a way that the net angular momentum vector evolves in time. From the perspective of angular momentum conservation, a change in the total angular momentum always requires an external torque. In the existing literature, such torques are usually attributed to inclined planets \citep{tanaka04, nealon19, zhu19, ballabio21}, binaries \citep{facchini13, facchini18, nealon19, fragner10, young23, rabago24}, stellar flybys \citep{cuello19b, nealon20, smallwood24}, or misaligned stellar magnetic fields \citep[][]{bouvier99, lai08}. In the context of black hole accretion disks, the torque can be due to the misalignment between the disk and the black hole spin (the Lense–Thirring effect, \citealt{bardeen75}).  

In contrast, a temperature perturbation induces a pressure perturbation, which is part of the internal force. The resulting torque must therefore be internal. Such internal torques can radially redistribute angular momentum throughout the disk, allowing warps to emerge from the non-equilibrium initial conditions. This appears to be the case in our radiation–hydrodynamical simulation (\texttt{R30}, Figure \ref{fig:warp_properties_evolution}) as well as in \texttt{H30} (Figure \ref{fig:hydro_warp_evolution}), where the outer disk warps in nearly the opposite direction. However, angular momentum is not strictly conserved: the relative twist between the inner ($\sim$ 80 au) and outer ($\sim$ 640 au) parts of the outer disk is not exactly $180^\circ$, so the $L_x$ and $L_y$ components of the two parts do not perfectly cancel. This non-conservation can be explained by a transport of angular momentum flux through our numerical boundaries. In fact, we find substantial $L_x$ and $L_y$ fluxes through the upper and lower $\theta$ boundaries, which dominate over the transport through the radial boundaries. Even with reflecting boundaries on radial and polar directions, $L_x$ and $L_y$ are not conserved. The non-conservation of angular momentum is more evident in the full-disk models without an exponential cutoff (\texttt{HF7R}, \texttt{HF30R}, \texttt{HFm1n1R0p054}, and \texttt{HFm1n1R0p01}). In this case, the whole disk develops the same twist (Figure \ref{fig:warp_prop_evolution_hydro_smaller}), so the domain-integrated angular momentum is certainly not conserved.

We will attempt to address this question in a future work, using domains that extend to the poles to prevent $L_x$ and $L_y$ flowing through these boundaries, as well as simulations in 3D Cartesian coordinates. This vertical flux of angular momentum presents a possible interpretation wherein the disk is effectively separated into two parts, the midplane region and the surface region, with angular momentum flowing between them. These surface layers can act as an environment through which the disk redistributes angular momentum, allowing the total to be conserved globally. One evidence that the surface and midplane have angular momenta with different directions is the strong shocks between them as in Figure \ref{fig:phi_theta_view}. Such oblique shocks may also be responsible for transferring angular momentum between them. An analogy may be drawn with the role of magnetohydrodynamic (MHD) winds, which remove angular momentum vertically and thereby enable disk accretion.

Following this reasoning, the condition for a disk to warp requires more than what we outlined in Section \ref{sec:condition}. In addition to excitation of the $m=1,n=1$ mode, the disk must be able to redistribute angular momentum between its inner and outer regions and/or exchange angular momentum vertically through its upper and lower surfaces (or with the surrounding environment). If the disk is restricted to a narrow radial annulus and confined vertically without any channel for angular momentum redistribution or exchange, the ring cannot tilt. Testing this requirement will be an important goal for future work.

The angular momentum constraint also leaves uncertain whether the disk can ultimately align with the shadow. In the absence of this constraint, the shadow ($m=1,n=1$ mode) would act like any other external torque and drive alignment, as we indeed see in the inner part of the outer disk in Figures \ref{fig:warp_properties_evolution}, \ref{fig:hydro_warp_evolution}, and \ref{fig:warp_prop_evolution_hydro_smaller}. However, angular momentum conservation forbids full alignment. The inner part of the outer disk never aligns exactly with the shadow, while the twist and tilt of the tapered region evolves to become even more misaligned with the shadow than in the initial condition.

\subsection{What leads to the initial shadow asymmetry?}
\label{sec:why2}
Observations leave little doubt that many disks exhibit asymmetric shadows \citep{benisty22}. Tens of such disks have been confirmed to host misaligned inner disks through optical/IR interferometry \citep{bohn22, codron25}. The prevalence of dippers also indicates that the inner and outer disks are often misaligned \citep{ansdell20}. A recent study comparing the stellar spin directions and the outer disks also indicates that at least one-third of sun-like stars are born with misaligned disks with $i\gtrsim 15^{\circ}$  \citep{biddle25}. All of the mechanisms listed in the previous subsection could contribute to the initial asymmetry. Of course, the warp can also be due to specific initial/boundary conditions, perhaps caused by environmental interactions such as late stage infall \citep{kuffmeier21, krieger24}.  Our results are largely agnostic to the exact cause of asymmetric irradiation. As noted in Section \ref{sec:condition}, however, the shadow-induced warp preferentially responds to stationary or slowly varying shadows.

Importantly, irradiation asymmetry does not necessarily require shadows. The stellar emission itself can be asymmetric (e.g., hotspots or cool spots). However, these variation happens on much shorter timescales, so it remains uncertain whether they will exert a persistent or time-averaged asymmetric thermal perturbation on the outer disk. 


A deeper question is whether an initially coplanar disk could spontaneously develop an asymmetric shadow and warp through some kind of instability. This would be analogous to the warp instability proposed by \citet{pringle96}, or to mechanisms similar to the thermal wave instability \citep[or irradiation instability,][]{watanabe08,ueda21b,wu21} that has been suggested to generate substructures in protoplanetary disks \citep{pavlyuchenkov22a,okuzumi22,kutra24}. If such an instability exists, it would represent a major discovery in accretion disk theory with profound implications for disk evolution. Caution is warranted, however, since many of these ideas are based on simplified analytic models, whereas radiative transfer is inherently a global process. These instabilities remain unconfirmed in radiation-hydrodynamical simulations \citep[e.g., for thermal wave instability, ][]{melonfuksman22, pavlyuchenkov22b}. 

Given the uncertainties surrounding the origin of the initial warp, a more tractable problem for future study is the feedback loop between the inner and outer disk as the interactions keep on coupling outwards. As we have briefly demonstrated in Figure \ref{fig:phi_theta_view} (middle right panel), thermal forcing from stellar irradiation acts over long ranges, tightly coupling the dynamics of the whole disk, from the star to the outer disk edge. The disk at a smaller radius $r_0$ influences the region behind it at $r_1$ through shadowing, which in turn affects the disk at $r_2$, $r_3$ and so on. The shadow in turn drives warping and accretion, which may then feed back onto $r_2$ and then $r_1$ and ultimately onto the star, changing the shadow orientation and geometry felt by greater radii. This feedback has not been fully explored in the present work, since our simulated outer disk is only marginally optically thick to stellar irradiation and therefore cannot further obscure the region behind it. An optically thicker disk would better capture how much the spiral and warp formation could cascade outward and feed back inward, potentially establishing a self-sustaining cycle. {This regime can only be studied using radiation-hydrodynamical simulations.}

\subsection{Unique Observational Opportunities for Forward Modeling}
\label{sec:observation}

A key characteristic of shadow-induced dynamics lies in its long-range forcing that couples the inner and outer disk with large scale separations. Processes on sub-au \citep[e.g., J1604,][]{sicilia-aguilar20} to 10 au \citep[e.g., SU Aur,][]{ginski21} scales in the very inner disk can influence the evolution of the entire disk through shadowing, extending out to outer disk sizes of 100–1000 au \citep[e.g., see a review from ][for typical disk sizes]{bae22}. By contrast, other types of perturbations are difficult to propagate to such large scales. For example, an inclined planet at (sub-)au scales can tilt the inner disk, but without the shadowing effect such perturbations would remain confined to the broken inner disk for typical protoplanetary disk viscosities \citep[$\alpha \lesssim 10^{-3}$,][]{zhu19}. Even the cavity carved by a stellar companion reaches only about five times the binary semi-major axis \citep{hirsh20}. Note that the interaction is not one-way: while the inner disk influences the outer disk through shadowing, the strong accretion in the outer disk (Figure \ref{fig:accretion}), triggered by spirals (Figure \ref{fig:warp_surface_evolution}) and shocks (Figure \ref{fig:phi_theta_view}), can in turn channel mass back into the inner disk.

While resolving the inner disk remains at the frontier of optical/IR interferometry \citep{bohn22, setterholm25, codron25}, the outer disks where shadows are cast have already been observed with dozens of high-resolution, high-sensitivity datasets from ALMA and SPHERE, including temporal monitoring in some cases \citep{pinilla18, debes23}. In these outer disks, we can already begin to distinguish shadow-induced signatures from those produced by other mechanisms, including planet signatures.

Shadow-induced perturbations differ from other mechanisms in that they enable a forward-modeling approach. Once the temperature structure (the cause) is constrained observationally, the resulting density and kinematic responses (the consequence) can be robustly predicted and tested from observations. By contrast, inferring the presence of planets or stellar companions from disk substructures is an inverse problem: the perturbers are often undetected, the mapping from morphology to perturber properties has degeneracies, and in many cases alternative mechanisms can reproduce the observations without invoking perturbers \citep{bae22, lesur22}. Observationally, detection limits often prevent planets from being either confidently confirmed or ruled out. In other words, the cause is difficult to identify. In rare cases, where planets are firmly confirmed, such as in PDS 70 \citep{keppler18, haffert19, benisty21}, forward modeling of planet–disk interactions has successfully reproduced the observations in detail \citep{bae19}. In the same spirit, shadow-induced perturbations constitute a well-posed forward problem.
In the following subsections, we outline how this framework can be applied in practice, while leaving the detailed modeling with synthetic observations to a subsequent paper. Since a tilted shadow always includes various modes, our discussions will be on all expected shadow-induced signatures, not just limited to the $m=1,n=1$ mode that induces warps.

\subsubsection{Observing azimuthal-vertical temperature structures}
\label{sec:observing_temperature}


We first focus on probing the temperature structure (the cause), since this alone is a sufficient condition for the disk to be dynamically perturbed. The $R-z$ temperature structures have been constrained using the brightness temperature of optically thick molecular tracers such as $^{12}$CO and $^{13}$CO ($^{18}$CO and CS in some cases) \citep{dartois03, pinte18, law21, galloway25, fehr25}. Because these lines are optically thick and typically arise from a thin emitting surface, their brightness temperature directly reflects the local thermal temperature. Since different molecules trace different emission heights, they together map out the vertical temperature profile: CS is closest to the midplane, $^{18}$CO and $^{13}$CO lies in between, and $^{12}$CO originates from the most elevated layers \citep{galloway25, fehr25}. In these works, the vertical dependence of temperature between these layers can be smoothly mapped out using a function in \citet{dartois03} (which is similar to Equation \ref{eq:angular_dependence} for coplanar shadow) that connects the superheated surface and cool midplane \citep{chiang97}. Even within a single isotopologue, higher-J transitions typically probe higher emission surfaces \citep{fehr25}. For instance, an azimuthal temperature asymmetry has been detected in $^{12}$CO J=2–1 but not in J=3–2 in TW Hya \citep{teague22}.

A natural next step beyond these standard temperature retrieval methods is to relax the assumption of azimuthal symmetry. As shown in Figure \ref{fig:shadow_decomposition_heatmap}, the temperature structure induced by an inclined shadow depends on both azimuth and height, i.e., $T=T(\phi, z)$. With a tilted shadow, the disk should also exhibits front–back temperature antisymmetry. Ideally, one would reconstruct the azimuthal temperature distribution across as many disk heights as possible. In practice, scattered-light observations can tightly constrain the tilt and shape of the shadow \citep{orihara25}, enabling the $\phi-z$ temperature structure to be reconstructed even from temperature measurements at just one or two emitting layers with the help of a shape function (Equation \ref{eq:angular_dependence}). For example, \citet{teague22} measured an $m=1$ temperature asymmetry in TW Hya, a system already known to host shadows with a small tilt \citep{debes23}. This suggests that the disk temperature structure contains an $m=1, n=1$ mode that induces a warp, even without directly probing the temperature asymmetry on the back side. Another example is J1604, where the inner disk inclination has even been inferred from disk kinematics to be $\sim$45$^\circ$ \citep{mayama18}. Its azimuthal gas intensity profile shows a dominant $m=2$ perturbation with a smaller $m=1$ component (see their Figure 3), consistent with the decomposition shown in our Figure \ref{fig:shadow_decomposition_heatmap}. More encouragingly, in disks with moderate inclinations and sufficiently high emission surfaces, recent studies have demonstrated that the front and back side contributions can be separated \citep[e.g.,][]{izquierdo25}, allowing the front–back temperature antisymmetry to be directly confirmed. Gas lines in edge-on disks also offer a promising probe of top–bottom temperature asymmetries \citep{dutrey17, dutrey25}.

The azimuthal temperature variation can be decomposed into low-order Fourier modes, such as $m=0,1,2,$ and $3$, as pioneered by \citet{teague22}, who fit the $m=1$ mode after subtracting $m=0$ mode (azimuthally averaged value). This approach offers two key advantages. First, the signal-to-noise ratio (SNR) is improved because data from the full azimuthal extent of the disk can be used to fit just two parameters (amplitude and azimuthal shift). Second, as we have shown, it is precisely these individual modes that drive the disk dynamics (Figures \ref{fig:decomposition_example}, \ref{fig:shadow_decomposition_heatmap}, \ref{fig:select_mode_amplitude}). Compared to detections that require localization, such as circumplanetary disks \citep{bae22b}, this fitting method is far less demanding in terms of SNR. Face-on disks like TW Hya are ideal for sampling the full azimuthal temperature profile $T(\phi)$. However, probing the vertical profile $T(z)$ is more challenging, since robustly constraining $z/r$ for different molecular tracers generally requires thermochemical modeling to predict their emitting heights \citep[e.g.,][]{cazzoletti18} . Highly inclined disks, on the other hand, demand careful treatment of flared emission surfaces and beam dilution. An intermediate inclination is therefore likely optimal, enabling mapping of the 3D temperature structure while minimizing beam dilution.

Once the $\phi-z$ temperature structure is mapped (and ideally, if data quality permits, fitted across multiple annuli so that the full 3D $(R, \phi, z)$ temperature structure can be reconstructed), the present work, together with others along this line \citep{montesinos16, montesinos18, cuello19, su24, qian24, zhang24b, ziampras25}, will have the utmost predictive power for the dynamical consequences, which could potentially be observed and confirmed. Even a null detection is valuable: in disks that exhibit shadows in NIR but no detectable temperature asymmetry in the gas, such measurements would help exclude other instabilities, such as the vertical shear instability (VSI), which requires even shorter cooling times \citep{lin15} than the cooling time it requires to have azimuthal temperature variation ($t_\mathrm{c}\lesssim 0.1$ for VSI vs. $t_\mathrm{c}\lesssim 2\pi$, see Section \ref{sec:condition}). Under such circumstances, a planet can become a more convincing driver of the observed disk substructures and kinematics.

\subsubsection{Observing dynamical consequences}
\label{sec:observing_dynamics}

The dynamical consequences of temperature perturbations can be studied on both population and individual levels. On a population level, \citet{curone25} find a strong positive correlation between the amplitude of the asymmetry in the dust continuum and the stellar accretion rate (and NIR excess). These trends can be naturally explained by shadow-induced perturbations. With a tilted shadow, stronger temperature perturbations excite stronger spirals and shocks. These dynamical features both enhance dust asymmetry and drive higher accretion onto the inner disk. With sufficiently large samples, one could further test the dependence of asymmetry and accretion on shadow inclination. For instance, our models predict that a $30^\circ$ tilted shadow produces stronger accretion and asymmetry than a $90^\circ$ tilted polar shadow (Figures \ref{fig:accretion} and \ref{fig:phi_theta_view}). In terms of resolved imaging and kinematics, this implies that disks with misaligned inner disks closer to coplanar configurations (corresponding to wide shadows in scattered-light images) should have distinct morphological and kinematic features than those closer to polar configurations (corresponding to narrow shadows). These morphological and kinematic differences can show up in the strength and number of spiral arms, the magnitude and shape of radial and vertical motions at different emission surfaces, the tilt, twist, and warp profiles, and the presence and location of shocks, as we will discuss below.

On the level of individual disks, dynamical consequences can be probed in several ways. The most direct and quantitative test of temperature perturbations is through disk kinematics. Non-Keplerian motions can be decomposed into low-order Fourier modes, just as with the temperature structure (Section \ref{sec:observing_temperature}). The amplitudes and azimuthal shifts of these modes can then be compared between temperature and velocity fields. A close match would confirm that the observed kinematics are driven by temperature asymmetries. By contrast, if the kinematic perturbations are much stronger than those expected from the measured temperature variations, this would point to other drivers such as disk instabilities or planet–disk interactions. With some current high-resolution high-sensitivity ALMA data, kinematic substructures such as spirals can be directly traced \citep[e.g.,][]{teague22b, izquierdo23} and compared against simulations without the need of Fourier transforms.

Other signatures of shadow-induced warps are promising but still require more quantitative study before they can be firmly attributed to temperature perturbations. One approach is to measure the radial profiles of relative tilt, twist, and warp amplitude, and compare them with Figure \ref{fig:warp_properties_evolution}. In observations, only the relative tilt and twist can be constrained, since the initial disk orientation is unknown. A distinctive prediction of an inclined shadow in a disk with an exponential cutoff is an outer-disk twist approaching 180$^\circ$, together with a large warp of $\sim$30$^\circ$ between the inner and outer disks (Figure \ref{fig:warp_properties_evolution}). Indeed, \citet{winter25} recently fit the velocity fields in the exoALMA sample to infer tilt, twist, and warp amplitudes, showing that warps might be common. 
Developing robust data retrieval methods will be critical for such comparisons \citep{aizawa25}.

Disks with both 30$^\circ$ and 90$^\circ$ tilted shadows (\texttt{R30}, \texttt{R90}) exhibit an $m=2$ variation in scale height, but the milder 30$^\circ$ inclination produces a much stronger variation up to a factor of 4, making it easier to identify (Figure \ref{fig:phi_theta_view}). In this case, the midplane region is significantly thinner than in the coplanar or polar cases and does not lie on a single plane. This highlights the need to fit emission surfaces that vary with azimuth $\phi$, rather than assuming axisymmetry as in current methods \citep{izquierdo21, disksurf, rosotti25}. The front and back sides of the disk at a given azimuth can also have different thicknesses. Because the variations are dominated by low-order modes ($m=0,1,2$), a parameterized model should be sufficient.

Shocks that raise the temperature near the midplane at sharp density transitions are also evident in our 30$^\circ$ tilted shadow case (Figure \ref{fig:phi_theta_view} black arrows). In contrast, in the 90$^\circ$ case the shocks occur much higher above the midplane. This suggests that shock heating should be easier to observe in disks with mildly inclined shadows. The shocks extend over the full azimuth and many of them occur within the shadowed regions. In each hemisphere, they typically span about half the azimuth ($\sim 180^\circ$). To date, nine protoplanetary disks show SO detections that may trace such shocks \citep[see][Section 4 for a detailed discussion and references therein]{zagaria25}. Synthetic observations of these shocks will be crucial to determine their direct observational signatures and to assess whether, and to what extent, shocks observed in protoplanetary disks can be attributed to shadow-induced dynamics. {Given the very high accretion rates and complex kinematics, one might expect disks with wide shadows to have relatively thick dust layers. However, the gas scale height in the \texttt{R30} model is actually much lower than that predicted from vertical hydrostatic equilibrium due to the nozzle shocks (Figure~\ref{fig:phi_theta_view}), indicating that dust coupled to the gas could also form a thinner layer than expected. A simulation that includes dust dynamics is needed to test these two competing effects.}

Shadow-induced perturbations shape both gas and dust substructures in morphology, which can be probed through ALMA millimeter continuum and molecular line observations, as well as scattered-light imaging of small grains that remain well coupled to the gas \citep{cuello19, su24, zhang24b, zhu25, ziampras25}. Our results show, however, that a 30$^\circ$ tilted shadow produces dynamics that differ substantially from those driven by a polar shadow. This means that previously proposed observational signatures need to be revisited to account for their dependence on shadow geometry. Further analysis is also required to establish which features are uniquely attributable to shadow-induced dynamics and which could be confused with other mechanisms, such as planet–disk interactions and gravitational instabilities. Again, a critical first step in this effort is to observationally confirm the azimuthally varying temperature structure (Sectiosn \ref{sec:observing_temperature}), which provides the foundation for forward modeling.

Finally, tracking the motion of shadows in scattered light (both their direction and speed) is critical. The relative velocity between the shadow and the disk rotation affects the induced dynamics, such as warping (Section \ref{sec:condition}). Measuring this motion would therefore provide a key test of whether resonances between disk rotation and shadow motion are at play. If the shadow rocks back and forth rather than rotating steadily in one direction \citep[e.g., in J1604,][]{pinilla18, nealon20b}, the time-averaged shadow shape and intensity would provide a more realistic description of the stellar irradiation than the instantaneous morphology captured in a single epoch.

\section{Conclusion}
\label{sec:conclusion}

Motivated by recent observations of shadows and lateral asymmetries in reflection nebulae of protoplanetary disks, we carry out radiation-hydrodynamical and pure hydrodynamical simulations (Table \ref{tab:models}) to investigate the dynamical consequences of shadows cast on the outer disk in generic configurations that are neither coplanar nor polar. Our main findings are summarized below.

\begin{itemize}

\item A transition disk irradiated by a 30$^\circ$ inclined shadow (\texttt{R30}) develops a warp. The inner disk tilts toward the shadow direction, while the exponentially tapered outer disk warps in a different azimuthal direction. Their mutual inclination can reach $\sim$32$^\circ$ (Figure \ref{fig:warp_surface_evolution}). 

\item The shadow also excites two-armed spirals (Figure \ref{fig:warp_surface_evolution}) and drives strong accretion, reaching $\alpha \sim 1$ (Figure \ref{fig:accretion}). The accretion is much stronger than in the polar shadow case (\texttt{R90}) with the same geometry and strength, due to a stronger $m=2, n=0$ mode (Figure \ref{fig:select_mode_amplitude}) and enhanced radial flows near the midplane (Figure \ref{fig:phi_theta_view}).

\item The disk in \texttt{R30} has a compressed midplane compared to \texttt{R90}, and its highest density regions do not lie on a perfect plane. {At 160 au, the azimuthal scale height variation of \texttt{R30} reaches a factor of 3.6, compared to only 1.125 in \texttt{R90}.} Shocks occur near the midplane, where the disk is vertically squeezed twice in the azimuthal direction (Figure \ref{fig:phi_theta_view}). Such dynamical midplane may be responsible for the high $\alpha$. This may explain some of the recent SO emission detected in ALMA observations.

\item We identify thermal forcing from the asymmetric temperature structure as the driver of the warp. Locally isothermal simulations with prescribed temperature (\texttt{H30}) almost reproduce the warp seen in full radiation-hydrodynamical runs (Figure \ref{fig:hydro_warp_evolution}).

\item Decomposing the temperature perturbation into azimuthal Fourier modes ($m$) and vertical Hermite modes ($n$), we find that the $m=1, n=1$ component is responsible for warping the disk (Figures \ref{fig:warp_prop_evolution_hydro_smaller} and \ref{fig:similarity_7p5_m1n1}).

\item For realistic shadow shapes, the $m=1, n=1$ mode peaks at a mutual inclination of $\sim 15^\circ$ between the shadow and outer disk, but its amplitude remains at least 25\% of the peak across the range $3^\circ$--30$^\circ$ (Figure \ref{fig:select_mode_amplitude}). This implies that shadow-induced warping can operate even in nearly coplanar disks without clear signs of shadow lanes.

\item We find that a 1\% temperature perturbation in the $m=1, n=1$ mode ($A_{0,11}$) produces a $\sim$1$^\circ$ disk tilt (Figure \ref{fig:inner_outer_twist}). This tilting brings the inner regions of the outer disk into closer alignment with the shadow, thereby decreasing the mutual inclination. This weakens the relative $m=1, n=1$ perturbation in this tilted frame (Figure \ref{fig:select_mode_amplitude}).

\item The full disk models ({radial power-law density without inner cavity and outer exponential cutoff}) with prescribed temperature perturbations have their tilts varying in periodicity on the order of 100–300 orbits (Figure \ref{fig:tilt_period}), which may be related to the global disk structure and the propagation of bending waves on large scales.

\item After subtracting the rigid tilt motion, substructures associated with sloshing and breathing of the bending wave ($m=1, n=1$ mode), as well as other non-bending wave modes are revealed, with the $n=0$ mode dominant in $v_r'$ and $n=1$ mode dominant in $v_\theta'$. For the $30^\circ$ inclined shadow, $m=2$ and $m=3$ spiral patterns emerge in $v_r'$ and $v_\theta'$, respectively (Figures~\ref{fig:cart_slices} and \ref{fig:residual_warp_surface_hydro}), consistent with the strong $m=2, n=0$ and $m=3, n=1$ components in the temperature perturbation (Figure~\ref{fig:select_mode_amplitude}). Similarly, for the $7.5^\circ$ inclined shadow, $m=2$ and $m=1$ patterns are prominent in $v_r'$ and $v_\theta'$, respectively (Figure \ref{fig:residual_warp_surface_hydro}), reflecting the presence of the $m=2, n=0$ component and the $m=1, n=1$ in the temperature perturbation (Figure~\ref{fig:select_mode_amplitude}).

\end{itemize}

The shadow-induced warp mechanism is robust (Section \ref{sec:condition}) but important questions remain regarding its fundamental workings (Sections \ref{sec:why1}) and its implications for disk evolution and planet formation (Section \ref{sec:why2}). What makes these perturbations particularly testable is that they constitute a forward-modeling problem (Section \ref{sec:observation}): we can directly measure their cause, the temperature perturbations (Section \ref{sec:observing_temperature}), and compare them with their dynamical consequences (Section \ref{sec:observing_dynamics}), both of which are accessible with current NIR and (sub)millimeter observations.

\begin{acknowledgments}

{We thank the anonymous reviewer for their careful, constructive, and thorough report, which has significantly improved the quality of our paper.} SZ thanks Jane Huang and Andr\'{e}s Izquierdo for helpful and constructive comments. Support for this work was provided by NASA through the NASA Hubble Fellowship grant
\#HST-HF2-51568 awarded by the Space Telescope Science Institute, which is operated by the
Association of Universities for Research in Astronomy, Inc., for NASA, under contract
NAS5-26555. ZZ acknowledge support from NASA award 80NSSC22K1413, 80NSSC25K7144 and NSF award 2429732 and 2408207. CWF acknowledges support from the W. M. Keck Foundation Fund and the IAS Fund for Memberships in the Natural Sciences.

\end{acknowledgments}

\begin{contribution}

SZ came up with the initial research concept and was responsible for developing, running, and analyzing simulations, and making figures, writing and submitting the manuscript. ZZ and CWF helped formalize the ideas, improve the figures and the manuscript.


\end{contribution}

%
\facilities{NASA High-End Computing Capability Systems}

\software{Athena++ \citep{stone20},
          Cartopy \citep{Cartopy},
          Matplotlib \citep{hunter07},
          Scipy \citep{2020SciPy-NMeth}
          }


\appendix

\section{Comparison of two definitions of the scale height}
\label{sec:scale_height_2}

\begin{figure}
    \centering
    \includegraphics[width=1.0\linewidth]{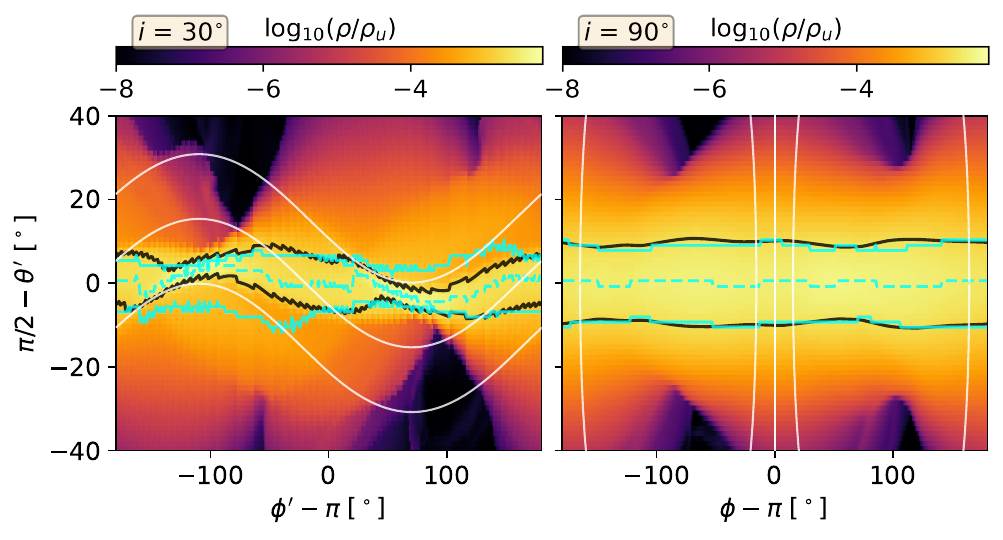}
    \caption{Density distribution in the $\phi'$–$\theta'$ plane for \texttt{R30} and \texttt{R90} at $t=500\ P_0$. These are the same density fields shown in the first row of Figure \ref{fig:phi_theta_view}, but they are zoomed in to the midplane regions, and displayed with linear $\phi'$–$\theta'$ projections. The solid cyan curve marks an alternative density-weighted definition of the gas scale height (Equation \ref{eq:density_weighted}), while the dashed cyan curve traces the vertical location of maximum density. The $\exp(-1/2)$ definition of the scale height is still shown in black curves, as in Figure \ref{fig:phi_theta_view}.}
    \label{fig:two_defination_scale_heights}
\end{figure}

In Figure \ref{fig:phi_theta_view}, we define the gas scale height as the location where the density drops to $\exp(-1/2)$ of the maximum density along $\theta'$. Since the vertical density structure is highly non-Gaussian, different definitions of the scale height yield different values. Here, we adopt an alternative definition following Equation 24 of \citet{fung19}. We define the scale height on one side of the disk as the location where the integrated density, measured from the density peak to the disk surface, reaches $\mathrm{erf}(-1/\sqrt{2}) \approx 0.68$ of the total surface density on that side. That is,
\begin{align}
    \frac{\int_{z'|_{\rho_{\max}}}^{z'|_{\rho_{\max}}+h_{\mathrm{u}}}\rho(z')\,dz'}{\int_{z'|_{\rho_{\max}}}^{+\infty} \rho(z')\,dz'} &= \mathrm{erf}\!\left(-\frac{1}{\sqrt{2}}\right), \nonumber \\
    \frac{\int_{z'|_{\rho_{\max}}}^{z'|_{\rho_{\max}}-h_{\mathrm{l}}}\rho(z')\,dz'}{\int_{z'|_{\rho_{\max}}}^{-\infty} \rho(z')\,dz'} &= \mathrm{erf}\!\left(-\frac{1}{\sqrt{2}}\right),
    \label{eq:density_weighted}
\end{align}
where $h_{\mathrm{u}}$ and $h_{\mathrm{l}}$ are the scale heights of the upper and lower surfaces. These surfaces are shown as solid cyan curves in Figure \ref{fig:two_defination_scale_heights}, while the dashed cyan curve shows the $\rho_{\max}$ surface. For reference, the black curves show the $\exp(-1/2)$ definition from Figure \ref{fig:phi_theta_view}.

For the \texttt{R30} models, the total thickness $h_{\mathrm{u}}+h_{\mathrm{l}}$ is larger under this density-weighted definition than under the $\exp(-1/2)$ definition. However, the thinner sides of the compressed regions yield nearly identical values in both definitions (e.g., the upper surface at $\phi'-\pi=-110^\circ$ and the lower surface at $\phi'-\pi = +80^\circ$). In the uncompressed regions, the density-weighted definition produces larger scale heights, so the azimuthal variation of $h_{\mathrm{u}}$ or $h_{\mathrm{l}}$ shows even greater contrast than in the first definition.  

For the \texttt{R90} model, the two definitions give very similar results. The density-weighted definition varies only by $\pm$1 grid cell, which is consistent with no azimuthal variation considering resolution effects. In fact, for a vertically Gaussian density distribution, the two definitions of the disk scale height should yield the same surface.

\section{Line of sight velocity for the radiation-hydrodynamical model \texttt{R30}}
\label{sec:los_r30}

While detailed line radiative transfer modeling is needed for full comparison with observations, here we calculate the line-of-sight velocity at the warped midplane ($z'/r=0$) and elevated surface ($z'/r=0.2$) at t = 500 $P_0$ to provide a more direct observational quantity that incorporates contributions from all three velocity components, $v_r'$, $v_\theta'$, and $v_\phi'$.

The procedure is as follows. We first take the 3D velocity field $(v_r', v_\theta', v_\phi'-v_K)$ that is transformed and projected to the rotated coordinates $(r,\theta',\phi')$ used in Figure \ref{fig:cart_slices}. This field is then transformed back to the simulation coordinates $(r,\theta,\phi)$ to obtain $(v_r, v_\theta-v_{K,\theta}, v_\phi-v_{K,\phi})$. The reason for subtracting the Keplerian component in the primed coordinates and then transforming back is that $v_\phi'$ is defined perpendicular to the local angular momentum vector $\hat{\mathbf{l}}(r)$ (Equation \ref{eq:beta_gamma_psi_defination}), which allows us to correctly subtract the Keplerian contribution from both $v_\theta$ and $v_\phi$.

We then define the line-of-sight direction as $\hat{n}$, with $i_\mathrm{ref}$ denoting its angle relative to the simulation $z$-axis ($i_\mathrm{ref} = \langle \hat{n}, \hat{z}\rangle$). The residual line-of-sight velocity is obtained by projecting $(v_r, v_\theta - v_{K,\theta}, v_\phi - v_{K,\phi})$ onto $\hat{n}$, giving $v_\mathrm{los} - v_{\mathrm{los},K}$. Because the local angular momentum vector $\hat{\mathbf{l}}(r)$ is generally misaligned with $\hat{z}$, the inclination of each annulus relative to $\hat{n}$ varies with $r$, which we define as $i_\mathrm{obs}(r) = \langle \hat{n}, \hat{\mathbf{l}}(r)\rangle$. A rotation of the line-of-sight vector $\hat{n}$ about the $z$-axis also changes these inclinations $i_\mathrm{obs}(r)$; we characterize this by the position angle PA$_0 = \arctan2(n_y, n_x)$.

\begin{figure}
    \centering
    \includegraphics[width=1.0\linewidth]{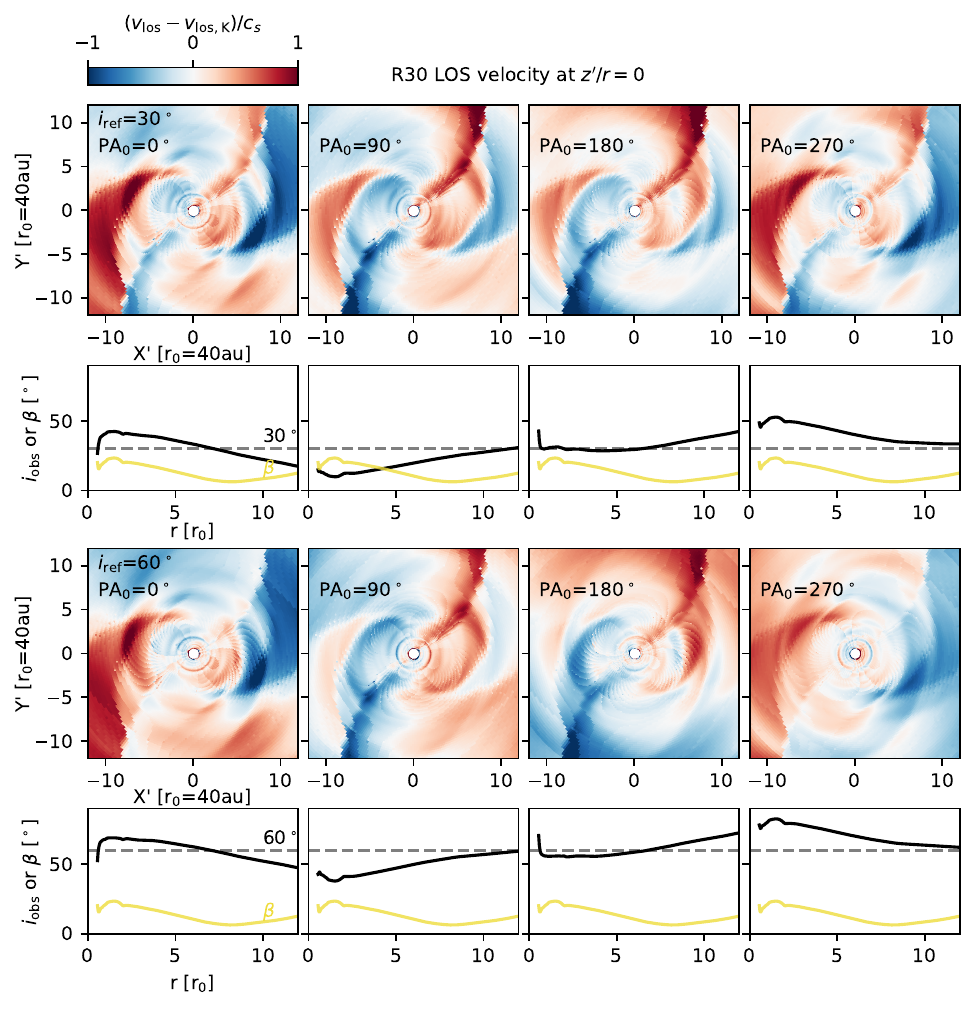}
    \caption{Residual line-of-sight velocity ($v_\mathrm{los}-v_{\mathrm{los},K}$) for \texttt{R30} at $z'/r=0$, shown in units of the local sound speed.
Top row: $i_\mathrm{ref}=30^\circ$.
Second row: corresponding $i_\mathrm{obs}(r)$ (black) with disk tilt $\beta$ (yellow) for reference.
Third row: $i_\mathrm{ref}=60^\circ$.
Fourth row: corresponding $i_\mathrm{obs}(r)$.
Columns from left to right show PA$_0=0^\circ$, $90^\circ$, $180^\circ$, and $270^\circ$.
These maps can be directly compared with the first row, fourth column of Figure \ref{fig:cart_slices}.}
    \label{fig:los_zr0}
\end{figure}

\begin{figure}
    \centering
    \includegraphics[width=1.0\linewidth]{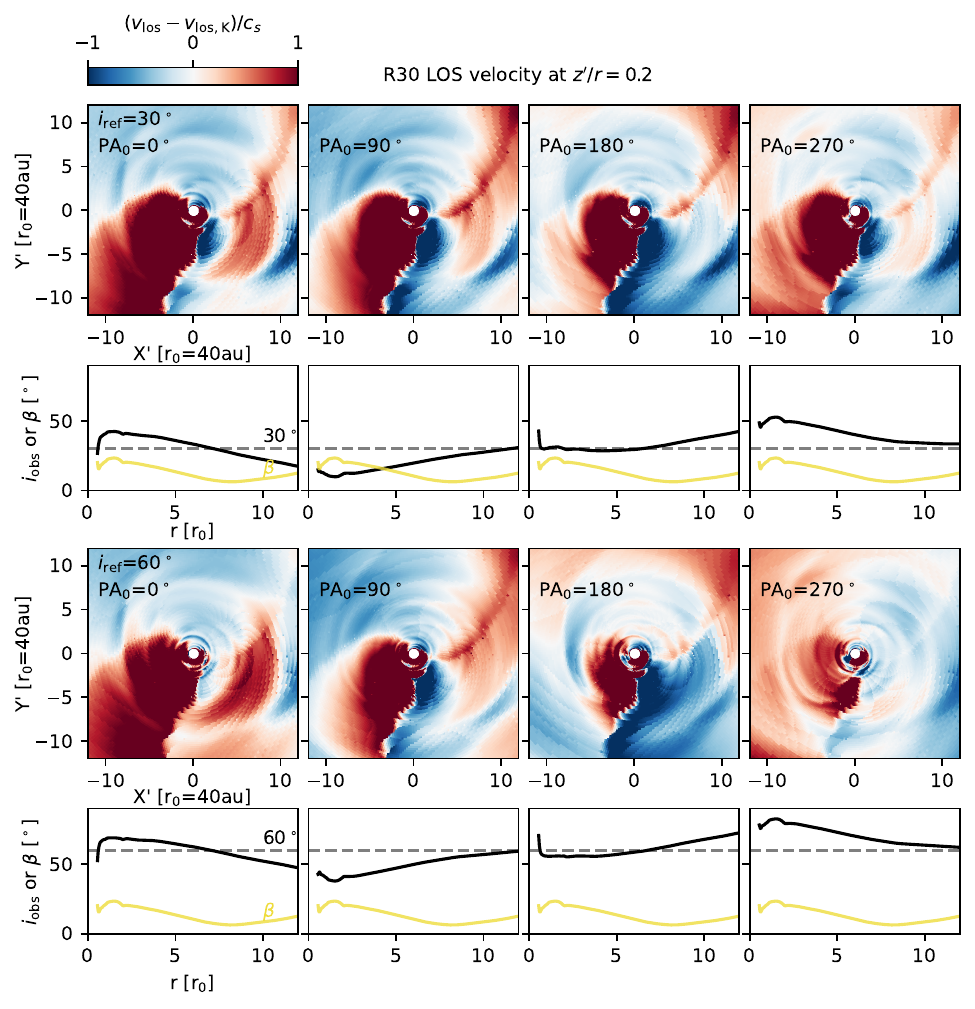}
    \caption{Similar to Figure \ref{fig:los_zr0}, but residual line-of-sight velocity at $z'/r=0.2$. These maps can be directly compared with the second row, fourth column of Figure \ref{fig:cart_slices}.}
    \label{fig:los_zr0p2}
\end{figure}

Finally, we transform the residual line-of-sight velocity back into the warped coordinates $(r,\theta',\phi')$, and present the maps at $z'/r=0$ and $z'/r=0.2$ in Figures \ref{fig:los_zr0} and \ref{fig:los_zr0p2}. Results are shown for $i_\mathrm{ref}=30^\circ$ (top two rows) and $60^\circ$ (bottom two rows), and for PA$_0=0^\circ$, $90^\circ$, $180^\circ$, and $270^\circ$ (from left to right), together with the radial profiles of the annulus inclinations relative to the line of sight, $i_\mathrm{obs}(r)$ (black solid curves in second and fourth rows).

In Figure \ref{fig:los_zr0}, at the midplane ($z'/r=0$), the $m=3$ patterns in $v_\theta'$ from Figure \ref{fig:cart_slices} are well preserved as long as $i_\mathrm{obs}(r)\lesssim50^\circ$. A similar result holds at $z'/r=0.2$ (Figure \ref{fig:los_zr0p2}), where the dominant $m=1$ mode and weaker $m=2$ modes are also retained. This gives us confidence that $v_\theta'$ perturbations can be reliably retrieved in observations up to inclinations of $\sim50^\circ$. We also note that for PA$_0=90^\circ$, $i_\mathrm{obs}(r) < i_\mathrm{ref}$ because this PA$_0$ aligns with the disk twist ($\gamma$) in the inner part of the outer disk (Figure \ref{fig:warp_properties_evolution}), whereas for PA$_0=180^\circ$, $i_\mathrm{obs}(r) > i_\mathrm{ref}$ because PA$_0$ anti-aligns with the twist.

\section{Vertical Slices of \texttt{R30} in the Code Coordinate}
\label{sec:cart_unprimed}

{The shadow appears bent in Figure~\ref{fig:cart_slices} when shown in the primed (rotated) coordinate frame, because different annuli have distinct angular momentum vectors and are thus rotated by different amounts. Figure~\ref{fig:compare_accretion_rad_hydro} provides a reference for the corresponding density and temperature structures in the original, unrotated (unprimed) simulation coordinates.}

\begin{figure}
    \centering
    \includegraphics[width=1.0\linewidth]{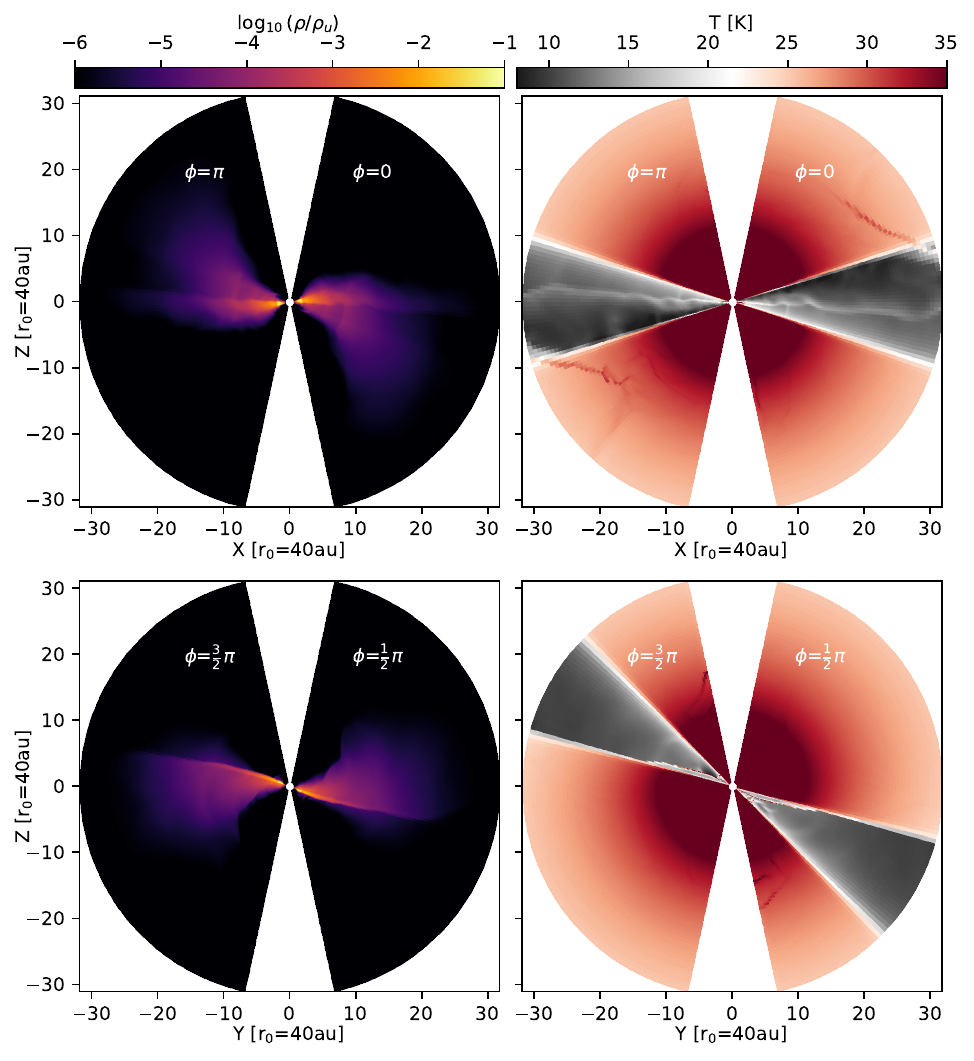}
    \caption{Vertical slices of the density and temperature for the \texttt{R30} model at $t = 500\ P_0$ ($P_0 \approx 253$~yr) are shown over the entire simulation domain in the unprimed code coordinates, in contrast to Figure~\ref{fig:cart_slices}.}
    \label{fig:cart_unprimed}
\end{figure}

\section{Comparison of the accretion history between the radiation-hydrodynamical and pure hydrodynamical models}

\begin{figure}
    \centering
    \includegraphics[width=1.0\linewidth]{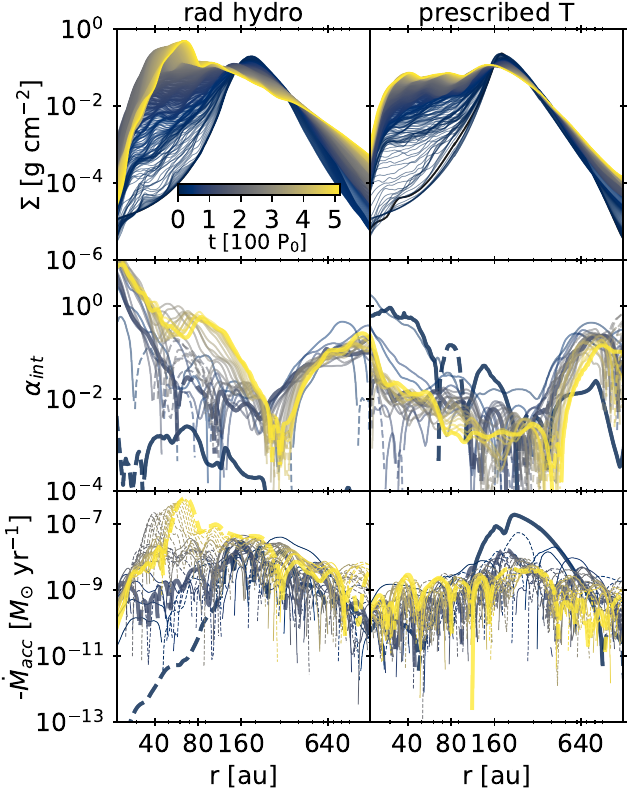}
    \caption{Similar to Figure \ref{fig:accretion}, the comparison of the accretion evolution for radiation hydrodynamical model (\texttt{R30}) and pure hydrodynamical model (\texttt{H30}) for a 30$^\circ$ inclined shadow.}
    \label{fig:compare_accretion_rad_hydro}
\end{figure}

Compared to the radiation–hydrodynamical model with a $30^\circ$ inclined shadow (\texttt{R30}), the pure hydrodynamical model with a prescribed temperature structure (\texttt{H30}) produces less accretion (Figure \ref{fig:compare_accretion_rad_hydro}). In this case, the inner disk remains less filled in. The resulting $\alpha_{\rm int}$ is on the order of $10^{-2}$ across the disk, corresponding to an accretion rate of $\sim 10^{-9}\ M_\odot\ \mathrm{yr}^{-1}$. {This difference might be partially attributed to the much smoother temperature gradient prescribed in Equation \ref{eq:angular_dependence} than the radiation-hydrodynamical simulations. {The qualitative similarity between pure hydrodynamical and radiation-hydrodynamical models is another example that including more realistic thermal physics can be incorporated through simplified prescriptions of temperature and cooling time with no additional cost, which has been benchmarked using more expensive radiation-hydrodynamical simulations \citep{melonfuksman22, muley24,zhang24, ziampras25b}.} } 

\section{Comparison of the boundary conditions on the disk evolution}
\label{sec:boundary_condition}

\begin{figure}
    \centering
    \includegraphics[width=1.0\linewidth]{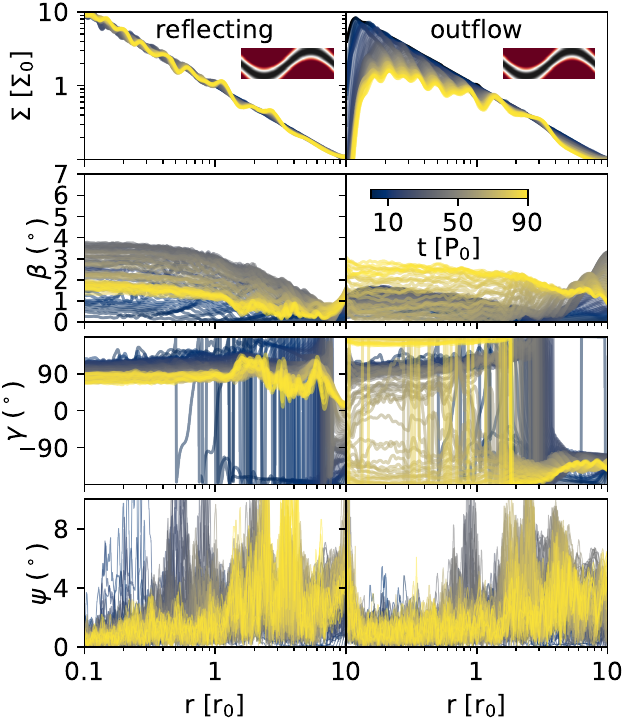}
    \caption{Time evolution of density (top row), tilt (second row), twist (third row), and warp (bottom row) in the pure hydrodynamical simulations with a $30^\circ$ inclined shadow. Results are shown for reflecting (left, \texttt{HF30R}) and modified outflow (right, \texttt{HF30Mo}) boundary conditions.}
    \label{fig:warp_evolution_30deg_different_bc}
\end{figure}

Figure \ref{fig:warp_evolution_30deg_different_bc} compares the full-disk models with a $30^\circ$ inclined shadow under reflecting (\texttt{HF30R}) and modified outflow (\texttt{HF30Mo}) boundary conditions. With reflecting boundaries, the surface density exhibits small-amplitude perturbations superimposed on the initial power-law profile, likely caused by the deposition of angular momentum flux from the $m=2$ spirals \citep{zhu25}. In contrast, the simulation with modified outflow boundaries also develops gaps and rings, but the inner disk is rapidly depleted since material can only flow outward at the boundary.

The tilt amplitudes are comparable between the two models, although their temporal evolution differs. The twist evolution is generally similar, except at late times ($\sim 90\ P_0$) when the modified outflow model shows the twist approaching zero. The warp evolution is broadly consistent between the two boundary conditions.

\section{Detailed Flow Structures of \texttt{HF30R} and \texttt{HF7R}}
\label{sec:detailed_flow_pure_hydro}

\begin{figure*}
    \centering
    \includegraphics[width=1.0\linewidth]{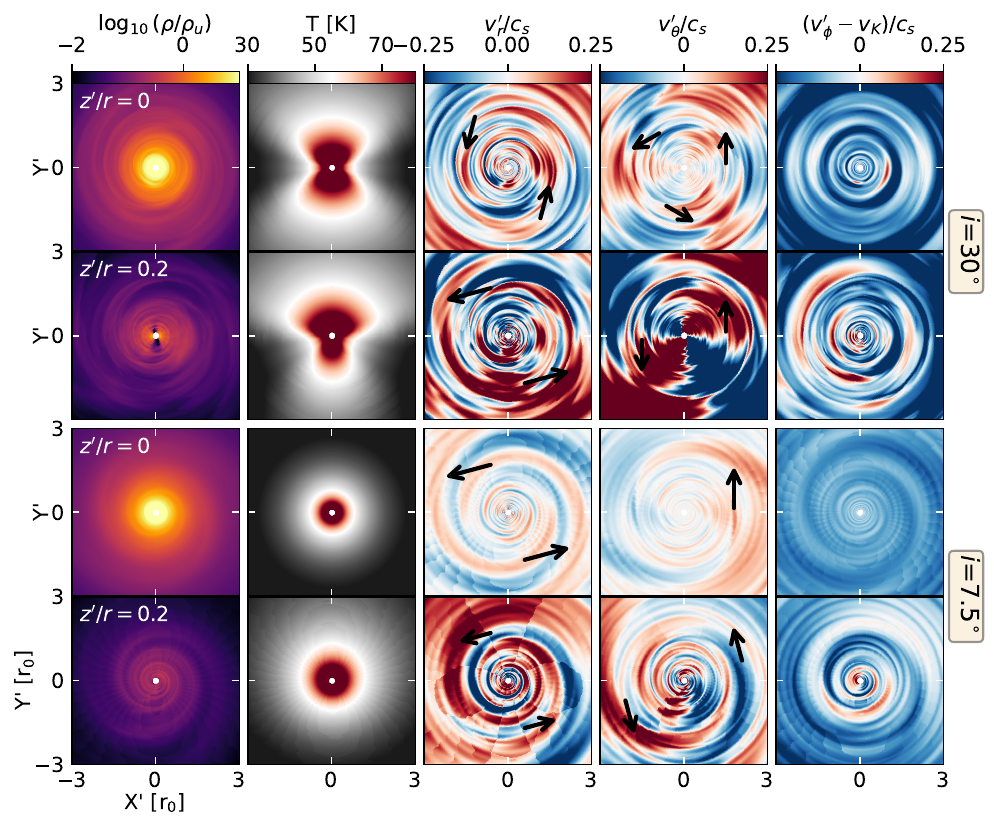}
    \caption{Various quantities on the rotated surfaces for the full-disk pure hydrodynamical simulations with a 20\% temperature perturbation. Top: $i=30^\circ$ (\texttt{HF30R}); Bottom: $i=7.5^\circ$ (\texttt{HF7R}). The layout is similar to Figure \ref{fig:cart_slices}. The arrows indicate the azimuthal locations of the local maxima in the velocity perturbations ($v_r'$ and $v_\theta'$), highlighting the dominant Fourier mode ($m=1,2,3$) at each layer. Similar to Figure \ref{fig:cart_slices}, some of the grid-like patterns arise from the nearest-neighbor interpolation used in the coordinate transformation.}
    \label{fig:residual_warp_surface_hydro}
\end{figure*}

In Figure \ref{fig:residual_warp_surface_hydro} we show the differences of the midplane and surface slices in the transformed frame (rigid tilt corrected) between 30$^\circ$ (\texttt{HF30R}) and 7.5$^\circ$ (\texttt{HF7R}) models. They can also be used to compare with the radiation hydrodynamical model (\texttt{R30}) from Figure \ref{fig:cart_slices}. The $i=30^\circ$ case (\texttt{HF30R}) shows stronger perturbations in density and velocities than \texttt{HF7R} due to its overall stronger azimuthal asymmetry in temperature (such as the $m=2,n=0$ mode). However, it has less tilt due to its weaker $m=1,n=1$ mode, as shown in Figure \ref{fig:warp_prop_evolution_hydro_smaller}. At the midplane, the $v_r'$ field shows $m=2$ spirals (black arrows) similar to the radiation-hydrodynamical simulation \texttt{R30}. This is because the $m=2, n=0$ perturbation dominates at the midplane. $v_\theta'$ also exhibits an $m=3$ symmetry (black arrows, even clearer than the radiation-hydrodynamical simulation \texttt{R30}'s $v_\theta'$ in Figure \ref{fig:cart_slices} first row), due to the $m=3, n=1$ mode in the shadow-induced temperature distribution. Since this model has much less tilt than the radiation-hydrodynamical model, the $z'/r=0.2$ surface still shows two shadows, although separated by less than $180^\circ$, similar to the image of HD 143006 \citep{benisty18}. The $v_r'$ and $v_\theta'$ fields in this model both show $m=2$ symmetry (black arrows). In contrast, in Figure \ref{fig:cart_slices}'s second panel, both the $v_r'$ and $v_\theta'$ of the \texttt{R30} model show $m=1$ perturbations due to the $m=1$ temperature perturbation on that surface.

The $i=7.5^\circ$ model shows fewer substructures and mainly $m=2$ spirals in both the midplane and the $z'/r=0.2$ density field, resembling MWC 758 \citep{grady13,benisty15,ren18}. At the midplane, the temperature asymmetry is no longer visible by eye, while at $z'/r=0.2$, the shadow-induced temperature drop becomes one-sided. The midplane $v_r'$ shows $m=2$ spirals (black arrows), but $v_\theta'$ instead shows an $m=1$ spiral (black arrow). At the surface, the $m=2$ spirals (black arrows in the last panel) are much stronger on the left side than the right side. In other words, there is a strong $m=1$ mode which is superimposed on top of the weaker $m=2$ structure which leads to the one armed feature dominating.

The different dominant modes between $v_r'$ and $v_\theta'$ observed in \texttt{R30}, \texttt{HF30R}, and \texttt{HF7R} can be explained by the Fourier–Hermite decomposition, which is also relevant for interpreting observations.
A linear analysis of perturbations in a disk shows that $n$-th order Hermite forcing by pressure perturbations (related to the temperature disturbance) readily excites corresponding $n$-th order variations in the horizontal $v_R$ and $v_\phi$ components \citep{fairbairn25}. However, owing to the $z$ derivatives in the vertical equation of motion, these $n$-th order pressure disturbances are essentially coupled with an $n-1$ response in the $v_z$ structure (essentially a ``derivative down'' in the Hermite expansion). Therefore, signatures observed at order $n$ in the horizontal velocity components ($v_R$ and $v_\phi$) are typically associated with structures at order $n-1$ in the vertical velocity, $v_z$
\citep[see Equation 12 in][]{fairbairn25}. 
For example, if the temperature perturbation is proportional to $\mathrm{He}_0(z/h)$, then $v_R$ shares the same form but $v_z=0$. If the temperature perturbation is proportional to $\mathrm{He}_1(z/h)$, then $v_R$ again follows the same form, but $v_z \propto \mathrm{He}_0(z/h) = 1$. This explains why the $m=2, n=0$ perturbations consistently appear in $v_r'$ but rarely in $v_\theta'$. An exception is the strong $m=2$ perturbation in $v_\theta'$ at $z'/r=0.2$ for \texttt{HF30R}. On the bottom surface ($z'/r=-0.2$), $v_\theta'$ flips sign (not shown, but nearly identical to the $z'/r=0.2$ surface in Figure \ref{fig:residual_warp_surface_hydro}, except with blue and red colors swapped), revealing a clear $m=2, n=1$ breathing mode. This indicates that the forcing is in the form of $m=2, n=2$. Such a pattern could arise either from the nonlinear product of two $m=1, n=1$ horizontal sloshing terms (see Figure \ref{fig:phi_theta_view}) or from coupling between the $m=2, n=0$ and $m=2, n=2$ modes \citep{fairbairn25}. Meanwhile, $n=1$ thermal forcing tends to excite an $n=0$  $v_z$ response, which explains the midplane appearance of the $m=3$ perturbations in $v_\theta'$ for the \texttt{HF30R} model and $m=1$ perturbations in \texttt{HF7R}, consistent with their dominant odd-$n$ modes ($m=3,n=1$ and $m=1,n=1$ modes, respectively).



\begin{thebibliography}{}
\expandafter\ifx\csname natexlab\endcsname\relax\def\natexlab#1{#1}\fi
\providecommand{\url}[1]{\href{#1}{#1}}
\providecommand{\dodoi}[1]{doi:~\href{http://doi.org/#1}{\nolinkurl{#1}}}
\providecommand{\doeprint}[1]{\href{http://ascl.net/#1}{\nolinkurl{http://ascl.net/#1}}}
\providecommand{\doarXiv}[1]{\href{https://arxiv.org/abs/#1}{\nolinkurl{https://arxiv.org/abs/#1}}}

\bibitem[{M. {Aizawa} \& R. {Orihara}(2025){Aizawa} \& {Orihara}}]{aizawa25}
{Aizawa}, M., \& {Orihara}, R. 2025, \bibinfo{title}{{On the Interpretation of Velocity Residuals in Protoplanetary Disks},} arXiv e-prints, arXiv:2507.23287, \dodoi{10.48550/arXiv.2507.23287}

\bibitem[{S.~M. {Andrews}(2020){Andrews}}]{andrews20}
{Andrews}, S.~M. 2020, \bibinfo{title}{{Observations of Protoplanetary Disk Structures},} \araa, 58, 483, \dodoi{10.1146/annurev-astro-031220-010302}

\bibitem[{M. {Ansdell} {et~al.}(2020){Ansdell}, {Gaidos}, {Hedges}, {Tazzari}, {Kraus}, {Wyatt}, {Kennedy}, {Williams}, {Mann}, {Angelo}, {D{\^u}chene}, {Mamajek}, {Carpenter}, {Esplin}, \& {Rizzuto}}]{ansdell20}
{Ansdell}, M., {Gaidos}, E., {Hedges}, C., {et~al.} 2020, \bibinfo{title}{{Are inner disc misalignments common? ALMA reveals an isotropic outer disc inclination distribution for young dipper stars},} \mnras, 492, 572, \dodoi{10.1093/mnras/stz3361}

\bibitem[{H. {Avenhaus} {et~al.}(2017){Avenhaus}, {Quanz}, {Schmid}, {Dominik}, {Stolker}, {Ginski}, {de Boer}, {Szul{\'a}gyi}, {Garufi}, {Zurlo}, {Hagelberg}, {Benisty}, {Henning}, {M{\'e}nard}, {Meyer}, {Baruffolo}, {Bazzon}, {Beuzit}, {Costille}, {Dohlen}, {Girard}, {Gisler}, {Kasper}, {Mouillet}, {Pragt}, {Roelfsema}, {Salasnich}, \& {Sauvage}}]{avenhaus17}
{Avenhaus}, H., {Quanz}, S.~P., {Schmid}, H.~M., {et~al.} 2017, \bibinfo{title}{{Exploring Dust around HD 142527 down to 0.″025 (4 au) Using SPHERE/ZIMPOL},} \aj, 154, 33, \dodoi{10.3847/1538-3881/aa7560}

\bibitem[{H. {Avenhaus} {et~al.}(2018){Avenhaus}, {Quanz}, {Garufi}, {Perez}, {Casassus}, {Pinte}, {Bertrang}, {Caceres}, {Benisty}, \& {Dominik}}]{avenhaus18}
{Avenhaus}, H., {Quanz}, S.~P., {Garufi}, A., {et~al.} 2018, \bibinfo{title}{{Disks around T Tauri Stars with SPHERE (DARTTS-S). I. SPHERE/IRDIS Polarimetric Imaging of Eight Prominent T Tauri Disks},} \apj, 863, 44, \dodoi{10.3847/1538-4357/aab846}

\bibitem[{J. {Bae} {et~al.}(2023){Bae}, {Isella}, {Zhu}, {Martin}, {Okuzumi}, \& {Suriano}}]{bae22}
{Bae}, J., {Isella}, A., {Zhu}, Z., {et~al.} 2023, in Astronomical Society of the Pacific Conference Series, Vol. 534, Protostars and Planets VII, ed. S.~{Inutsuka}, Y.~{Aikawa}, T.~{Muto}, K.~{Tomida}, \& M.~{Tamura}, 423, \dodoi{10.48550/arXiv.2210.13314}

\bibitem[{J. {Bae} {et~al.}(2019){Bae}, {Zhu}, {Baruteau}, {Benisty}, {Dullemond}, {Facchini}, {Isella}, {Keppler}, {P{\'e}rez}, \& {Teague}}]{bae19}
{Bae}, J., {Zhu}, Z., {Baruteau}, C., {et~al.} 2019, \bibinfo{title}{{An Ideal Testbed for Planet-Disk Interaction: Two Giant Protoplanets in Resonance Shaping the PDS 70 Protoplanetary Disk},} \apjl, 884, L41, \dodoi{10.3847/2041-8213/ab46b0}

\bibitem[{J. {Bae} {et~al.}(2022){Bae}, {Teague}, {Andrews}, {Benisty}, {Facchini}, {Galloway-Sprietsma}, {Loomis}, {Aikawa}, {Alarc{\'o}n}, {Bergin}, {Bergner}, {Booth}, {Cataldi}, {Cleeves}, {Czekala}, {Guzm{\'a}n}, {Huang}, {Ilee}, {Kurtovic}, {Law}, {Le Gal}, {Liu}, {Long}, {M{\'e}nard}, {{\"O}berg}, {P{\'e}rez}, {Qi}, {Schwarz}, {Sierra}, {Walsh}, {Wilner}, \& {Zhang}}]{bae22b}
{Bae}, J., {Teague}, R., {Andrews}, S.~M., {et~al.} 2022, \bibinfo{title}{{Molecules with ALMA at Planet-forming Scales (MAPS): A Circumplanetary Disk Candidate in Molecular-line Emission in the AS 209 Disk},} \apjl, 934, L20, \dodoi{10.3847/2041-8213/ac7fa3}

\bibitem[{G. {Ballabio} {et~al.}(2021){Ballabio}, {Nealon}, {Alexander}, {Cuello}, {Pinte}, \& {Price}}]{ballabio21}
{Ballabio}, G., {Nealon}, R., {Alexander}, R.~D., {et~al.} 2021, \bibinfo{title}{{HD 143006: circumbinary planet or misaligned disc?},} \mnras, 504, 888, \dodoi{10.1093/mnras/stab922}

\bibitem[{J.~M. {Bardeen} \& J.~A. {Petterson}(1975){Bardeen} \& {Petterson}}]{bardeen75}
{Bardeen}, J.~M., \& {Petterson}, J.~A. 1975, \bibinfo{title}{{The Lense-Thirring Effect and Accretion Disks around Kerr Black Holes},} \apjl, 195, L65, \dodoi{10.1086/181711}

\bibitem[{M.~R. {Bate} {et~al.}(2002){Bate}, {Ogilvie}, {Lubow}, \& {Pringle}}]{bate02}
{Bate}, M.~R., {Ogilvie}, G.~I., {Lubow}, S.~H., \& {Pringle}, J.~E. 2002, \bibinfo{title}{{The excitation, propagation and dissipation of waves in accretion discs: the non-linear axisymmetric case},} \mnras, 332, 575, \dodoi{10.1046/j.1365-8711.2002.05289.x}

\bibitem[{M. {Benisty} {et~al.}(2015){Benisty}, {Juhasz}, {Boccaletti}, {Avenhaus}, {Milli}, {Thalmann}, {Dominik}, {Pinilla}, {Buenzli}, {Pohl}, {Beuzit}, {Birnstiel}, {de Boer}, {Bonnefoy}, {Chauvin}, {Christiaens}, {Garufi}, {Grady}, {Henning}, {Huelamo}, {Isella}, {Langlois}, {M{\'e}nard}, {Mouillet}, {Olofsson}, {Pantin}, {Pinte}, \& {Pueyo}}]{benisty15}
{Benisty}, M., {Juhasz}, A., {Boccaletti}, A., {et~al.} 2015, \bibinfo{title}{{Asymmetric features in the protoplanetary disk MWC 758},} \aap, 578, L6, \dodoi{10.1051/0004-6361/201526011}

\bibitem[{M. {Benisty} {et~al.}(2017){Benisty}, {Stolker}, {Pohl}, {de Boer}, {Lesur}, {Dominik}, {Dullemond}, {Langlois}, {Min}, {Wagner}, {Henning}, {Juhasz}, {Pinilla}, {Facchini}, {Apai}, {van Boekel}, {Garufi}, {Ginski}, {M{\'e}nard}, {Pinte}, {Quanz}, {Zurlo}, {Boccaletti}, {Bonnefoy}, {Beuzit}, {Chauvin}, {Cudel}, {Desidera}, {Feldt}, {Fontanive}, {Gratton}, {Kasper}, {Lagrange}, {LeCoroller}, {Mouillet}, {Mesa}, {Sissa}, {Vigan}, {Antichi}, {Buey}, {Fusco}, {Gisler}, {Llored}, {Magnard}, {Moeller-Nilsson}, {Pragt}, {Roelfsema}, {Sauvage}, \& {Wildi}}]{benisty17}
{Benisty}, M., {Stolker}, T., {Pohl}, A., {et~al.} 2017, \bibinfo{title}{{Shadows and spirals in the protoplanetary disk HD 100453},} \aap, 597, A42, \dodoi{10.1051/0004-6361/201629798}

\bibitem[{M. {Benisty} {et~al.}(2018){Benisty}, {Juh{\'a}sz}, {Facchini}, {Pinilla}, {de Boer}, {P{\'e}rez}, {Keppler}, {Muro-Arena}, {Villenave}, {Andrews}, {Dominik}, {Dullemond}, {Gallenne}, {Garufi}, {Ginski}, \& {Isella}}]{benisty18}
{Benisty}, M., {Juh{\'a}sz}, A., {Facchini}, S., {et~al.} 2018, \bibinfo{title}{{Shadows and asymmetries in the T Tauri disk HD 143006: evidence for a misaligned inner disk},} \aap, 619, A171, \dodoi{10.1051/0004-6361/201833913}

\bibitem[{M. {Benisty} {et~al.}(2021){Benisty}, {Bae}, {Facchini}, {Keppler}, {Teague}, {Isella}, {Kurtovic}, {P{\'e}rez}, {Sierra}, {Andrews}, {Carpenter}, {Czekala}, {Dominik}, {Henning}, {Menard}, {Pinilla}, \& {Zurlo}}]{benisty21}
{Benisty}, M., {Bae}, J., {Facchini}, S., {et~al.} 2021, \bibinfo{title}{{A Circumplanetary Disk around PDS70c},} \apjl, 916, L2, \dodoi{10.3847/2041-8213/ac0f83}

\bibitem[{M. {Benisty} {et~al.}(2023){Benisty}, {Dominik}, {Follette}, {Garufi}, {Ginski}, {Hashimoto}, {Keppler}, {Kley}, \& {Monnier}}]{benisty22}
{Benisty}, M., {Dominik}, C., {Follette}, K., {et~al.} 2023, in Astronomical Society of the Pacific Conference Series, Vol. 534, Protostars and Planets VII, ed. S.~{Inutsuka}, Y.~{Aikawa}, T.~{Muto}, K.~{Tomida}, \& M.~{Tamura}, 605, \dodoi{10.48550/arXiv.2203.09991}

\bibitem[{G.~H.~M. {Bertrang} {et~al.}(2018){Bertrang}, {Avenhaus}, {Casassus}, {Montesinos}, {Kirchschlager}, {Perez}, {Cieza}, \& {Wolf}}]{bertrang18}
{Bertrang}, G.~H.~M., {Avenhaus}, H., {Casassus}, S., {et~al.} 2018, \bibinfo{title}{{HD 169142 in the eyes of ZIMPOL/SPHERE},} \mnras, 474, 5105, \dodoi{10.1093/mnras/stx3052}

\bibitem[{L.~I. {Biddle} {et~al.}(2025){Biddle}, {Bowler}, {Morgan}, {Tran}, \& {Wu}}]{biddle25}
{Biddle}, L.~I., {Bowler}, B.~P., {Morgan}, M., {Tran}, Q.~H., \& {Wu}, Y.-L. 2025, \bibinfo{title}{{One-third of Sun-like stars are born with misaligned planet-forming disks},} \nat, 644, 356, \dodoi{10.1038/s41586-025-09324-0}

\bibitem[{T. {Birnstiel} {et~al.}(2018){Birnstiel}, {Dullemond}, {Zhu}, {Andrews}, {Bai}, {Wilner}, {Carpenter}, {Huang}, {Isella}, {Benisty}, {P{\'e}rez}, \& {Zhang}}]{birnstiel18}
{Birnstiel}, T., {Dullemond}, C.~P., {Zhu}, Z., {et~al.} 2018, \bibinfo{title}{{The Disk Substructures at High Angular Resolution Project (DSHARP). V. Interpreting ALMA Maps of Protoplanetary Disks in Terms of a Dust Model},} \apjl, 869, L45, \dodoi{10.3847/2041-8213/aaf743}

\bibitem[{A.~J. {Bohn} {et~al.}(2022){Bohn}, {Benisty}, {Perraut}, {van der Marel}, {W{\"o}lfer}, {van Dishoeck}, {Facchini}, {Manara}, {Teague}, {Francis}, {Berger}, {Garcia-Lopez}, {Ginski}, {Henning}, {Kenworthy}, {Kraus}, {M{\'e}nard}, {M{\'e}rand}, \& {P{\'e}rez}}]{bohn22}
{Bohn}, A.~J., {Benisty}, M., {Perraut}, K., {et~al.} 2022, \bibinfo{title}{{Probing inner and outer disk misalignments in transition disks. Constraints from VLTI/GRAVITY and ALMA observations},} \aap, 658, A183, \dodoi{10.1051/0004-6361/202142070}

\bibitem[{J. {Bouvier} {et~al.}(1999){Bouvier}, {Chelli}, {Allain}, {Carrasco}, {Costero}, {Cruz-Gonzalez}, {Dougados}, {Fern{\'a}ndez}, {Mart{\'\i}n}, {M{\'e}nard}, {Mennessier}, {Mujica}, {Recillas}, {Salas}, {Schmidt}, \& {Wichmann}}]{bouvier99}
{Bouvier}, J., {Chelli}, A., {Allain}, S., {et~al.} 1999, \bibinfo{title}{{Magnetospheric accretion onto the T Tauri star AA Tauri. I. Constraints from multisite spectrophotometric monitoring},} \aap, 349, 619

\bibitem[{C.~J. {Burrows} {et~al.}(1996){Burrows}, {Stapelfeldt}, {Watson}, {Krist}, {Ballester}, {Clarke}, {Crisp}, {Gallagher}, {Griffiths}, {Hester}, {Hoessel}, {Holtzman}, {Mould}, {Scowen}, {Trauger}, \& {Westphal}}]{Burrows96}
{Burrows}, C.~J., {Stapelfeldt}, K.~R., {Watson}, A.~M., {et~al.} 1996, \bibinfo{title}{{Hubble Space Telescope Observations of the Disk and Jet of HH 30},} \apj, 473, 437, \dodoi{10.1086/178156}

\bibitem[{N. {Calvet} {et~al.}(1991){Calvet}, {Patino}, {Magris}, \& {D'Alessio}}]{calvet91}
{Calvet}, N., {Patino}, A., {Magris}, G.~C., \& {D'Alessio}, P. 1991, \bibinfo{title}{{Irradiation of Accretion Disks around Young Objects. I. Near-Infrared CO Bands},} \apj, 380, 617, \dodoi{10.1086/170618}

\bibitem[{S. {Casassus} {et~al.}(2019){Casassus}, {P{\'e}rez}, {Osses}, \& {Marino}}]{cassasus18}
{Casassus}, S., {P{\'e}rez}, S., {Osses}, A., \& {Marino}, S. 2019, \bibinfo{title}{{Cooling in the shade of warped transition discs},} \mnras, 486, L58, \dodoi{10.1093/mnrasl/slz059}

\bibitem[{P. {Cazzoletti} {et~al.}(2018){Cazzoletti}, {van Dishoeck}, {Visser}, {Facchini}, \& {Bruderer}}]{cazzoletti18}
{Cazzoletti}, P., {van Dishoeck}, E.~F., {Visser}, R., {Facchini}, S., \& {Bruderer}, S. 2018, \bibinfo{title}{{CN rings in full protoplanetary disks around young stars as probes of disk structure},} \aap, 609, A93, \dodoi{10.1051/0004-6361/201731457}

\bibitem[{E.~I. {Chiang} \& P. {Goldreich}(1997){Chiang} \& {Goldreich}}]{chiang97}
{Chiang}, E.~I., \& {Goldreich}, P. 1997, \bibinfo{title}{{Spectral Energy Distributions of T Tauri Stars with Passive Circumstellar Disks},} \apj, 490, 368, \dodoi{10.1086/304869}

\bibitem[{I. {Codron} {et~al.}(2025){Codron}, {Kraus}, {Monnier}, {Marino}, {Davies}, {Anugu}, {Gardner}, {Ibrahim}, {Lanthermann}, \& {Le Bouquin}}]{codron25}
{Codron}, I., {Kraus}, S., {Monnier}, J.~D., {et~al.} 2025, \bibinfo{title}{{HD 143006: interferometric confirmation of misaligned protoplanetary disc with CHARA/MIRCX and VLTI/PIONIER},} \mnras, 541, 1600, \dodoi{10.1093/mnras/staf1033}

\bibitem[{N. {Cuello} {et~al.}(2019{\natexlab{a}}){Cuello}, {Montesinos}, {Stammler}, {Louvet}, \& {Cuadra}}]{cuello19}
{Cuello}, N., {Montesinos}, M., {Stammler}, S.~M., {Louvet}, F., \& {Cuadra}, J. 2019{\natexlab{a}}, \bibinfo{title}{{Dusty spirals triggered by shadows in transition discs},} \aap, 622, A43, \dodoi{10.1051/0004-6361/201731732}

\bibitem[{N. {Cuello} {et~al.}(2019{\natexlab{b}}){Cuello}, {Dipierro}, {Mentiplay}, {Price}, {Pinte}, {Cuadra}, {Laibe}, {M{\'e}nard}, {Poblete}, \& {Montesinos}}]{cuello19b}
{Cuello}, N., {Dipierro}, G., {Mentiplay}, D., {et~al.} 2019{\natexlab{b}}, \bibinfo{title}{{Flybys in protoplanetary discs: I. Gas and dust dynamics},} \mnras, 483, 4114, \dodoi{10.1093/mnras/sty3325}

\bibitem[{P. {Curone} {et~al.}(2025){Curone}, {Facchini}, {Andrews}, {Testi}, {Benisty}, {Czekala}, {Huang}, {Ilee}, {Isella}, {Lodato}, {Loomis}, {Stadler}, {Winter}, {Bae}, {Barraza-Alfaro}, {Cataldi}, {Cuello}, {Fasano}, {Flock}, {Fukagawa}, {Galloway-Sprietsma}, {Garg}, {Hall}, {Izquierdo}, {Kanagawa}, {Lesur}, {Longarini}, {Menard}, {Orihara}, {Pinte}, {Price}, {Rosotti}, {Teague}, {Wafflard-Fernandez}, {Wilner}, {W{\"o}lfer}, {Yen}, {Yoshida}, \& {Zawadzki}}]{curone25}
{Curone}, P., {Facchini}, S., {Andrews}, S.~M., {et~al.} 2025, \bibinfo{title}{{exoALMA. IV. Substructures, Asymmetries, and the Faint Outer Disk in Continuum Emission},} \apjl, 984, L9, \dodoi{10.3847/2041-8213/adc438}

\bibitem[{P. {D'Alessio} {et~al.}(1998){D'Alessio}, {Cant{\"o}}, {Calvet}, \& {Lizano}}]{dalessio98}
{D'Alessio}, P., {Cant{\"o}}, J., {Calvet}, N., \& {Lizano}, S. 1998, \bibinfo{title}{{Accretion Disks around Young Objects. I. The Detailed Vertical Structure},} \apj, 500, 411, \dodoi{10.1086/305702}

\bibitem[{E. {Dartois} {et~al.}(2003){Dartois}, {Dutrey}, \& {Guilloteau}}]{dartois03}
{Dartois}, E., {Dutrey}, A., \& {Guilloteau}, S. 2003, \bibinfo{title}{{Structure of the DM Tau Outer Disk: Probing the vertical kinetic temperature gradient},} \aap, 399, 773, \dodoi{10.1051/0004-6361:20021638}

\bibitem[{J. {Debes} {et~al.}(2023){Debes}, {Nealon}, {Alexander}, {Weinberger}, {Wolff}, {Hines}, {Kastner}, {Jang-Condell}, {Pinte}, {Plavchan}, \& {Pueyo}}]{debes23}
{Debes}, J., {Nealon}, R., {Alexander}, R., {et~al.} 2023, \bibinfo{title}{{The Surprising Evolution of the Shadow on the TW Hya Disk},} \apj, 948, 36, \dodoi{10.3847/1538-4357/acbdf1}

\bibitem[{A. {Dutrey} {et~al.}(2017){Dutrey}, {Guilloteau}, {Pi{\'e}tu}, {Chapillon}, {Wakelam}, {Di Folco}, {Stoecklin}, {Denis-Alpizar}, {Gorti}, {Teague}, {Henning}, {Semenov}, \& {Grosso}}]{dutrey17}
{Dutrey}, A., {Guilloteau}, S., {Pi{\'e}tu}, V., {et~al.} 2017, \bibinfo{title}{{The Flying Saucer: Tomography of the thermal and density gas structure of an edge-on protoplanetary disk},} \aap, 607, A130, \dodoi{10.1051/0004-6361/201730645}

\bibitem[{A. {Dutrey} {et~al.}(2025){Dutrey}, {Denis-Alpizar}, {Guilloteau}, {Foucher}, {Gavino}, {Semenov}, {Pietu}, {Chapillon}, {Testi}, {Dartois}, {DiFolco}, {Furuya}, {Gorti}, {Grosso}, {Henning}, {Hur{\'e}}, {K{\'o}sp{\'a}l}, {Le Petit}, {Majumdar}, {Meshaka}, {Nomura}, {Phuong}, {Ruaud}, {Tang}, \& {Wolf}}]{dutrey25}
{Dutrey}, A., {Denis-Alpizar}, O., {Guilloteau}, S., {et~al.} 2025, \bibinfo{title}{{Edge-On Disk Study (EODS) III: Molecular Stratification in the Flying Saucer Disk},} arXiv e-prints, arXiv:2509.26033, \dodoi{10.48550/arXiv.2509.26033}

\bibitem[{S. {Facchini} {et~al.}(2018){Facchini}, {Juh{\'a}sz}, \& {Lodato}}]{facchini18}
{Facchini}, S., {Juh{\'a}sz}, A., \& {Lodato}, G. 2018, \bibinfo{title}{{Signatures of broken protoplanetary discs in scattered light and in sub-millimetre observations},} \mnras, 473, 4459, \dodoi{10.1093/mnras/stx2523}

\bibitem[{S. {Facchini} {et~al.}(2013){Facchini}, {Lodato}, \& {Price}}]{facchini13}
{Facchini}, S., {Lodato}, G., \& {Price}, D.~J. 2013, \bibinfo{title}{{Wave-like warp propagation in circumbinary discs - I. Analytic theory and numerical simulations},} \mnras, 433, 2142, \dodoi{10.1093/mnras/stt877}

\bibitem[{C.~W. {Fairbairn}(2025){Fairbairn}}]{fairbairn25}
{Fairbairn}, C.~W. 2025, \bibinfo{title}{{Linear Bending Wave Propagation in Laminar and Turbulent Disks},} \apj, 979, 156, \dodoi{10.3847/1538-4357/ad9c73}

\bibitem[{C.~W. {Fairbairn} \& G.~I. {Ogilvie}(2021){Fairbairn} \& {Ogilvie}}]{fairbairn21}
{Fairbairn}, C.~W., \& {Ogilvie}, G.~I. 2021, \bibinfo{title}{{Non-linear dynamics of hydrodynamic tori as a model of oscillations and bending waves in astrophysical discs},} \mnras, 505, 4906, \dodoi{10.1093/mnras/stab1554}

\bibitem[{F. {Farago} \& J. {Laskar}(2010){Farago} \& {Laskar}}]{farago10}
{Farago}, F., \& {Laskar}, J. 2010, \bibinfo{title}{{High-inclination orbits in the secular quadrupolar three-body problem},} \mnras, 401, 1189, \dodoi{10.1111/j.1365-2966.2009.15711.x}

\bibitem[{A.~J. {Fehr} \& S.~M. {Andrews}(2025){Fehr} \& {Andrews}}]{fehr25}
{Fehr}, A.~J., \& {Andrews}, S.~M. 2025, \bibinfo{title}{{Measuring the Two-Dimensional Thermal Structures of Protoplanetary Disks},} arXiv e-prints, arXiv:2509.15196.
\newblock \doarXiv{2509.15196}

\bibitem[{M.~M. {Fragner} \& R.~P. {Nelson}(2010){Fragner} \& {Nelson}}]{fragner10}
{Fragner}, M.~M., \& {Nelson}, R.~P. 2010, \bibinfo{title}{{Evolution of warped and twisted accretion discs in close binary systems},} \aap, 511, A77, \dodoi{10.1051/0004-6361/200913088}

\bibitem[{L. {Francis} \& N. {van der Marel}(2020){Francis} \& {van der Marel}}]{francis20}
{Francis}, L., \& {van der Marel}, N. 2020, \bibinfo{title}{{Dust-depleted Inner Disks in a Large Sample of Transition Disks through Long-baseline ALMA Observations},} \apj, 892, 111, \dodoi{10.3847/1538-4357/ab7b63}

\bibitem[{J. {Fung} {et~al.}(2019){Fung}, {Zhu}, \& {Chiang}}]{fung19}
{Fung}, J., {Zhu}, Z., \& {Chiang}, E. 2019, \bibinfo{title}{{Circumplanetary Disk Dynamics in the Isothermal and Adiabatic Limits},} \apj, 887, 152, \dodoi{10.3847/1538-4357/ab53da}

\bibitem[{M. {Galloway-Sprietsma} {et~al.}(2025){Galloway-Sprietsma}, {Bae}, {Izquierdo}, {Stadler}, {Longarini}, {Teague}, {Andrews}, {Winter}, {Benisty}, {Facchini}, {Rosotti}, {Zawadzki}, {Pinte}, {Fasano}, {Barraza-Alfaro}, {Cataldi}, {Cuello}, {Curone}, {Czekala}, {Flock}, {Fukagawa}, {Gardner}, {Garg}, {Hall}, {Huang}, {Ilee}, {Kanagawa}, {Lesur}, {Lodato}, {Loomis}, {Menard}, {Orihara}, {Price}, {Wafflard-Fernandez}, {Wilner}, {W{\"o}lfer}, {Yen}, \& {Yoshida}}]{galloway25}
{Galloway-Sprietsma}, M., {Bae}, J., {Izquierdo}, A.~F., {et~al.} 2025, \bibinfo{title}{{exoALMA. V. Gaseous Emission Surfaces and Temperature Structures},} \apjl, 984, L10, \dodoi{10.3847/2041-8213/adc437}

\bibitem[{C. {Ginski} {et~al.}(2021){Ginski}, {Facchini}, {Huang}, {Benisty}, {Vaendel}, {Stapper}, {Dominik}, {Bae}, {M{\'e}nard}, {Muro-Arena}, {Hogerheijde}, {McClure}, {van Holstein}, {Birnstiel}, {Boehler}, {Bohn}, {Flock}, {Mamajek}, {Manara}, {Pinilla}, {Pinte}, \& {Ribas}}]{ginski21}
{Ginski}, C., {Facchini}, S., {Huang}, J., {et~al.} 2021, \bibinfo{title}{{Disk Evolution Study Through Imaging of Nearby Young Stars (DESTINYS): Late Infall Causing Disk Misalignment and Dynamic Structures in SU Aur},} \apjl, 908, L25, \dodoi{10.3847/2041-8213/abdf57}

\bibitem[{C. {Ginski} {et~al.}(2024){Ginski}, {Garufi}, {Benisty}, {Tazaki}, {Dominik}, {Ribas}, {Engler}, {Birnstiel}, {Chauvin}, {Columba}, {Facchini}, {Goncharov}, {Hagelberg}, {Henning}, {Hogerheijde}, {van Holstein}, {Huang}, {Muto}, {Pinilla}, {Kanagawa}, {Kim}, {Kurtovic}, {Langlois}, {Manara}, {Milli}, {Momose}, {Orihara}, {Pawellek}, {Pinte}, {Rab}, {Schmidt}, {Snik}, {Wahhaj}, {Williams}, \& {Zurlo}}]{ginski24}
{Ginski}, C., {Garufi}, A., {Benisty}, M., {et~al.} 2024, \bibinfo{title}{{The SPHERE view of the Chamaeleon I star-forming region. The full census of planet-forming disks with GTO and DESTINYS programs},} \aap, 685, A52, \dodoi{10.1051/0004-6361/202244005}

\bibitem[{U. {Gorti} {et~al.}(2011){Gorti}, {Hollenbach}, {Najita}, \& {Pascucci}}]{gorti11}
{Gorti}, U., {Hollenbach}, D., {Najita}, J., \& {Pascucci}, I. 2011, \bibinfo{title}{{Emission Lines from the Gas Disk around TW Hydra and the Origin of the Inner Hole},} \apj, 735, 90, \dodoi{10.1088/0004-637X/735/2/90}

\bibitem[{C.~A. {Grady} {et~al.}(2013){Grady}, {Muto}, {Hashimoto}, {Fukagawa}, {Currie}, {Biller}, {Thalmann}, {Sitko}, {Russell}, {Wisniewski}, {Dong}, {Kwon}, {Sai}, {Hornbeck}, {Schneider}, {Hines}, {Moro Mart{\'\i}n}, {Feldt}, {Henning}, {Pott}, {Bonnefoy}, {Bouwman}, {Lacour}, {Mueller}, {Juh{\'a}sz}, {Crida}, {Chauvin}, {Andrews}, {Wilner}, {Kraus}, {Dahm}, {Robitaille}, {Jang-Condell}, {Abe}, {Akiyama}, {Brandner}, {Brandt}, {Carson}, {Egner}, {Follette}, {Goto}, {Guyon}, {Hayano}, {Hayashi}, {Hayashi}, {Hodapp}, {Ishii}, {Iye}, {Janson}, {Kandori}, {Knapp}, {Kudo}, {Kusakabe}, {Kuzuhara}, {Mayama}, {McElwain}, {Matsuo}, {Miyama}, {Morino}, {Nishimura}, {Pyo}, {Serabyn}, {Suto}, {Suzuki}, {Takami}, {Takato}, {Terada}, {Tomono}, {Turner}, {Watanabe}, {Yamada}, {Takami}, {Usuda}, \& {Tamura}}]{grady13}
{Grady}, C.~A., {Muto}, T., {Hashimoto}, J., {et~al.} 2013, \bibinfo{title}{{Spiral Arms in the Asymmetrically Illuminated Disk of MWC 758 and Constraints on Giant Planets},} \apj, 762, 48, \dodoi{10.1088/0004-637X/762/1/48}

\bibitem[{S.~Y. {Haffert} {et~al.}(2019){Haffert}, {Bohn}, {de Boer}, {Snellen}, {Brinchmann}, {Girard}, {Keller}, \& {Bacon}}]{haffert19}
{Haffert}, S.~Y., {Bohn}, A.~J., {de Boer}, J., {et~al.} 2019, \bibinfo{title}{{Two accreting protoplanets around the young star PDS 70},} Nature Astronomy, 3, 749, \dodoi{10.1038/s41550-019-0780-5}

\bibitem[{J. {Hashimoto} {et~al.}(2024){Hashimoto}, {Dong}, {Muto}, {Liu}, \& {Terada}}]{hashimoto24}
{Hashimoto}, J., {Dong}, R., {Muto}, T., {Liu}, H.~B., \& {Terada}, Y. 2024, \bibinfo{title}{{Shadowing in the Protoplanetary Disk of ZZ Tau IRS with HST},} \aj, 167, 75, \dodoi{10.3847/1538-3881/ad1b5e}

\bibitem[{K. {Hirsh} {et~al.}(2020){Hirsh}, {Price}, {Gonzalez}, {Ubeira-Gabellini}, \& {Ragusa}}]{hirsh20}
{Hirsh}, K., {Price}, D.~J., {Gonzalez}, J.-F., {Ubeira-Gabellini}, M.~G., \& {Ragusa}, E. 2020, \bibinfo{title}{{On the cavity size in circumbinary discs},} \mnras, 498, 2936, \dodoi{10.1093/mnras/staa2536}

\bibitem[{J.~D. Hunter(2007)Hunter}]{hunter07}
Hunter, J.~D. 2007, \bibinfo{title}{Matplotlib: A 2D graphics environment,} Computing in Science \& Engineering, 9, 90, \dodoi{10.1109/MCSE.2007.55}

\bibitem[{S. {Hunziker} {et~al.}(2021){Hunziker}, {Schmid}, {Ma}, {Menard}, {Avenhaus}, {Boccaletti}, {Beuzit}, {Chauvin}, {Dohlen}, {Dominik}, {Engler}, {Ginski}, {Gratton}, {Henning}, {Langlois}, {Milli}, {Mouillet}, {Tschudi}, {van Holstein}, \& {Vigan}}]{hunziker21}
{Hunziker}, S., {Schmid}, H.~M., {Ma}, J., {et~al.} 2021, \bibinfo{title}{{HD 142527: quantitative disk polarimetry with SPHERE},} \aap, 648, A110, \dodoi{10.1051/0004-6361/202040166}

\bibitem[{A.~F. {Izquierdo} {et~al.}(2021){Izquierdo}, {Testi}, {Facchini}, {Rosotti}, \& {van Dishoeck}}]{izquierdo21}
{Izquierdo}, A.~F., {Testi}, L., {Facchini}, S., {Rosotti}, G.~P., \& {van Dishoeck}, E.~F. 2021, \bibinfo{title}{{The Disc Miner. I. A statistical framework to detect and quantify kinematical perturbations driven by young planets in discs},} \aap, 650, A179, \dodoi{10.1051/0004-6361/202140779}

\bibitem[{A.~F. {Izquierdo} {et~al.}(2023){Izquierdo}, {Testi}, {Facchini}, {Rosotti}, {van Dishoeck}, {W{\"o}lfer}, \& {Paneque-Carre{\~n}o}}]{izquierdo23}
{Izquierdo}, A.~F., {Testi}, L., {Facchini}, S., {et~al.} 2023, \bibinfo{title}{{The Disc Miner. II. Revealing gas substructures and kinematic signatures from planet-disc interaction through line profile analysis},} \aap, 674, A113, \dodoi{10.1051/0004-6361/202245425}

\bibitem[{A.~F. {Izquierdo} {et~al.}(2025){Izquierdo}, {Stadler}, {Galloway-Sprietsma}, {Benisty}, {Pinte}, {Bae}, {Teague}, {Facchini}, {W{\"o}lfer}, {Longarini}, {Curone}, {Andrews}, {Barraza-Alfaro}, {Cataldi}, {Cuello}, {Czekala}, {Fasano}, {Flock}, {Fukagawa}, {Garg}, {Hall}, {Hammond}, {Hilder}, {Huang}, {Ilee}, {Isella}, {Kanagawa}, {Lesur}, {Lodato}, {Loomis}, {Orihara}, {Price}, {Rosotti}, {Testi}, {Yen}, {Wafflard-Fernandez}, {Wilner}, {Winter}, {Yoshida}, \& {Zawadzki}}]{izquierdo25}
{Izquierdo}, A.~F., {Stadler}, J., {Galloway-Sprietsma}, M., {et~al.} 2025, \bibinfo{title}{{exoALMA. III. Line-intensity Modeling and System Property Extraction from Protoplanetary Disks},} \apjl, 984, L8, \dodoi{10.3847/2041-8213/adc439}

\bibitem[{Y.-F. {Jiang}(2021){Jiang}}]{jiang21}
{Jiang}, Y.-F. 2021, \bibinfo{title}{{An Implicit Finite Volume Scheme to Solve the Time-dependent Radiation Transport Equation Based on Discrete Ordinates},} \apjs, 253, 49, \dodoi{10.3847/1538-4365/abe303}

\bibitem[{Y.-F. {Jiang} {et~al.}(2014){Jiang}, {Stone}, \& {Davis}}]{jiang14}
{Jiang}, Y.-F., {Stone}, J.~M., \& {Davis}, S.~W. 2014, \bibinfo{title}{{An Algorithm for Radiation Magnetohydrodynamics Based on Solving the Time-dependent Transfer Equation},} \apjs, 213, 7, \dodoi{10.1088/0067-0049/213/1/7}

\bibitem[{A. {Juh{\'a}sz} \& S. {Facchini}(2017){Juh{\'a}sz} \& {Facchini}}]{juhasz17}
{Juh{\'a}sz}, A., \& {Facchini}, S. 2017, \bibinfo{title}{{Observational signatures of linear warps in circumbinary discs},} \mnras, 466, 4053, \dodoi{10.1093/mnras/stw3389}

\bibitem[{N. {Kaaz} {et~al.}(2023){Kaaz}, {Liska}, {Jacquemin-Ide}, {Andalman}, {Musoke}, {Tchekhovskoy}, \& {Porth}}]{kaaz23}
{Kaaz}, N., {Liska}, M. T.~P., {Jacquemin-Ide}, J., {et~al.} 2023, \bibinfo{title}{{Nozzle Shocks, Disk Tearing, and Streamers Drive Rapid Accretion in 3D GRMHD Simulations of Warped Thin Disks},} \apj, 955, 72, \dodoi{10.3847/1538-4357/ace051}

\bibitem[{N. {Kaaz} {et~al.}(2025){Kaaz}, {Lithwick}, {Liska}, \& {Tchekhovskoy}}]{kaaz25}
{Kaaz}, N., {Lithwick}, Y., {Liska}, M., \& {Tchekhovskoy}, A. 2025, \bibinfo{title}{{Extreme Scale Height Variations and Nozzle Shocks in Warped Disks},} \apj, 979, 192, \dodoi{10.3847/1538-4357/ad9a85}

\bibitem[{M. {Keppler} {et~al.}(2018){Keppler}, {Benisty}, {M{\"u}ller}, {Henning}, {van Boekel}, {Cantalloube}, {Ginski}, {van Holstein}, {Maire}, {Pohl}, {Samland}, {Avenhaus}, {Baudino}, {Boccaletti}, {de Boer}, {Bonnefoy}, {Chauvin}, {Desidera}, {Langlois}, {Lazzoni}, {Marleau}, {Mordasini}, {Pawellek}, {Stolker}, {Vigan}, {Zurlo}, {Birnstiel}, {Brandner}, {Feldt}, {Flock}, {Girard}, {Gratton}, {Hagelberg}, {Isella}, {Janson}, {Juhasz}, {Kemmer}, {Kral}, {Lagrange}, {Launhardt}, {Matter}, {M{\'e}nard}, {Milli}, {Molli{\`e}re}, {Olofsson}, {P{\'e}rez}, {Pinilla}, {Pinte}, {Quanz}, {Schmidt}, {Udry}, {Wahhaj}, {Williams}, {Buenzli}, {Cudel}, {Dominik}, {Galicher}, {Kasper}, {Lannier}, {Mesa}, {Mouillet}, {Peretti}, {Perrot}, {Salter}, {Sissa}, {Wildi}, {Abe}, {Antichi}, {Augereau}, {Baruffolo}, {Baudoz}, {Bazzon}, {Beuzit}, {Blanchard}, {Brems}, {Buey}, {De Caprio}, {Carbillet}, {Carle}, {Cascone}, {Cheetham}, {Claudi}, {Costille}, {Delboulb{\'e}}, {Dohlen}, {Fantinel}, {Feautrier}, {Fusco}, {Giro},
  {Gluck}, {Gry}, {Hubin}, {Hugot}, {Jaquet}, {Le Mignant}, {Llored}, {Madec}, {Magnard}, {Martinez}, {Maurel}, {Meyer}, {M{\"o}ller-Nilsson}, {Moulin}, {Mugnier}, {Orign{\'e}}, {Pavlov}, {Perret}, {Petit}, {Pragt}, {Puget}, {Rabou}, {Ramos}, {Rigal}, {Rochat}, {Roelfsema}, {Rousset}, {Roux}, {Salasnich}, {Sauvage}, {Sevin}, {Soenke}, {Stadler}, {Suarez}, {Turatto}, \& {Weber}}]{keppler18}
{Keppler}, M., {Benisty}, M., {M{\"u}ller}, A., {et~al.} 2018, \bibinfo{title}{{Discovery of a planetary-mass companion within the gap of the transition disk around PDS 70},} \aap, 617, A44, \dodoi{10.1051/0004-6361/201832957}

\bibitem[{M. {Keppler} {et~al.}(2020){Keppler}, {Penzlin}, {Benisty}, {van Boekel}, {Henning}, {van Holstein}, {Kley}, {Garufi}, {Ginski}, {Brandner}, {Bertrang}, {Boccaletti}, {de Boer}, {Bonavita}, {Brown Sevilla}, {Chauvin}, {Dominik}, {Janson}, {Langlois}, {Lodato}, {Maire}, {M{\'e}nard}, {Pantin}, {Pinte}, {Stolker}, {Szul{\'a}gyi}, {Thebault}, {Villenave}, {Zurlo}, {Rabou}, {Feautrier}, {Feldt}, {Madec}, \& {Wildi}}]{keppler20}
{Keppler}, M., {Penzlin}, A., {Benisty}, M., {et~al.} 2020, \bibinfo{title}{{Gap, shadows, spirals, and streamers: SPHERE observations of binary-disk interactions in GG Tauri A},} \aap, 639, A62, \dodoi{10.1051/0004-6361/202038032}

\bibitem[{C.~N. {Kimmig} \& C.~P. {Dullemond}(2024){Kimmig} \& {Dullemond}}]{kimmig24}
{Kimmig}, C.~N., \& {Dullemond}, C.~P. 2024, \bibinfo{title}{{Warped disk evolution in grid-based simulations},} \aap, 689, A45, \dodoi{10.1051/0004-6361/202348660}

\bibitem[{C.~N. {Kimmig} \& M. {Villenave}(2025){Kimmig} \& {Villenave}}]{kimmig25}
{Kimmig}, C.~N., \& {Villenave}, M. 2025, \bibinfo{title}{{Asymmetric signatures of warps in edge-on disks},} \aap, 698, A146, \dodoi{10.1051/0004-6361/202453313}

\bibitem[{S. {Kraus} {et~al.}(2017){Kraus}, {Kreplin}, {Fukugawa}, {Muto}, {Sitko}, {Young}, {Bate}, {Grady}, {Harries}, {Monnier}, {Willson}, \& {Wisniewski}}]{kraus17}
{Kraus}, S., {Kreplin}, A., {Fukugawa}, M., {et~al.} 2017, \bibinfo{title}{{Dust-trapping Vortices and a Potentially Planet-triggered Spiral Wake in the Pre-transitional Disk of V1247 Orionis},} \apjl, 848, L11, \dodoi{10.3847/2041-8213/aa8edc}

\bibitem[{A. {Krieger} {et~al.}(2024){Krieger}, {Kuffmeier}, {Reissl}, {Dullemond}, {Ginski}, \& {Wolf}}]{krieger24}
{Krieger}, A., {Kuffmeier}, M., {Reissl}, S., {et~al.} 2024, \bibinfo{title}{{Feasibility of detecting shadows in disks induced by infall},} \aap, 686, A111, \dodoi{10.1051/0004-6361/202348354}

\bibitem[{M. {Kuffmeier} {et~al.}(2021){Kuffmeier}, {Dullemond}, {Reissl}, \& {Goicovic}}]{kuffmeier21}
{Kuffmeier}, M., {Dullemond}, C.~P., {Reissl}, S., \& {Goicovic}, F.~G. 2021, \bibinfo{title}{{Misaligned disks induced by infall},} \aap, 656, A161, \dodoi{10.1051/0004-6361/202039614}

\bibitem[{T. {Kutra} {et~al.}(2024){Kutra}, {Wu}, \& {Lithwick}}]{kutra24}
{Kutra}, T., {Wu}, Y., \& {Lithwick}, Y. 2024, \bibinfo{title}{{Irradiated Disks May Settle into Staircases},} \apj, 964, 165, \dodoi{10.3847/1538-4357/ad26e5}

\bibitem[{D. {Lai}(2014){Lai}}]{lai14}
{Lai}, D. 2014, \bibinfo{title}{{Star-disc-binary interactions in protoplanetary disc systems and primordial spin-orbit misalignments},} \mnras, 440, 3532, \dodoi{10.1093/mnras/stu485}

\bibitem[{D. {Lai} \& H. {Zhang}(2008){Lai} \& {Zhang}}]{lai08}
{Lai}, D., \& {Zhang}, H. 2008, \bibinfo{title}{{Wave Excitation in Disks around Rotating Magnetic Stars},} \apj, 683, 949, \dodoi{10.1086/589822}

\bibitem[{C.~J. {Law} {et~al.}(2021){Law}, {Teague}, {Loomis}, {Bae}, {{\"O}berg}, {Czekala}, {Andrews}, {Aikawa}, {Alarc{\'o}n}, {Bergin}, {Bergner}, {Booth}, {Bosman}, {Calahan}, {Cataldi}, {Cleeves}, {Furuya}, {Guzm{\'a}n}, {Huang}, {Ilee}, {Le Gal}, {Liu}, {Long}, {M{\'e}nard}, {Nomura}, {P{\'e}rez}, {Qi}, {Schwarz}, {Soto}, {Tsukagoshi}, {Yamato}, {van't Hoff}, {Walsh}, {Wilner}, \& {Zhang}}]{law21}
{Law}, C.~J., {Teague}, R., {Loomis}, R.~A., {et~al.} 2021, \bibinfo{title}{{Molecules with ALMA at Planet-forming Scales (MAPS). IV. Emission Surfaces and Vertical Distribution of Molecules},} \apjs, 257, 4, \dodoi{10.3847/1538-4365/ac1439}

\bibitem[{G. {Lesur} {et~al.}(2023){Lesur}, {Flock}, {Ercolano}, {Lin}, {Yang}, {Barranco}, {Benitez-Llambay}, {Goodman}, {Johansen}, {Klahr}, {Laibe}, {Lyra}, {Marcus}, {Nelson}, {Squire}, {Simon}, {Turner}, {Umurhan}, \& {Youdin}}]{lesur22}
{Lesur}, G., {Flock}, M., {Ercolano}, B., {et~al.} 2023, in Astronomical Society of the Pacific Conference Series, Vol. 534, Protostars and Planets VII, ed. S.~{Inutsuka}, Y.~{Aikawa}, T.~{Muto}, K.~{Tomida}, \& M.~{Tamura}, 465

\bibitem[{M.-K. {Lin} \& A.~N. {Youdin}(2015){Lin} \& {Youdin}}]{lin15}
{Lin}, M.-K., \& {Youdin}, A.~N. 2015, \bibinfo{title}{{Cooling Requirements for the Vertical Shear Instability in Protoplanetary Disks},} \apj, 811, 17, \dodoi{10.1088/0004-637X/811/1/17}

\bibitem[{G. {Lodato} {et~al.}(2023){Lodato}, {Rampinelli}, {Viscardi}, {Longarini}, {Izquierdo}, {Paneque-Carre{\~n}o}, {Testi}, {Facchini}, {Miotello}, {Veronesi}, \& {Hall}}]{lodato23}
{Lodato}, G., {Rampinelli}, L., {Viscardi}, E., {et~al.} 2023, \bibinfo{title}{{Dynamical mass measurements of two protoplanetary discs},} \mnras, 518, 4481, \dodoi{10.1093/mnras/stac3223}

\bibitem[{C. {Longarini} {et~al.}(2025){Longarini}, {Lodato}, {Rosotti}, {Andrews}, {Winter}, {Stadler}, {Izquierdo}, {Galloway-Sprietsma}, {Facchini}, {Curone}, {Benisty}, {Teague}, {Bae}, {Barraza-Alfaro}, {Cataldi}, {Czekala}, {Cuello}, {Fasano}, {Flock}, {Fukagawa}, {Garg}, {Hall}, {Hammond}, {Hardiman}, {Hilder}, {Huang}, {Ilee}, {Isella}, {Kanagawa}, {Lesur}, {Loomis}, {M{\'e}nard}, {Orihara}, {Pinte}, {Price}, {Testi}, {Fernandez}, {W{\"o}lfer}, {Yen}, {Yoshida}, \& {Zawadzki}}]{longarini25}
{Longarini}, C., {Lodato}, G., {Rosotti}, G., {et~al.} 2025, \bibinfo{title}{{exoALMA. XII. Weighing and Sizing exoALMA Disks with Rotation Curve Modelling},} \apjl, 984, L17, \dodoi{10.3847/2041-8213/adc431}

\bibitem[{S.~H. {Lubow} \& R.~G. {Martin}(2018){Lubow} \& {Martin}}]{lubow18}
{Lubow}, S.~H., \& {Martin}, R.~G. 2018, \bibinfo{title}{{Linear analysis of the evolution of nearly polar low-mass circumbinary discs},} \mnras, 473, 3733, \dodoi{10.1093/mnras/stx2643}

\bibitem[{S.~H. {Lubow} \& G.~I. {Ogilvie}(2000){Lubow} \& {Ogilvie}}]{lubow00}
{Lubow}, S.~H., \& {Ogilvie}, G.~I. 2000, \bibinfo{title}{{On the Tilting of Protostellar Disks by Resonant Tidal Effects},} \apj, 538, 326, \dodoi{10.1086/309101}

\bibitem[{S. {Marino} {et~al.}(2015){Marino}, {Perez}, \& {Casassus}}]{marino15}
{Marino}, S., {Perez}, S., \& {Casassus}, S. 2015, \bibinfo{title}{{Shadows Cast by a Warp in the HD 142527 Protoplanetary Disk},} \apjl, 798, L44, \dodoi{10.1088/2041-8205/798/2/L44}

\bibitem[{P. {Martire} {et~al.}(2024){Martire}, {Longarini}, {Lodato}, {Rosotti}, {Winter}, {Facchini}, {Hardiman}, {Benisty}, {Stadler}, {Izquierdo}, \& {Testi}}]{martire24}
{Martire}, P., {Longarini}, C., {Lodato}, G., {et~al.} 2024, \bibinfo{title}{{Rotation curves in protoplanetary disks with thermal stratification},} arXiv e-prints, arXiv:2402.12236, \dodoi{10.48550/arXiv.2402.12236}

\bibitem[{J.~S. {Mathis} {et~al.}(1977){Mathis}, {Rumpl}, \& {Nordsieck}}]{mathis77}
{Mathis}, J.~S., {Rumpl}, W., \& {Nordsieck}, K.~H. 1977, \bibinfo{title}{{The size distribution of interstellar grains.},} \apj, 217, 425, \dodoi{10.1086/155591}

\bibitem[{S. {Mayama} {et~al.}(2018){Mayama}, {Akiyama}, {Pani{\'c}}, {Miley}, {Tsukagoshi}, {Muto}, {Dong}, {de Leon}, {Mizuki}, {Oh}, {Hashimoto}, {Sai}, {Currie}, {Takami}, {Grady}, {Hayashi}, {Tamura}, \& {Inutsuka}}]{mayama18}
{Mayama}, S., {Akiyama}, E., {Pani{\'c}}, O., {et~al.} 2018, \bibinfo{title}{{ALMA Reveals a Misaligned Inner Gas Disk inside the Large Cavity of a Transitional Disk},} \apjl, 868, L3, \dodoi{10.3847/2041-8213/aae88b}

\bibitem[{M.~J. {McCaughrean} \& C.~R. {O'Dell}(1996){McCaughrean} \& {O'Dell}}]{McCaughrean96}
{McCaughrean}, M.~J., \& {O'Dell}, C.~R. 1996, \bibinfo{title}{{Direct Imaging of Circumstellar Disks in the Orion Nebula},} \aj, 111, 1977, \dodoi{10.1086/117934}

\bibitem[{J.~D. {Melon Fuksman} \& H. {Klahr}(2022){Melon Fuksman} \& {Klahr}}]{melonfuksman22}
{Melon Fuksman}, J.~D., \& {Klahr}, H. 2022, \bibinfo{title}{{No Self-shadowing Instability in 2D Radiation Hydrodynamical Models of Irradiated Protoplanetary Disks},} \apj, 936, 16, \dodoi{10.3847/1538-4357/ac7fee}

\bibitem[{ {Met Office}(2010 - 2015){Met Office}}]{Cartopy}
{Met Office}. 2010 - 2015, Cartopy: a cartographic {Python} library with a {Matplotlib} interface, Exeter, Devon.
\newblock \url{https://scitools.org.uk/cartopy}

\bibitem[{M. {Montesinos} \& N. {Cuello}(2018){Montesinos} \& {Cuello}}]{montesinos18}
{Montesinos}, M., \& {Cuello}, N. 2018, \bibinfo{title}{{Planetary-like spirals caused by moving shadows in transition discs},} \mnras, 475, L35, \dodoi{10.1093/mnrasl/sly001}

\bibitem[{M. {Montesinos} {et~al.}(2016){Montesinos}, {Perez}, {Casassus}, {Marino}, {Cuadra}, \& {Christiaens}}]{montesinos16}
{Montesinos}, M., {Perez}, S., {Casassus}, S., {et~al.} 2016, \bibinfo{title}{{Spiral Waves Triggered by Shadows in Transition Disks},} \apjl, 823, L8, \dodoi{10.3847/2041-8205/823/1/L8}

\bibitem[{D. {Muley} {et~al.}(2024){Muley}, {Melon Fuksman}, \& {Klahr}}]{muley24}
{Muley}, D., {Melon Fuksman}, J.~D., \& {Klahr}, H. 2024, \bibinfo{title}{{Three-temperature radiation hydrodynamics with PLUTO: Thermal and kinematic signatures of accreting protoplanets},} \aap, 687, A213, \dodoi{10.1051/0004-6361/202449739}

\bibitem[{G.~A. {Muro-Arena} {et~al.}(2020){Muro-Arena}, {Benisty}, {Ginski}, {Dominik}, {Facchini}, {Villenave}, {van Boekel}, {Chauvin}, {Garufi}, {Henning}, {Janson}, {Keppler}, {Matter}, {M{\'e}nard}, {Stolker}, {Zurlo}, {Blanchard}, {Maurel}, {Moeller-Nilsson}, {Petit}, {Roux}, {Sevin}, \& {Wildi}}]{muro-arena20}
{Muro-Arena}, G.~A., {Benisty}, M., {Ginski}, C., {et~al.} 2020, \bibinfo{title}{{Shadowing and multiple rings in the protoplanetary disk of HD 139614},} \aap, 635, A121, \dodoi{10.1051/0004-6361/201936509}

\bibitem[{R. {Nealon} {et~al.}(2020{\natexlab{a}}){Nealon}, {Cuello}, \& {Alexander}}]{nealon20}
{Nealon}, R., {Cuello}, N., \& {Alexander}, R. 2020{\natexlab{a}}, \bibinfo{title}{{Flyby-induced misalignments in planet-hosting discs},} \mnras, 491, 4108, \dodoi{10.1093/mnras/stz3186}

\bibitem[{R. {Nealon} {et~al.}(2019){Nealon}, {Pinte}, {Alexander}, {Mentiplay}, \& {Dipierro}}]{nealon19}
{Nealon}, R., {Pinte}, C., {Alexander}, R., {Mentiplay}, D., \& {Dipierro}, G. 2019, \bibinfo{title}{{Scattered light shadows in warped protoplanetary discs},} \mnras, 484, 4951, \dodoi{10.1093/mnras/stz346}

\bibitem[{R. {Nealon} {et~al.}(2020{\natexlab{b}}){Nealon}, {Price}, \& {Pinte}}]{nealon20b}
{Nealon}, R., {Price}, D.~J., \& {Pinte}, C. 2020{\natexlab{b}}, \bibinfo{title}{{Rocking shadows in broken circumbinary discs},} \mnras, 493, L143, \dodoi{10.1093/mnrasl/slaa026}

\bibitem[{Y. {Ohta} {et~al.}(2016){Ohta}, {Fukagawa}, {Sitko}, {Muto}, {Kraus}, {Grady}, {Wisniewski}, {Swearingen}, {Shibai}, {Sumi}, {Hashimoto}, {Kudo}, {Kusakabe}, {Momose}, {Okamoto}, {Kotani}, {Takami}, {Currie}, {Thalmann}, {Janson}, {Akiyama}, {Follette}, {Mayama}, {Abe}, {Brandner}, {Brandt}, {Carson}, {Egner}, {Feldt}, {Goto}, {Guyon}, {Hayano}, {Hayashi}, {Hayashi}, {Henning}, {Hodapp}, {Ishii}, {Iye}, {Kandori}, {Knapp}, {Kuzuhara}, {Kwon}, {Matsuo}, {McElwain}, {Miyama}, {Morino}, {Moro-Mart{\'\i}n}, {Nishimura}, {Pyo}, {Serabyn}, {Suenaga}, {Suto}, {Suzuki}, {Takahashi}, {Takami}, {Takato}, {Terada}, {Tomono}, {Turner}, {Usuda}, {Watanabe}, {Yamada}, \& {Tamura}}]{ohta16}
{Ohta}, Y., {Fukagawa}, M., {Sitko}, M.~L., {et~al.} 2016, \bibinfo{title}{{Extreme asymmetry in the polarized disk of V1247 Orionis*},} \pasj, 68, 53, \dodoi{10.1093/pasj/psw051}

\bibitem[{S. {Okuzumi} {et~al.}(2022){Okuzumi}, {Ueda}, \& {Turner}}]{okuzumi22}
{Okuzumi}, S., {Ueda}, T., \& {Turner}, N.~J. 2022, \bibinfo{title}{{A global two-layer radiative transfer model for axisymmetric, shadowed protoplanetary disks},} \pasj, 74, 828, \dodoi{10.1093/pasj/psac040}

\bibitem[{R. {Orihara} \& M. {Momose}(2025){Orihara} \& {Momose}}]{orihara25}
{Orihara}, R., \& {Momose}, M. 2025, \bibinfo{title}{{Shadow-based Framework for Estimating Transition Disk Geometries},} \apj, 986, 215, \dodoi{10.3847/1538-4357/add890}

\bibitem[{J.~C.~B. {Papaloizou} \& D.~N.~C. {Lin}(1995){Papaloizou} \& {Lin}}]{papaloizou95}
{Papaloizou}, J.~C.~B., \& {Lin}, D.~N.~C. 1995, \bibinfo{title}{{On the Dynamics of Warped Accretion Disks},} \apj, 438, 841, \dodoi{10.1086/175127}

\bibitem[{Y.~N. {Pavlyuchenkov} {et~al.}(2022{\natexlab{a}}){Pavlyuchenkov}, {Maksimova}, \& {Akimkin}}]{pavlyuchenkov22a}
{Pavlyuchenkov}, Y.~N., {Maksimova}, L.~A., \& {Akimkin}, V.~V. 2022{\natexlab{a}}, \bibinfo{title}{{Simulation of Thermal Surface Waves in a Protoplanetary Disk in 1+1D Approximation},} Astronomy Reports, 66, 321, \dodoi{10.1134/S1063772922050055}

\bibitem[{Y.~N. {Pavlyuchenkov} {et~al.}(2022{\natexlab{b}}){Pavlyuchenkov}, {Maksimova}, \& {Akimkin}}]{pavlyuchenkov22b}
{Pavlyuchenkov}, Y.~N., {Maksimova}, L.~A., \& {Akimkin}, V.~V. 2022{\natexlab{b}}, \bibinfo{title}{{Simulation of Thermal Surface Waves in a Protoplanetary Disk in a Two-Dimensional Approximation},} Astronomy Reports, 66, 800, \dodoi{10.1134/S1063772922100110}

\bibitem[{P. {Pinilla} {et~al.}(2015){Pinilla}, {de Boer}, {Benisty}, {Juh{\'a}sz}, {de Juan Ovelar}, {Dominik}, {Avenhaus}, {Birnstiel}, {Girard}, {Huelamo}, {Isella}, \& {Milli}}]{pinilla15}
{Pinilla}, P., {de Boer}, J., {Benisty}, M., {et~al.} 2015, \bibinfo{title}{{Variability and dust filtration in the transition disk J160421.7-213028 observed in optical scattered light},} \aap, 584, L4, \dodoi{10.1051/0004-6361/201526981}

\bibitem[{P. {Pinilla} {et~al.}(2018){Pinilla}, {Benisty}, {de Boer}, {Manara}, {Bouvier}, {Dominik}, {Ginski}, {Loomis}, \& {Sicilia Aguilar}}]{pinilla18}
{Pinilla}, P., {Benisty}, M., {de Boer}, J., {et~al.} 2018, \bibinfo{title}{{Variable Outer Disk Shadowing around the Dipper Star RXJ1604.3-2130},} \apj, 868, 85, \dodoi{10.3847/1538-4357/aae824}

\bibitem[{C. {Pinte} {et~al.}(2018){Pinte}, {M{\'e}nard}, {Duch{\^e}ne}, {Hill}, {Dent}, {Woitke}, {Maret}, {van der Plas}, {Hales}, {Kamp}, {Thi}, {de Gregorio-Monsalvo}, {Rab}, {Quanz}, {Avenhaus}, {Carmona}, \& {Casassus}}]{pinte18}
{Pinte}, C., {M{\'e}nard}, F., {Duch{\^e}ne}, G., {et~al.} 2018, \bibinfo{title}{{Direct mapping of the temperature and velocity gradients in discs. Imaging the vertical CO snow line around IM Lupi},} \aap, 609, A47, \dodoi{10.1051/0004-6361/201731377}

\bibitem[{J.~E. {Pringle}(1996){Pringle}}]{pringle96}
{Pringle}, J.~E. 1996, \bibinfo{title}{{Self-induced warping of accretion discs},} \mnras, 281, 357, \dodoi{10.1093/mnras/281.1.357}

\bibitem[{Y. {Qian} \& Y. {Wu}(2024){Qian} \& {Wu}}]{qian24}
{Qian}, Y., \& {Wu}, Y. 2024, \bibinfo{title}{{Shadows Wreak Havocs in Transition Disks},} arXiv e-prints, arXiv:2407.09613, \dodoi{10.48550/arXiv.2407.09613}

\bibitem[{I. {Rabago} {et~al.}(2024){Rabago}, {Zhu}, {Lubow}, \& {Martin}}]{rabago24}
{Rabago}, I., {Zhu}, Z., {Lubow}, S., \& {Martin}, R.~G. 2024, \bibinfo{title}{{Warps and breaks in circumbinary discs},} \mnras, \dodoi{10.1093/mnras/stae1787}

\bibitem[{B. {Ren} {et~al.}(2018){Ren}, {Dong}, {Esposito}, {Pueyo}, {Debes}, {Poteet}, {Choquet}, {Benisty}, {Chiang}, {Grady}, {Hines}, {Schneider}, \& {Soummer}}]{ren18}
{Ren}, B., {Dong}, R., {Esposito}, T.~M., {et~al.} 2018, \bibinfo{title}{{A Decade of MWC 758 Disk Images: Where Are the Spiral-arm-driving Planets?},} \apjl, 857, L9, \dodoi{10.3847/2041-8213/aab7f5}

\bibitem[{E.~A. {Rich} {et~al.}(2019){Rich}, {Wisniewski}, {Currie}, {Fukagawa}, {Grady}, {Sitko}, {Pikhartova}, {Hashimoto}, {Abe}, {Brandner}, {Brandt}, {Carson}, {Chilcote}, {Dong}, {Feldt}, {Goto}, {Groff}, {Guyon}, {Hayano}, {Hayashi}, {Hayashi}, {Henning}, {Hodapp}, {Ishii}, {Iye}, {Janson}, {Jovanovic}, {Kandori}, {Kasdin}, {Knapp}, {Kudo}, {Kusakabe}, {Kuzuhara}, {Kwon}, {Lozi}, {Martinache}, {Matsuo}, {Mayama}, {McElwain}, {Miyama}, {Morino}, {Moro-Martin}, {Nakagawa}, {Nishimura}, {Pyo}, {Serabyn}, {Suto}, {Russel}, {Suzuki}, {Takami}, {Takato}, {Terada}, {Thalmann}, {Turner}, {Uyama}, {Wagner}, {Watanabe}, {Yamada}, {Takami}, {Usuda}, \& {Tamura}}]{rich19}
{Rich}, E.~A., {Wisniewski}, J.~P., {Currie}, T., {et~al.} 2019, \bibinfo{title}{{Multi-epoch Direct Imaging and Time-variable Scattered Light Morphology of the HD 163296 Protoplanetary Disk},} \apj, 875, 38, \dodoi{10.3847/1538-4357/ab0f3b}

\bibitem[{G.~P. {Rosotti} {et~al.}(2025){Rosotti}, {Longarini}, {Paneque-Carre{\~n}o}, {Cataldi}, {Galloway-Sprietsma}, {Andrews}, {Bae}, {Barraza-Alfaro}, {Benisty}, {Curone}, {Czekala}, {Facchini}, {Fasano}, {Flock}, {Fukagawa}, {Garg}, {Hall}, {Huang}, {Ilee}, {Izquierdo}, {Kanagawa}, {Lesur}, {Lodato}, {Loomis}, {Orihara}, {Pinte}, {Price}, {Stadler}, {Teague}, {Fernandez}, {Winter}, {W{\"o}lfer}, {Yen}, {Yoshida}, \& {Zawadzki}}]{rosotti25}
{Rosotti}, G.~P., {Longarini}, C., {Paneque-Carre{\~n}o}, T., {et~al.} 2025, \bibinfo{title}{{exoALMA. XV. Interpreting the Height of CO Emission Layer},} \apjl, 984, L20, \dodoi{10.3847/2041-8213/adc42e}

\bibitem[{B.~S. {Safonov} {et~al.}(2022){Safonov}, {Strakhov}, {Goliguzova}, \& {Voziakova}}]{safonov22}
{Safonov}, B.~S., {Strakhov}, I.~A., {Goliguzova}, M.~V., \& {Voziakova}, O.~V. 2022, \bibinfo{title}{{Apparent Motion of the Circumstellar Envelope of CQ Tau in Scattered Light},} \aj, 163, 31, \dodoi{10.3847/1538-3881/ac36cb}

\bibitem[{B.~R. {Setterholm} {et~al.}(2025){Setterholm}, {Monnier}, {Baron}, {Bae}, {Kluska}, {Kraus}, {Calvet}, {Ibrahim}, {Rich}, {Anugu}, {Davies}, {Ennis}, {Gardner}, {Labdon}, {Lanthermann}, \& {Schaefer}}]{setterholm25}
{Setterholm}, B.~R., {Monnier}, J.~D., {Baron}, F., {et~al.} 2025, \bibinfo{title}{{The Dynamic Inner Disk of a Planet-forming Star},} \aj, 169, 318, \dodoi{10.3847/1538-3881/adcd68}

\bibitem[{A. {Sicilia-Aguilar} {et~al.}(2020){Sicilia-Aguilar}, {Manara}, {de Boer}, {Benisty}, {Pinilla}, \& {Bouvier}}]{sicilia-aguilar20}
{Sicilia-Aguilar}, A., {Manara}, C.~F., {de Boer}, J., {et~al.} 2020, \bibinfo{title}{{Time-resolved photometry of the young dipper RX J1604.3-2130A. Unveiling the structure and mass transport through the innermost disk},} \aap, 633, A37, \dodoi{10.1051/0004-6361/201936565}

\bibitem[{J.~L. {Smallwood} {et~al.}(2024){Smallwood}, {Nealon}, {Cuello}, {Dong}, \& {Booth}}]{smallwood24}
{Smallwood}, J.~L., {Nealon}, R., {Cuello}, N., {Dong}, R., \& {Booth}, R.~A. 2024, \bibinfo{title}{{Formation of misaligned second-generation discs through fly-by encounters},} \mnras, 527, 2094, \dodoi{10.1093/mnras/stad3057}

\bibitem[{T. {Stolker} {et~al.}(2016){Stolker}, {Dominik}, {Avenhaus}, {Min}, {de Boer}, {Ginski}, {Schmid}, {Juhasz}, {Bazzon}, {Waters}, {Garufi}, {Augereau}, {Benisty}, {Boccaletti}, {Henning}, {Langlois}, {Maire}, {M{\'e}nard}, {Meyer}, {Pinte}, {Quanz}, {Thalmann}, {Beuzit}, {Carbillet}, {Costille}, {Dohlen}, {Feldt}, {Gisler}, {Mouillet}, {Pavlov}, {Perret}, {Petit}, {Pragt}, {Rochat}, {Roelfsema}, {Salasnich}, {Soenke}, \& {Wildi}}]{stolker16}
{Stolker}, T., {Dominik}, C., {Avenhaus}, H., {et~al.} 2016, \bibinfo{title}{{Shadows cast on the transition disk of HD 135344B. Multiwavelength VLT/SPHERE polarimetric differential imaging},} \aap, 595, A113, \dodoi{10.1051/0004-6361/201528039}

\bibitem[{J.~M. {Stone} {et~al.}(2020){Stone}, {Tomida}, {White}, \& {Felker}}]{stone20}
{Stone}, J.~M., {Tomida}, K., {White}, C.~J., \& {Felker}, K.~G. 2020, \bibinfo{title}{{The Athena++ Adaptive Mesh Refinement Framework: Design and Magnetohydrodynamic Solvers},} \apjs, 249, 4, \dodoi{10.3847/1538-4365/ab929b}

\bibitem[{Z. {Su} \& X.-N. {Bai}(2024){Su} \& {Bai}}]{su24}
{Su}, Z., \& {Bai}, X.-N. 2024, \bibinfo{title}{{Dynamical Consequence of Shadows Cast to the Outer Protoplanetary Disks: I. Two-dimensional Simulations},} arXiv e-prints, arXiv:2407.12659, \dodoi{10.48550/arXiv.2407.12659}

\bibitem[{H. {Tanaka} \& K. {Okada}(2024){Tanaka} \& {Okada}}]{tanaka24}
{Tanaka}, H., \& {Okada}, K. 2024, \bibinfo{title}{{Three-dimensional Interaction between a Planet and an Isothermal Gaseous Disk. III. Locally Isothermal Cases},} \apj, 968, 28, \dodoi{10.3847/1538-4357/ad410d}

\bibitem[{H. {Tanaka} \& W.~R. {Ward}(2004){Tanaka} \& {Ward}}]{tanaka04}
{Tanaka}, H., \& {Ward}, W.~R. 2004, \bibinfo{title}{{Three-dimensional Interaction between a Planet and an Isothermal Gaseous Disk. II. Eccentricity Waves and Bending Waves},} \apj, 602, 388, \dodoi{10.1086/380992}

\bibitem[{R. {Teague} {et~al.}(2022{\natexlab{a}}){Teague}, {Bae}, {Benisty}, {Andrews}, {Facchini}, {Huang}, \& {Wilner}}]{teague22}
{Teague}, R., {Bae}, J., {Benisty}, M., {et~al.} 2022{\natexlab{a}}, \bibinfo{title}{{Gas and Dust Shadows in the TW Hydrae Disk},} \apj, 930, 144, \dodoi{10.3847/1538-4357/ac67a3}

\bibitem[{R. {Teague} {et~al.}(2022{\natexlab{b}}){Teague}, {Law}, {Huang}, \& {Meng}}]{disksurf}
{Teague}, R., {Law}, C.~J., {Huang}, J., \& {Meng}, F. 2022{\natexlab{b}}, \bibinfo{title}{{disksurf: Measure the molecular emission surface of protoplanetary disks},}, Astrophysics Source Code Library, record ascl:2207.028 \doeprint{2207.028}

\bibitem[{R. {Teague} {et~al.}(2022{\natexlab{c}}){Teague}, {Bae}, {Andrews}, {Benisty}, {Bergin}, {Facchini}, {Huang}, {Longarini}, \& {Wilner}}]{teague22b}
{Teague}, R., {Bae}, J., {Andrews}, S.~M., {et~al.} 2022{\natexlab{c}}, \bibinfo{title}{{Mapping the Complex Kinematic Substructure in the TW Hya Disk},} \apj, 936, 163, \dodoi{10.3847/1538-4357/ac88ca}

\bibitem[{R. {Teague} {et~al.}(2025){Teague}, {Benisty}, {Facchini}, {Fukagawa}, {Pinte}, {Andrews}, {Bae}, {Barraza-Alfaro}, {Cataldi}, {Cuello}, {Curone}, {Czekala}, {Fasano}, {Flock}, {Galloway-Sprietsma}, {Garg}, {Hall}, {Hammond}, {Hilder}, {Huang}, {Ilee}, {Izquierdo}, {Kanagawa}, {Lesur}, {Lodato}, {Longarini}, {Loomis}, {Masset}, {Menard}, {Orihara}, {Price}, {Rosotti}, {Stadler}, {Testi}, {Yen}, {Wafflard-Fernandez}, {Wilner}, {Winter}, {W{\"o}lfer}, {Yoshida}, \& {Zawadzki}}]{teague25}
{Teague}, R., {Benisty}, M., {Facchini}, S., {et~al.} 2025, \bibinfo{title}{{exoALMA. I. Science Goals, Project Design, and Data Products},} \apjl, 984, L6, \dodoi{10.3847/2041-8213/adc43b}

\bibitem[{C.~E.~J.~M.~L.~J. {Terquem}(1998){Terquem}}]{terquem98}
{Terquem}, C. E.~J.~M.~L.~J. 1998, \bibinfo{title}{{The Response of Accretion Disks to Bending Waves: Angular Momentum Transport and Resonances},} \apj, 509, 819, \dodoi{10.1086/306505}

\bibitem[{L. {Trapman} {et~al.}(2025){Trapman}, {Zhang}, {Rosotti}, {Pinilla}, {Tabone}, {Pascucci}, {Agurto-Gangas}, {Anania}, {Carpenter}, {Cieza}, {Deng}, {Gonz{\'a}lez-Ruilova}, {Hogerheijde}, {Kurtovic}, {Kuznetsova}, {Miley}, {P{\'e}rez}, {Ruiz-Rodriguez}, {Schwarz}, {Sierra}, {TorresVillanueva}, \& {Vioque}}]{trapman25}
{Trapman}, L., {Zhang}, K., {Rosotti}, G.~P., {et~al.} 2025, \bibinfo{title}{{The ALMA Survey of Gas Evolution of PROtoplanetary Disks (AGE-PRO). V. Protoplanetary Gas Disk Masses},} \apj, 989, 5, \dodoi{10.3847/1538-4357/adcd6e}

\bibitem[{T. {Ueda} {et~al.}(2021){Ueda}, {Flock}, \& {Birnstiel}}]{ueda21b}
{Ueda}, T., {Flock}, M., \& {Birnstiel}, T. 2021, \bibinfo{title}{{Thermal Wave Instability as an Origin of Gap and Ring Structures in Protoplanetary Disks},} \apjl, 914, L38, \dodoi{10.3847/2041-8213/ac0631}

\bibitem[{T. {Uyama} {et~al.}(2020){Uyama}, {Muto}, {Mawet}, {Christiaens}, {Hashimoto}, {Kudo}, {Kuzuhara}, {Ruane}, {Beichman}, {Absil}, {Akiyama}, {Bae}, {Bottom}, {Choquet}, {Currie}, {Dong}, {Follette}, {Fukagawa}, {Guidi}, {Huby}, {Kwon}, {Mayama}, {Meshkat}, {Reggiani}, {Ricci}, {Serabyn}, {Tamura}, {Testi}, {Wallack}, {Williams}, \& {Zhu}}]{uyama20}
{Uyama}, T., {Muto}, T., {Mawet}, D., {et~al.} 2020, \bibinfo{title}{{Near-infrared Imaging of a Spiral in the CQ Tau Disk},} \aj, 159, 118, \dodoi{10.3847/1538-3881/ab7006}

\bibitem[{N. {van der Marel}(2023){van der Marel}}]{vandermarel23}
{van der Marel}, N. 2023, \bibinfo{title}{{Transition disks: the observational revolution from SEDs to imaging},} European Physical Journal Plus, 138, 225, \dodoi{10.1140/epjp/s13360-022-03628-0}

\bibitem[{M. {Villenave} {et~al.}(2024){Villenave}, {Stapelfeldt}, {Duch{\^e}ne}, {M{\'e}nard}, {Wolff}, {Perrin}, {Pinte}, {Tazaki}, \& {Padgett}}]{villenave24}
{Villenave}, M., {Stapelfeldt}, K.~R., {Duch{\^e}ne}, G., {et~al.} 2024, \bibinfo{title}{{JWST Imaging of Edge-on Protoplanetary Disks. II. Appearance of Edge-on Disks with a Tilted Inner Region: Case Study of IRAS04302+2247},} \apj, 961, 95, \dodoi{10.3847/1538-4357/ad0c4b}

\bibitem[{P. Virtanen {et~al.}(2020)Virtanen, Gommers, Oliphant, Haberland, Reddy, Cournapeau, Burovski, Peterson, Weckesser, Bright, {van der Walt}, Brett, Wilson, Millman, Mayorov, Nelson, Jones, Kern, Larson, Carey, Polat, Feng, Moore, {VanderPlas}, Laxalde, Perktold, Cimrman, Henriksen, Quintero, Harris, Archibald, Ribeiro, Pedregosa, {van Mulbregt}, \& {SciPy 1.0 Contributors}}]{2020SciPy-NMeth}
Virtanen, P., Gommers, R., Oliphant, T.~E., {et~al.} 2020, \bibinfo{title}{{{SciPy} 1.0: Fundamental Algorithms for Scientific Computing in Python},} Nature Methods, 17, 261, \dodoi{10.1038/s41592-019-0686-2}

\bibitem[{S.-i. {Watanabe} \& D.~N.~C. {Lin}(2008){Watanabe} \& {Lin}}]{watanabe08}
{Watanabe}, S.-i., \& {Lin}, D.~N.~C. 2008, \bibinfo{title}{{Thermal Waves in Irradiated Protoplanetary Disks},} \apj, 672, 1183, \dodoi{10.1086/523347}

\bibitem[{J.~P. {Williams} {et~al.}(2025){Williams}, {Benisty}, {Ginski}, {Lodato}, \& {Vincent}}]{williams25}
{Williams}, J.~P., {Benisty}, M., {Ginski}, C., {Lodato}, G., \& {Vincent}, M. 2025, \bibinfo{title}{{Radiative Transfer Modeling of a Shadowed Protoplanetary Disk Assisted by a Neural Network},} \apj, 991, 176, \dodoi{10.3847/1538-4357/adfa15}

\bibitem[{A.~J. {Winter} {et~al.}(2025){Winter}, {Benisty}, {Izquierdo}, {Lodato}, {Teague}, {Kimmig}, {Andrews}, {Bae}, {Barraza-Alfaro}, {Cuello}, {Curone}, {Czekala}, {Facchini}, {Fasano}, {Hall}, {Hardiman}, {Hilder}, {Ilee}, {Fukagawa}, {Longarini}, {M{\'e}nard}, {Orihara}, {Pinte}, {Price}, {Rosotti}, {Stadler}, {Wilner}, {W{\"o}lfer}, {Yen}, {Yoshida}, \& {Zawadzki}}]{winter25}
{Winter}, A.~J., {Benisty}, M., {Izquierdo}, A.~F., {et~al.} 2025, \bibinfo{title}{{exoALMA. XVIII. Interpreting large scale kinematic structures as moderate warping},} arXiv e-prints, arXiv:2507.11669, \dodoi{10.48550/arXiv.2507.11669}

\bibitem[{S.~G. {Wolff} {et~al.}(2016){Wolff}, {Perrin}, {Millar-Blanchaer}, {Nielsen}, {Wang}, {Cardwell}, {Chilcote}, {Dong}, {Draper}, {Duch{\^e}ne}, {Fitzgerald}, {Goodsell}, {Grady}, {Graham}, {Greenbaum}, {Hartung}, {Hibon}, {Hines}, {Hung}, {Kalas}, {Macintosh}, {Marchis}, {Marois}, {Pueyo}, {Rantakyr{\"o}}, {Schneider}, {Sivaramakrishnan}, \& {Wiktorowicz}}]{wolff16}
{Wolff}, S.~G., {Perrin}, M., {Millar-Blanchaer}, M.~A., {et~al.} 2016, \bibinfo{title}{{The PDS 66 Circumstellar Disk as Seen in Polarized Light with the Gemini Planet Imager},} \apjl, 818, L15, \dodoi{10.3847/2041-8205/818/1/L15}

\bibitem[{K. {Wood} \& B. {Whitney}(1998){Wood} \& {Whitney}}]{wood98}
{Wood}, K., \& {Whitney}, B. 1998, \bibinfo{title}{{Scattered Light Signatures of Magnetic Accretion in Classical T Tauri Stars},} \apjl, 506, L43, \dodoi{10.1086/311643}

\bibitem[{Y. {Wu} \& Y. {Lithwick}(2021){Wu} \& {Lithwick}}]{wu21}
{Wu}, Y., \& {Lithwick}, Y. 2021, \bibinfo{title}{{The Irradiation Instability of Protoplanetary Disks},} \apj, 923, 123, \dodoi{10.3847/1538-4357/ac2b9c}

\bibitem[{A.~K. {Young} {et~al.}(2023){Young}, {Stevenson}, {Nixon}, \& {Rice}}]{young23}
{Young}, A.~K., {Stevenson}, S., {Nixon}, C.~J., \& {Rice}, K. 2023, \bibinfo{title}{{On the conditions for warping and breaking protoplanetary discs},} \mnras, 525, 2616, \dodoi{10.1093/mnras/stad2451}

\bibitem[{H.-G. {Yun} {et~al.}(2025{\natexlab{a}}){Yun}, {Kim}, {Bae}, \& {Han}}]{yun25a}
{Yun}, H.-G., {Kim}, W.-T., {Bae}, J., \& {Han}, C. 2025{\natexlab{a}}, \bibinfo{title}{{Vertical Shear Instability in Thermally Stratified Protoplanetary Disks. I. A Linear Stability Analysis},} \apj, 980, 14, \dodoi{10.3847/1538-4357/ad9f41}

\bibitem[{H.-G. {Yun} {et~al.}(2025{\natexlab{b}}){Yun}, {Kim}, {Bae}, \& {Han}}]{yun25b}
{Yun}, H.-G., {Kim}, W.-T., {Bae}, J., \& {Han}, C. 2025{\natexlab{b}}, \bibinfo{title}{{Vertical Shear Instability in Thermally Stratified Protoplanetary Disks. II. Hydrodynamic Simulations and Observability},} \apj, 980, 15, \dodoi{10.3847/1538-4357/ad9f42}

\bibitem[{F. {Zagaria} {et~al.}(2025){Zagaria}, {Jiang}, {Cataldi}, {Facchini}, {Benisty}, {Aikawa}, {Andrews}, {Bae}, {Barraza-Alfaro}, {Curone}, {Czekala}, {Fasano}, {Hall}, {Hammond}, {Huang}, {Ilee}, {Izquierdo}, {Lawrence}, {Lodato}, {M{\'e}nard}, {Pinte}, {Rosotti}, {Stadler}, {Teague}, {Testi}, {Wilner}, {Winter}, \& {Yoshida}}]{zagaria25}
{Zagaria}, F., {Jiang}, H., {Cataldi}, G., {et~al.} 2025, \bibinfo{title}{{SO Emission in the Dynamically Perturbed Protoplanetary Disks around CQ Tau and MWC 758},} \apj, 989, 30, \dodoi{10.3847/1538-4357/ade683}

\bibitem[{H. {Zhang} \& D. {Lai}(2006){Zhang} \& {Lai}}]{zhang06}
{Zhang}, H., \& {Lai}, D. 2006, \bibinfo{title}{{Wave excitation in three-dimensional discs by external potential},} \mnras, 368, 917, \dodoi{10.1111/j.1365-2966.2006.10167.x}

\bibitem[{K. {Zhang} {et~al.}(2025){Zhang}, {P{\'e}rez}, {Pascucci}, {Pinilla}, {Cieza}, {Carpenter}, {Trapman}, {Deng}, {Agurto-Gangas}, {Sierra}, {Kurtovic}, {Ruiz-Rodriguez}, {Vioque}, {Miley}, {Tabone}, {Gonz{\'a}lez-Ruilova}, {Anania}, {Rosotti}, {TorresVillanueva}, {Hogerheijde}, {Schwarz}, \& {Kuznetsova}}]{zhangk25}
{Zhang}, K., {P{\'e}rez}, L.~M., {Pascucci}, I., {et~al.} 2025, \bibinfo{title}{{The ALMA Survey of Gas Evolution of PROtoplanetary Disks (AGE-PRO). I. Program Overview and Summary of First Results},} \apj, 989, 1, \dodoi{10.3847/1538-4357/addebe}

\bibitem[{S. {Zhang} \& Z. {Zhu}(2024){Zhang} \& {Zhu}}]{zhang24b}
{Zhang}, S., \& {Zhu}, Z. 2024, \bibinfo{title}{{3D Radiation-hydrodynamical Simulations of Shadows on Transition Disks},} \apjl, 974, L38, \dodoi{10.3847/2041-8213/ad815f}

\bibitem[{S. {Zhang} {et~al.}(2024){Zhang}, {Zhu}, \& {Jiang}}]{zhang24}
{Zhang}, S., {Zhu}, Z., \& {Jiang}, Y.-F. 2024, \bibinfo{title}{{Thermal Structure Determines Kinematics: Vertical Shear Instability in Stellar Irradiated Protoplanetary Disks},} \apj, 968, 29, \dodoi{10.3847/1538-4357/ad4109}

\bibitem[{Z. {Zhu}(2019){Zhu}}]{zhu19}
{Zhu}, Z. 2019, \bibinfo{title}{{Inclined massive planets in a protoplanetary disc: gap opening, disc breaking, and observational signatures},} \mnras, 483, 4221, \dodoi{10.1093/mnras/sty3358}

\bibitem[{Z. {Zhu} {et~al.}(2015){Zhu}, {Dong}, {Stone}, \& {Rafikov}}]{zhu15}
{Zhu}, Z., {Dong}, R., {Stone}, J.~M., \& {Rafikov}, R.~R. 2015, \bibinfo{title}{{The Structure of Spiral Shocks Excited by Planetary-mass Companions},} \apj, 813, 88, \dodoi{10.1088/0004-637X/813/2/88}

\bibitem[{Z. {Zhu} {et~al.}(2012){Zhu}, {Nelson}, {Dong}, {Espaillat}, \& {Hartmann}}]{zhu12}
{Zhu}, Z., {Nelson}, R.~P., {Dong}, R., {Espaillat}, C., \& {Hartmann}, L. 2012, \bibinfo{title}{{Dust Filtration by Planet-induced Gap Edges: Implications for Transitional Disks},} \apj, 755, 6, \dodoi{10.1088/0004-637X/755/1/6}

\bibitem[{Z. {Zhu} {et~al.}(2025){Zhu}, {Zhang}, \& {Johnson}}]{zhu25}
{Zhu}, Z., {Zhang}, S., \& {Johnson}, T.~M. 2025, \bibinfo{title}{{Asymmetric Temperature Variations In Protoplanetary Disks. I. Linear Theory, Corotating Spirals, and Ring Formation},} \apj, 980, 259, \dodoi{10.3847/1538-4357/adae0d}

\bibitem[{A. {Ziampras} {et~al.}(2025{\natexlab{a}}){Ziampras}, {Cordwell}, {Rafikov}, \& {Nelson}}]{ziampras25b}
{Ziampras}, A., {Cordwell}, A.~J., {Rafikov}, R.~R., \& {Nelson}, R.~P. 2025{\natexlab{a}}, \bibinfo{title}{{How two-dimensional are planet-disc interactions? II. Radiation hydrodynamics and suitable cooling prescriptions},} arXiv e-prints, arXiv:2509.20464, \dodoi{10.48550/arXiv.2509.20464}

\bibitem[{A. {Ziampras} {et~al.}(2025{\natexlab{b}}){Ziampras}, {Dullemond}, {Birnstiel}, {Benisty}, \& {Nelson}}]{ziampras25}
{Ziampras}, A., {Dullemond}, C.~P., {Birnstiel}, T., {Benisty}, M., \& {Nelson}, R.~P. 2025{\natexlab{b}}, \bibinfo{title}{{Spirals, rings, and vortices shaped by shadows in protoplanetary discs: from radiative hydrodynamical simulations to observable signatures},} \mnras, 540, 1185, \dodoi{10.1093/mnras/staf785}

\end{thebibliography}
\end{document}